\documentclass[floats,prb,twocolumn]{revtex4}
\usepackage{enumerate}
\usepackage{xcolor}
\definecolor{DarkGray}{rgb}{0.1,0.1,0.5}
\usepackage[colorlinks=true,breaklinks, linkcolor= DarkGray,citecolor= DarkGray,urlcolor= DarkGray]{hyperref}
\usepackage{url}
\usepackage{verbatim}
\usepackage{amsmath}
\usepackage{latexsym}
\usepackage{revsymb}
\usepackage{epsfig}
\usepackage{pstricks}
\usepackage{ifthen}
\usepackage{mathrsfs}

\usepackage{natbib}
\usepackage{amsfonts}
\usepackage{amsmath}
\usepackage{amssymb}
\usepackage{subfigure}
\usepackage{amsthm}
\usepackage{graphicx}
\newcommand{\appref}[1]{\hyperref[#1]{{Appendix~\ref*{#1}}}}
\newcommand{\be}{\begin{eqnarray} \begin{aligned}}
\newcommand{\ee}{\end{aligned} \end{eqnarray} }
\newcommand{\benn}{\begin{eqnarray*} \begin{aligned}}
\newcommand{\eenn}{\end{aligned} \end{eqnarray*} }

\newcommand*{\textfrac}[2]{{{#1}/{#2}}}

\newcommand*{\cE}{\mathcal{E}}
\newcommand*{\cF}{\mathcal{F}}

\newcommand*{\cH}{\mathcal{H}}

\newcommand*{\cD}{\mathscr{D}}

\newcommand*{\End}{\mathsf{End}}

\newcommand*{\tr}{\mathop{\mathrm{tr}}\nolimits}

\newcommand{\bc}{\begin{center}}
\newcommand{\ec}{\end{center}}

\newcommand{\id}{\mathbb{I}}

\usepackage{amsfonts}

\def\id{\mathbb{I}}

\def\01{\{0,1\}}

\newcommand{\ket}[1]{|#1\rangle}
\newcommand{\bra}[1]{\langle#1|}

\newcommand{\proj}[1]{|#1\rangle\langle#1|}

\newcommand*{\Xsingleoperator}{{\bf O}}

\newcommand*{\Xisometry}{{\bf W}}
\newcommand*{\Xunitary}{{\bf U}}

\newcommand*{\merafirst}{\raisebox{-85.626pt}{\epsfig{file=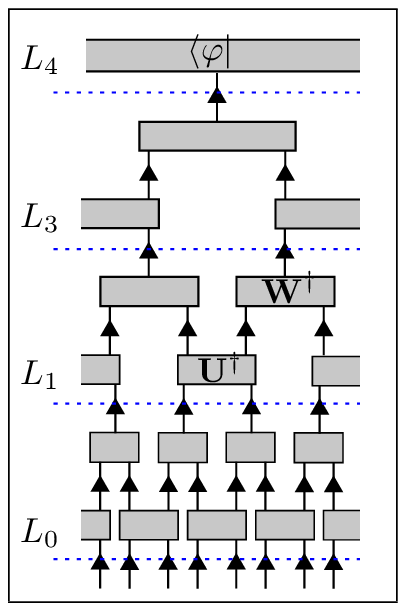,clip=}}}

\newcommand*{\twochain}{\raisebox{-23.283pt}{\epsfig{file=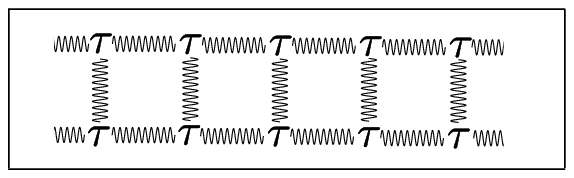,clip=}}}

\newcommand*{\twochainnumbered}{\raisebox{-23.283pt}{\epsfig{file=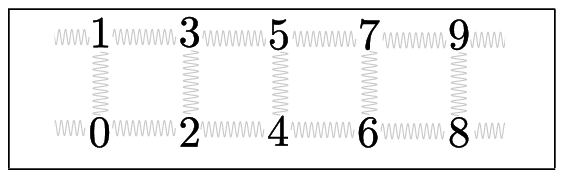,clip=}}}

\newcommand*{\merathird}{\raisebox{-54.486pt}{\epsfig{file=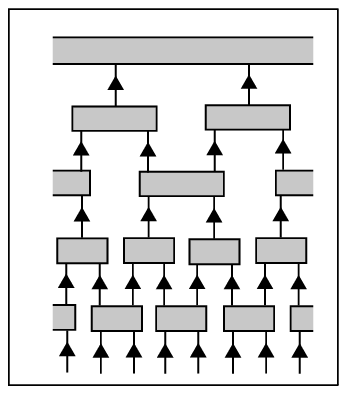,clip=}}}

\newcommand*{\merafourth}{\raisebox{-31.806pt}{\epsfig{file=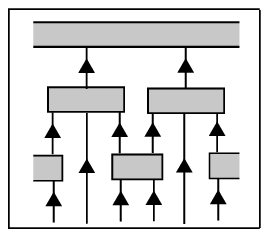,clip=}}}

\newcommand*{\goshbbasis}{\raisebox{-7.260pt}{\epsfig{file=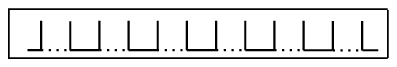,clip=}}}

\newcommand*{\goshwiggle}{\raisebox{-4.500pt}{\epsfig{file=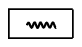,clip=}}}

\newcommand*{\goshamove}{\raisebox{-7.260pt}{\epsfig{file=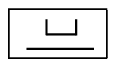,clip=}}}

\newcommand*{\goshbmove}{\raisebox{-7.260pt}{\epsfig{file=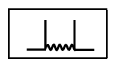,clip=}}}

\newcommand*{\goshstraight}{\raisebox{-4.500pt}{\epsfig{file=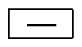,clip=}}}

\newcommand*{\goshdotted}{\raisebox{-4.500pt}{\epsfig{file=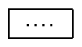,clip=}}}

\newcommand*{\goshdbasis}{\raisebox{-7.260pt}{\epsfig{file=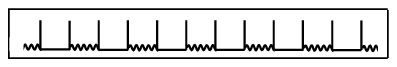,clip=}}}

\newcommand*{\goshdbasismodified}{\raisebox{-7.260pt}{\epsfig{file=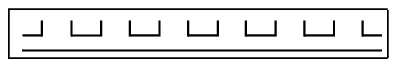,clip=}}}

\newcommand*{\goshb}{\raisebox{-19.746pt}{\epsfig{file=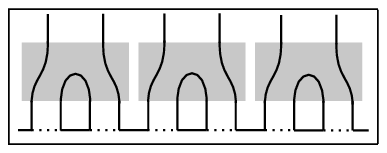,clip=}}}

\newcommand*{\goshbY}{\raisebox{-19.746pt}{\epsfig{file=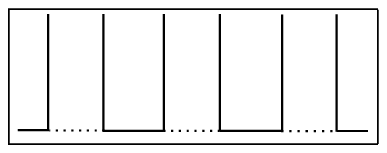,clip=}}}

\newcommand*{\goshisom}{\raisebox{-14.043pt}{\epsfig{file=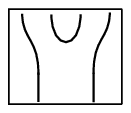,clip=}}}

\newcommand*{\ffiba}{\raisebox{-25.383pt}{\epsfig{file=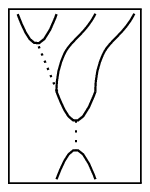,clip=}}}

\newcommand*{\ffibb}{\raisebox{-25.383pt}{\epsfig{file=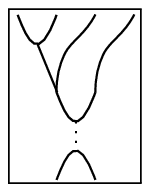,clip=}}}

\newcommand*{\ffibc}{\raisebox{-25.383pt}{\epsfig{file=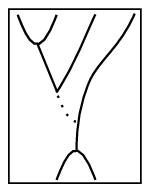,clip=}}}

\newcommand*{\ffibd}{\raisebox{-25.383pt}{\epsfig{file=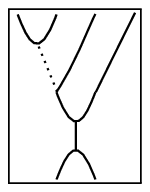,clip=}}}

\newcommand*{\ffibe}{\raisebox{-25.383pt}{\epsfig{file=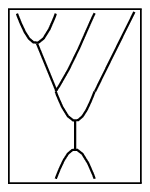,clip=}}}

\newcommand*{\braida}{\raisebox{-21.123pt}{\epsfig{file=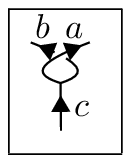,clip=}}}

\newcommand*{\braidproja}{\raisebox{-32.463pt}{\epsfig{file=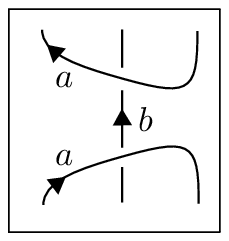,clip=}}}

\newcommand*{\braidprojb}{\raisebox{-32.463pt}{\epsfig{file=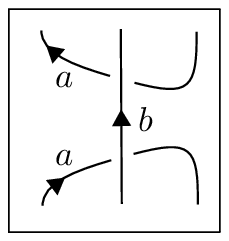,clip=}}}

\newcommand*{\braidprojc}{\raisebox{-32.463pt}{\epsfig{file=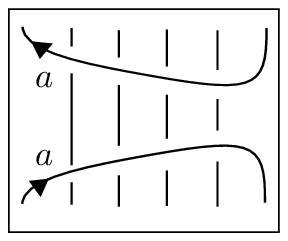,clip=}}}

\newcommand*{\braidb}{\raisebox{-21.123pt}{\epsfig{file=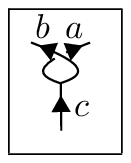,clip=}}}

\newcommand*{\trivalent}{\raisebox{-21.123pt}{\epsfig{file=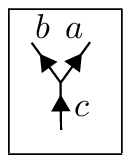,clip=}}}

\newcommand*{\crossinga}{\raisebox{-24.003pt}{\epsfig{file=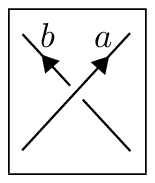,clip=}}}

\newcommand*{\crossingb}{\raisebox{-24.003pt}{\epsfig{file=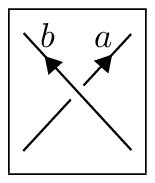,clip=}}}

\newcommand*{\resolvedcrossing}{\raisebox{-24.003pt}{\epsfig{file=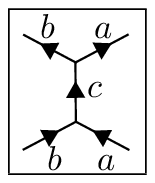,clip=}}}

\newcommand*{\merasecond}{\raisebox{-74.286pt}{\epsfig{file=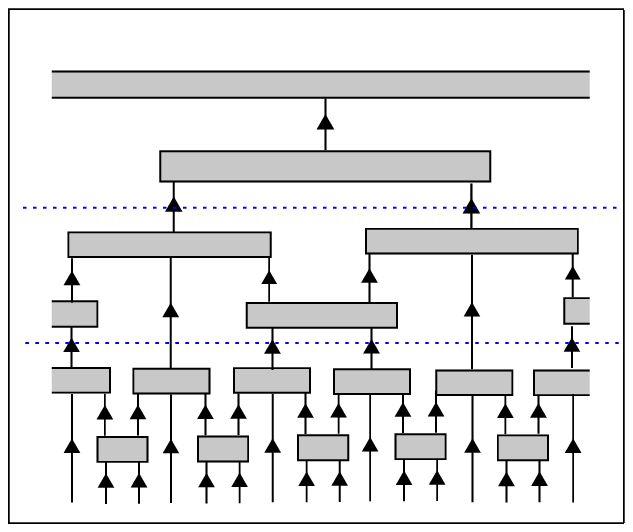,clip=}}}

\newcommand*{\unitarysecond}{\raisebox{-37.443pt}{\epsfig{file=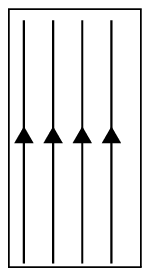,clip=}}}

\newcommand*{\isometryfirst}{\raisebox{-37.443pt}{\epsfig{file=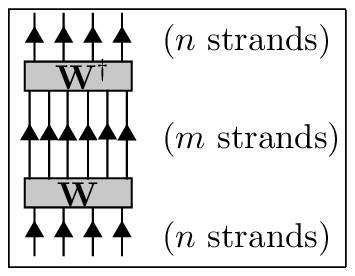,clip=}}}

\newcommand*{\merastate}{\raisebox{-170.703pt}{\epsfig{file=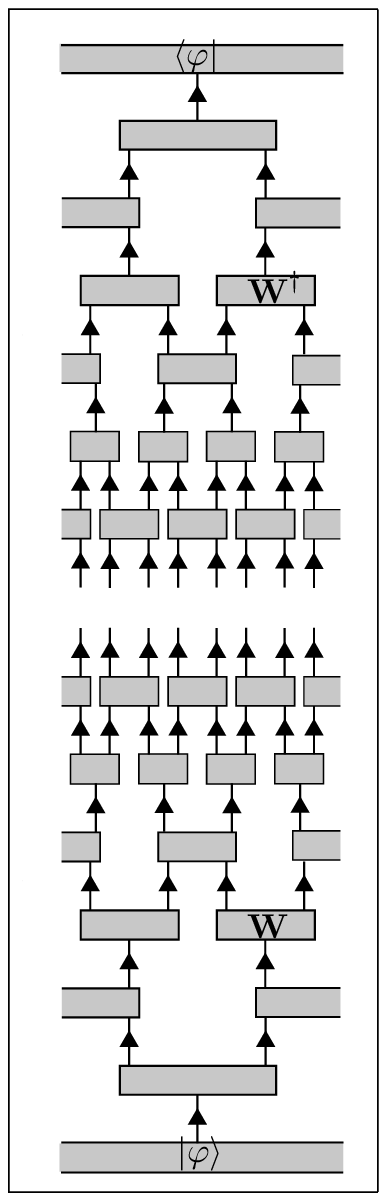,clip=}}}

\newcommand*{\meraexpectation}{\raisebox{-170.703pt}{\epsfig{file=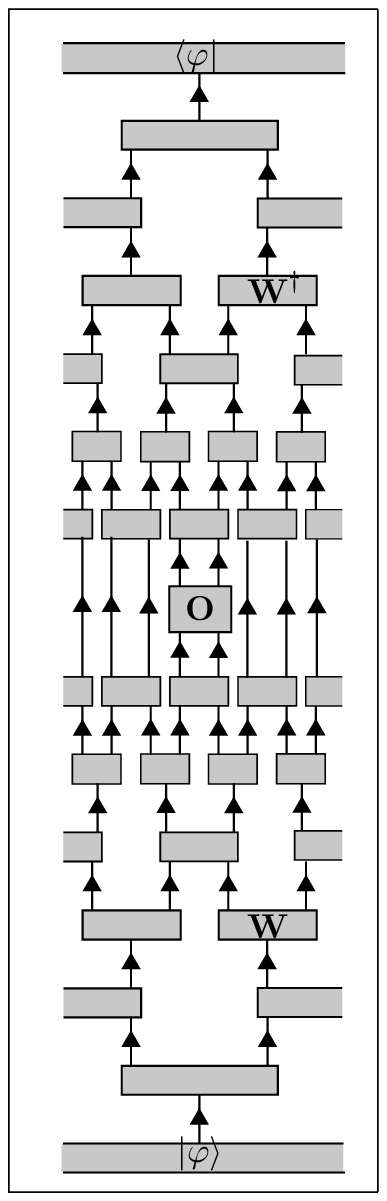,clip=}}}

\newcommand*{\meraexpectationsimplified}{\raisebox{-170.703pt}{\epsfig{file=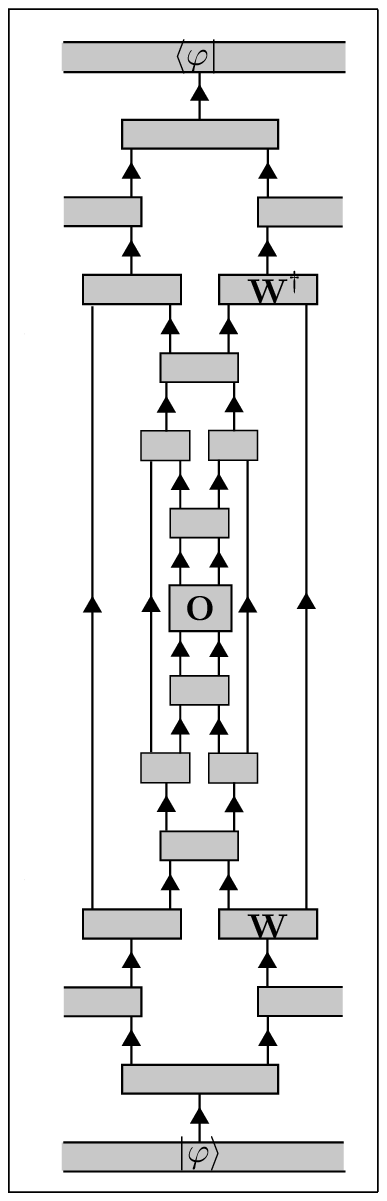,clip=}}}

\newcommand*{\supopzero}{\raisebox{-21.123pt}{\epsfig{file=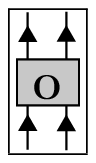,clip=}}}

\newcommand*{\supopsecond}{\raisebox{-44.523pt}{\epsfig{file=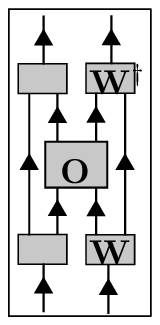,clip=}}}

\newcommand*{\supopthird}{\raisebox{-44.523pt}{\epsfig{file=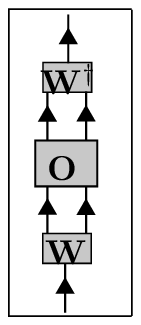,clip=}}}

\newcommand*{\supopfirst}{\raisebox{-45.243pt}{\epsfig{file=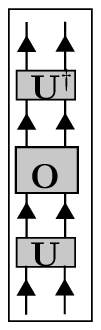,clip=}}}

\newcommand*{\twoanyonsecond}{\raisebox{-19.305pt}{\epsfig{file=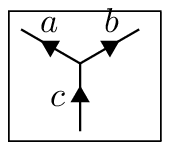,clip=}}}

\newcommand*{\pusha}{\raisebox{-24.663pt}{\epsfig{file=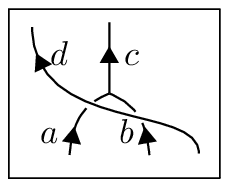,clip=}}}

\newcommand*{\pushab}{\raisebox{-24.663pt}{\epsfig{file=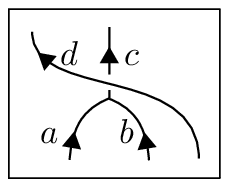,clip=}}}

\newcommand*{\twoanyonfirst}{\raisebox{-19.026pt}{\epsfig{file=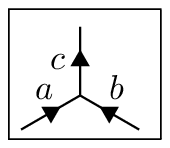,clip=}}}

\newcommand*{\standardbasisvector}{\raisebox{-44.523pt}{\epsfig{file=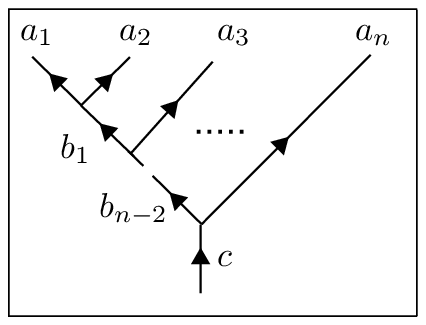,clip=}}}

\newcommand*{\threeanyonsecond}{\raisebox{-31.806pt}{\epsfig{file=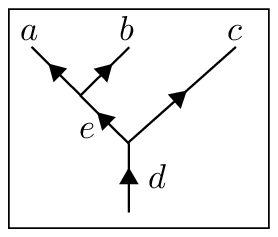,clip=}}}

\newcommand*{\threeanyonthird}{\raisebox{-31.806pt}{\epsfig{file=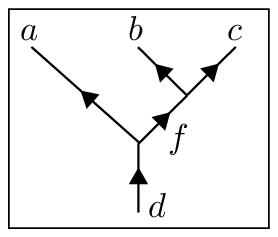,clip=}}}

\newcommand*{\bubblefirst}{\raisebox{-31.806pt}{\epsfig{file=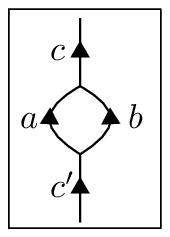,clip=}}}

\newcommand*{\bubblesecond}{\raisebox{-31.806pt}{\epsfig{file=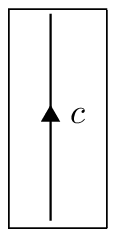,clip=}}}

\newcommand*{\vacuumfirst}{\raisebox{-28.926pt}{\epsfig{file=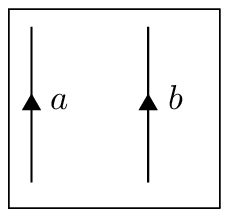,clip=}}}

\newcommand*{\vacuumsecond}{\raisebox{-28.926pt}{\epsfig{file=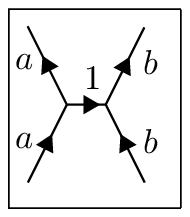,clip=}}}

\newcommand*{\totalchargeprojection}{\raisebox{-68.643pt}{\epsfig{file=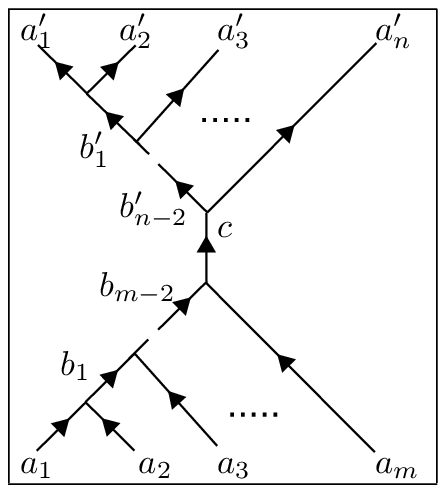,clip=}}}

\newcommand*{\identityoperator}{\raisebox{-20.466pt}{\epsfig{file=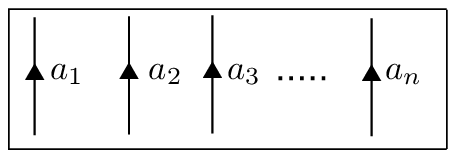,clip=}}}

\newcommand*{\factoropthird}{\raisebox{-23.283pt}{\epsfig{file=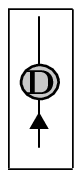,clip=}}}

\newcommand*{\factoropsecond}{\raisebox{-23.283pt}{\epsfig{file=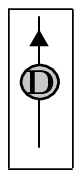,clip=}}}

\newcommand*{\factoropfirst}{\raisebox{-23.283pt}{\epsfig{file=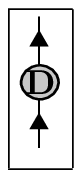,clip=}}}

\newcommand*{\operatorfirst}{\raisebox{-28.926pt}{\epsfig{file=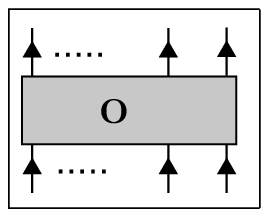,clip=}}}

\newcommand*{\operatorthird}{\raisebox{-28.926pt}{\epsfig{file=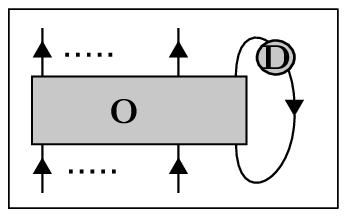,clip=}}}

\newcommand*{\operatorsecond}{\raisebox{-57.303pt}{\epsfig{file=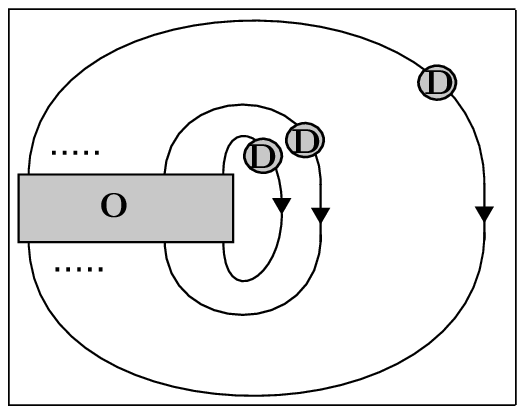,clip=}}}

\newcommand*{\chainfirst}{\raisebox{-21.123pt}{\epsfig{file=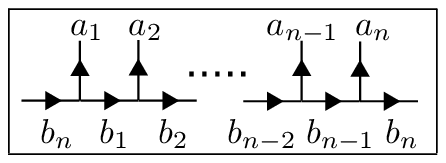,clip=}}}

\newcommand*{\chainbra}{\raisebox{-20.466pt}{\epsfig{file=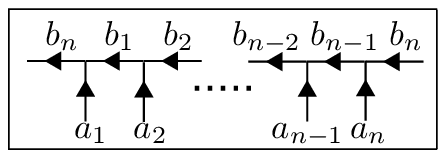,clip=}}}

\newcommand*{\chainoperatorsecond}{\raisebox{-20.466pt}{\epsfig{file=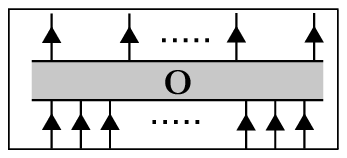,clip=}}}

\newcommand*{\chainoperator}{\raisebox{-26.103pt}{\epsfig{file=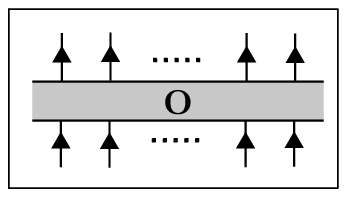,clip=}}}

\newcommand*{\chainoperatorsecondtrace}{\raisebox{-29.646pt}{\epsfig{file=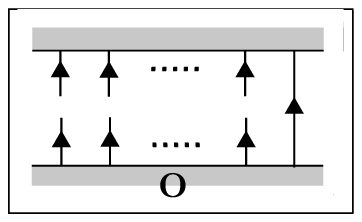,clip=}}}

\newcommand*{\chainoperatorthirdtrace}{\raisebox{-29.646pt}{\epsfig{file=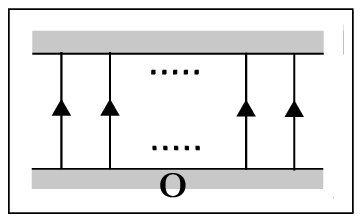,clip=}}}

\newcommand*{\topbox}{\raisebox{-21.123pt}{\epsfig{file=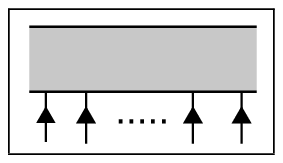,clip=}}}

\newcommand*{\isometrybox}{\raisebox{-26.103pt}{\epsfig{file=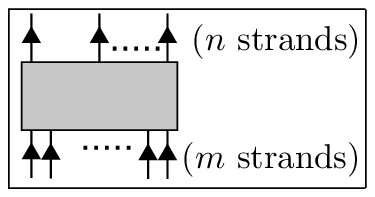,clip=}}}

\newcommand*{\unitarythird}{\raisebox{-36.003pt}{\epsfig{file=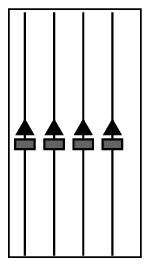,clip=}}}

\newcommand*{\traceexpectation}{\raisebox{-28.926pt}{\epsfig{file=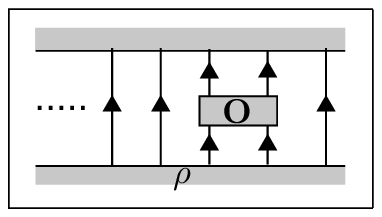,clip=}}}

\newcommand*{\statea}{\raisebox{-9.780pt}{\epsfig{file=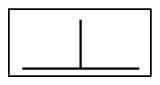,clip=}}}

\newcommand*{\stateb}{\raisebox{-9.780pt}{\epsfig{file=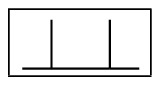,clip=}}}

\newcommand*{\statec}{\raisebox{-9.780pt}{\epsfig{file=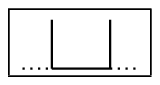,clip=}}}

\newcommand*{\stated}{\raisebox{-9.780pt}{\epsfig{file=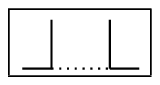,clip=}}}

\newcommand*{\fiba}{\raisebox{-10.503pt}{\epsfig{file=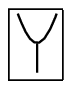,clip=}}}

\newcommand*{\fibb}{\raisebox{-14.763pt}{\epsfig{file=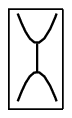,clip=}}}

\newcommand*{\fibc}{\raisebox{-14.763pt}{\epsfig{file=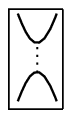,clip=}}}

\newcommand*{\fibd}{\raisebox{-14.043pt}{\epsfig{file=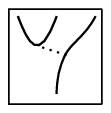,clip=}}}

\newcommand*{\fibe}{\raisebox{-14.043pt}{\epsfig{file=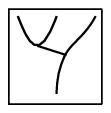,clip=}}}

\newcommand*{\fibf}{\raisebox{-17.226pt}{\epsfig{file=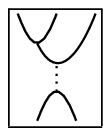,clip=}}}

\newcommand*{\fibg}{\raisebox{-17.226pt}{\epsfig{file=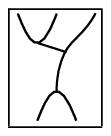,clip=}}}

\newcommand*{\fibh}{\raisebox{-17.226pt}{\epsfig{file=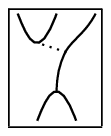,clip=}}}

\newcommand*{\fibm}{\raisebox{-14.043pt}{\epsfig{file=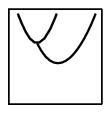,clip=}}}

\newcommand*{\fibn}{\raisebox{-14.043pt}{\epsfig{file=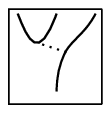,clip=}}}

\newcommand*{\fibo}{\raisebox{-14.043pt}{\epsfig{file=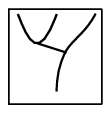,clip=}}}

\newcommand*{\fibp}{\raisebox{-14.043pt}{\epsfig{file=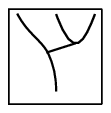,clip=}}}

\newcommand*{\fibq}{\raisebox{-14.043pt}{\epsfig{file=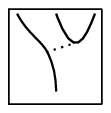,clip=}}}

\newcommand*{\phasediagram}{\raisebox{-79.263pt}{\epsfig{file=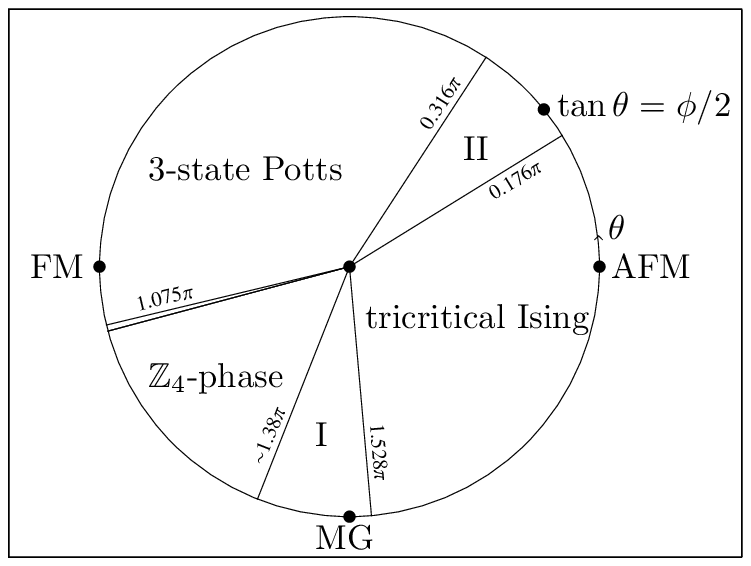,clip=}}}

\newcommand*{\chainsecond}{\raisebox{-31.806pt}{\epsfig{file=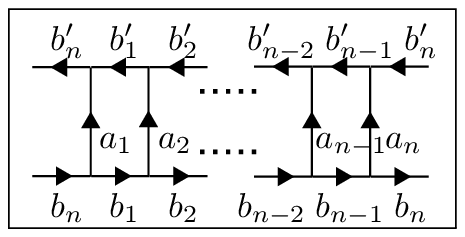,clip=}}}

\newcommand*{\parallelfirst}{\raisebox{-27.543pt}{\epsfig{file=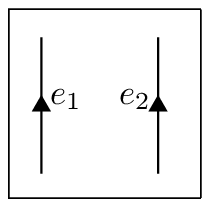,clip=}}}

\newcommand*{\parallelsecond}{\raisebox{-28.926pt}{\epsfig{file=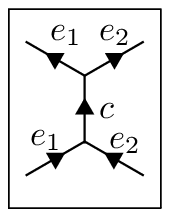,clip=}}}

\newcommand*{\chainthird}{\raisebox{-24.663pt}{\epsfig{file=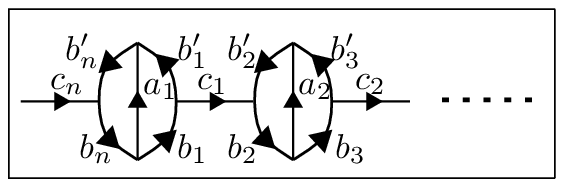,clip=}}}

\newcommand*{\chainfourth}{\raisebox{-21.843pt}{\epsfig{file=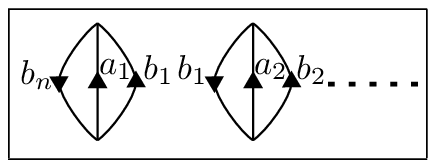,clip=}}}

\newcommand*{\chainfifth}{\raisebox{-21.843pt}{\epsfig{file=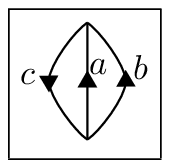,clip=}}}

\newcommand*{\fourtotwomera}{\raisebox{-106.923pt}{\epsfig{file=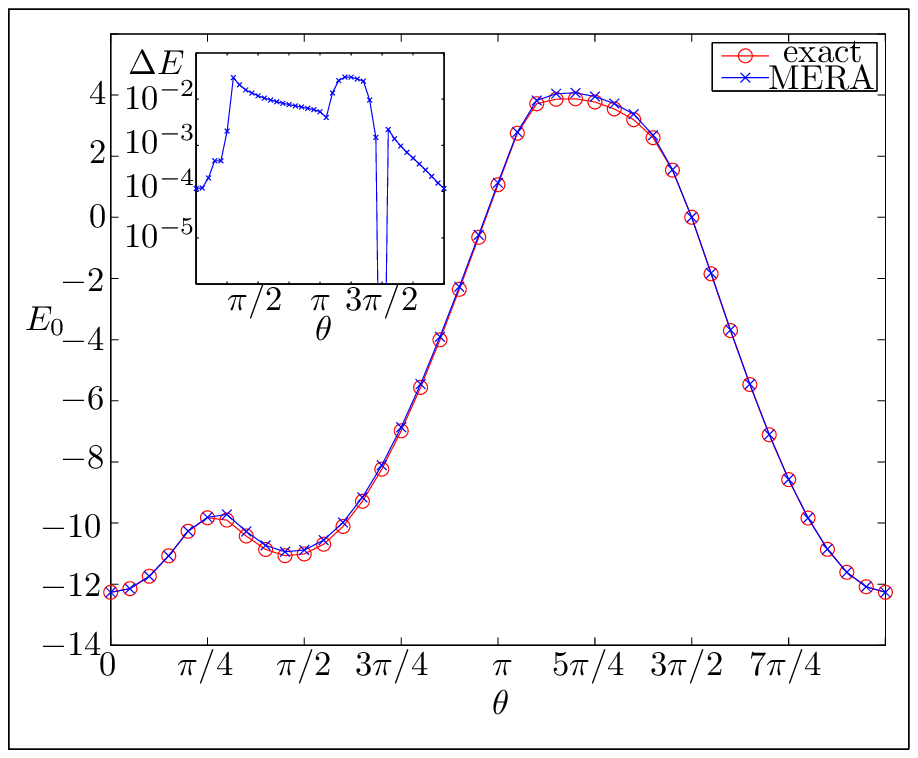,clip=}}}

\newcommand*{\sixtotwomera}{\raisebox{-106.923pt}{\epsfig{file=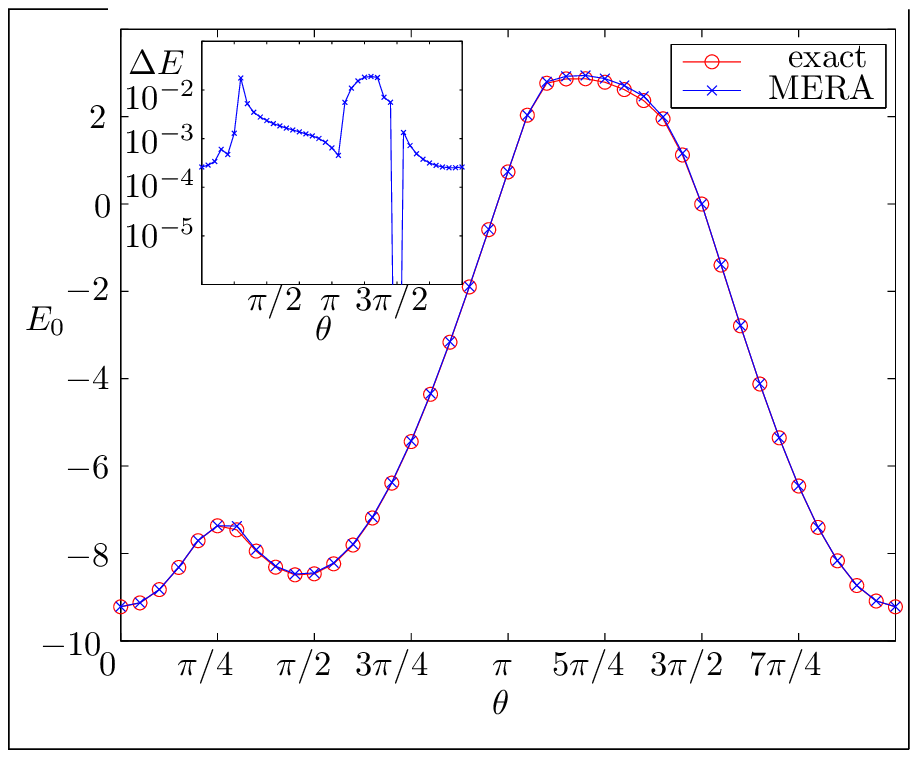,clip=}}}

\newcommand*{\fourtwomerapt}{\raisebox{-106.923pt}{\epsfig{file=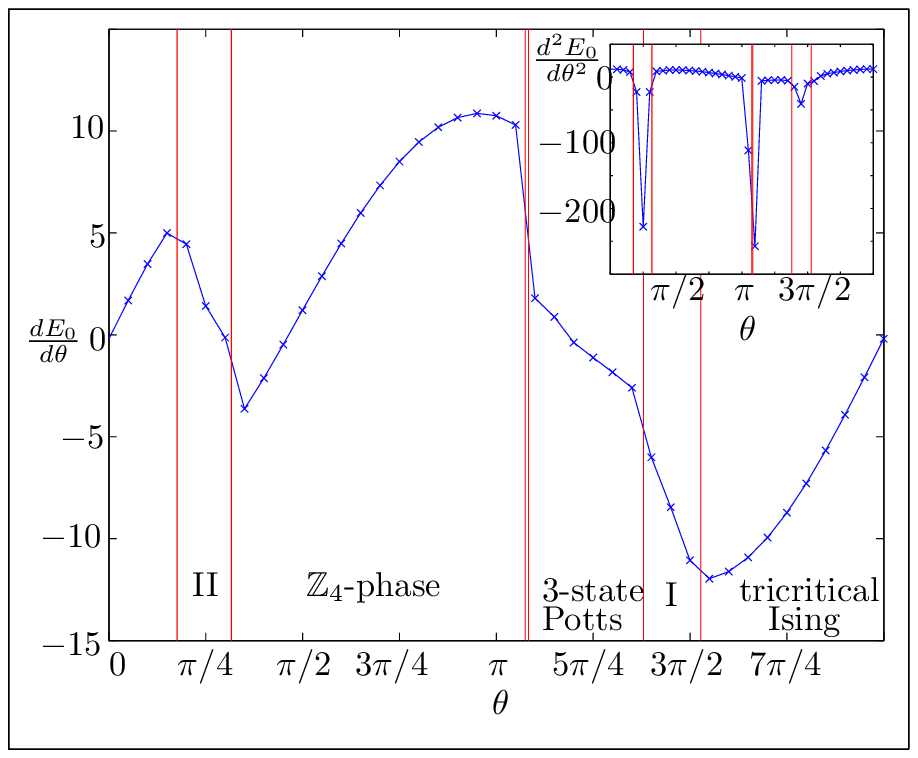,clip=}}}

\newcommand*{\oneeightcorrfourtwo}{\raisebox{-78.543pt}{\epsfig{file=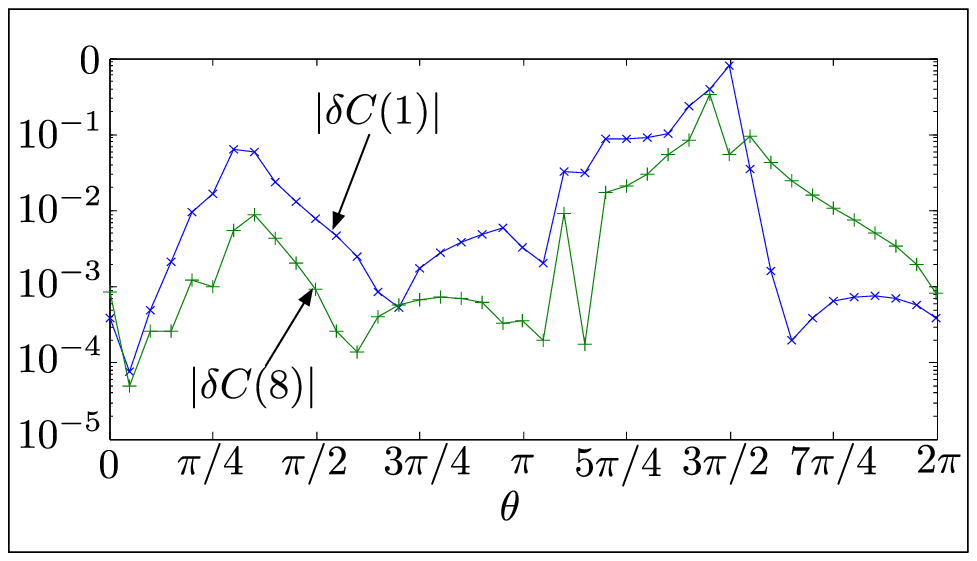,clip=}}}

\newcommand*{\onesixcorrsixtwo}{\raisebox{-78.543pt}{\epsfig{file=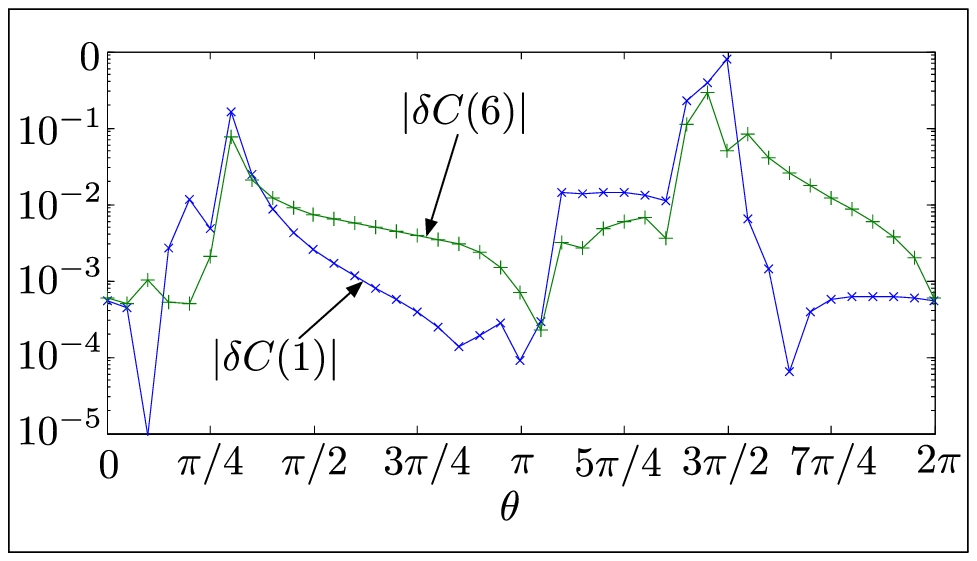,clip=}}}

\newcommand*{\allcorrsixtwo}{\raisebox{-78.543pt}{\epsfig{file=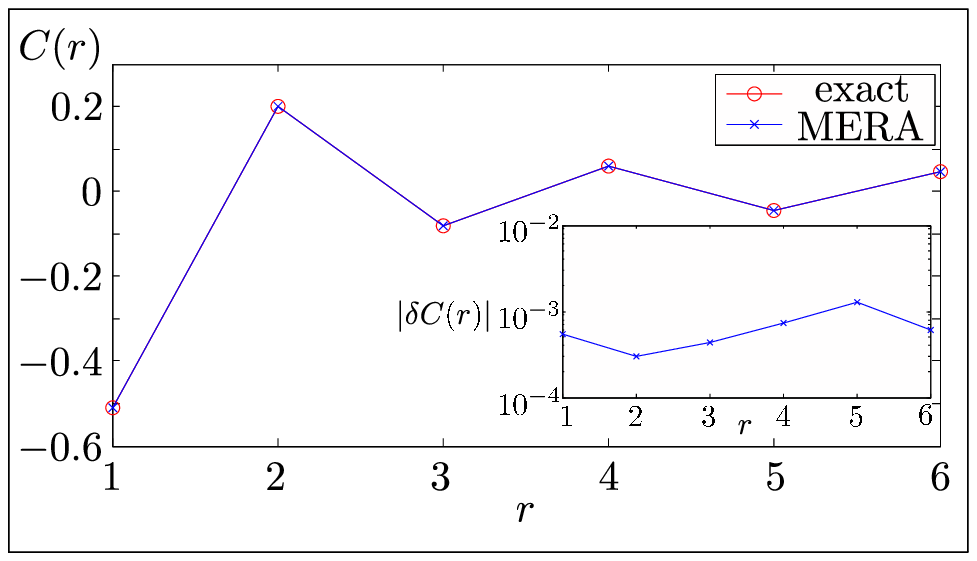,clip=}}}

\newcommand*{\allcorrsixtwoFM}{\raisebox{-84.243pt}{\epsfig{file=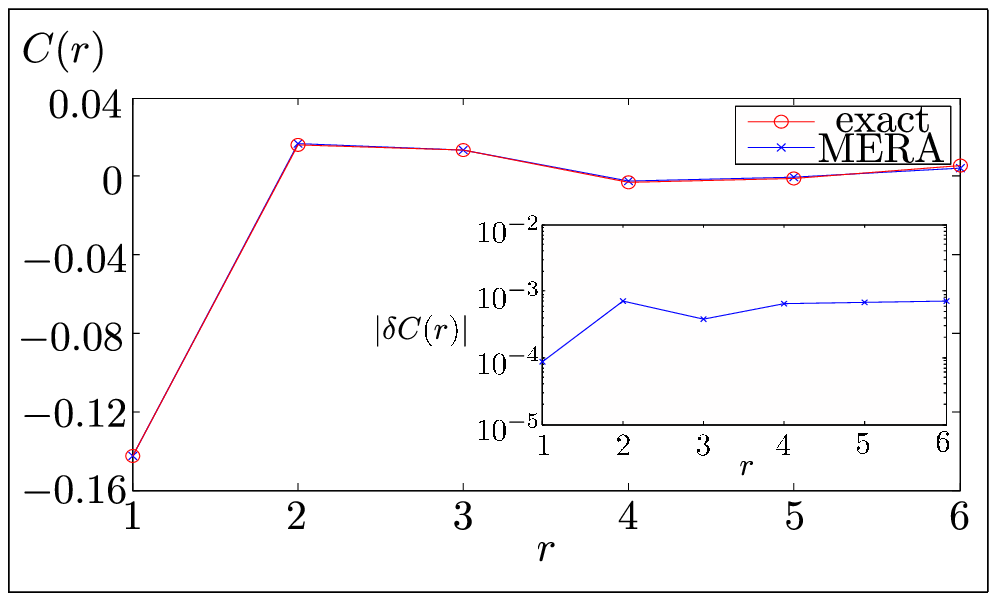,clip=}}}

\newcommand*{\allcorrfourtwoFM}{\raisebox{-84.243pt}{\epsfig{file=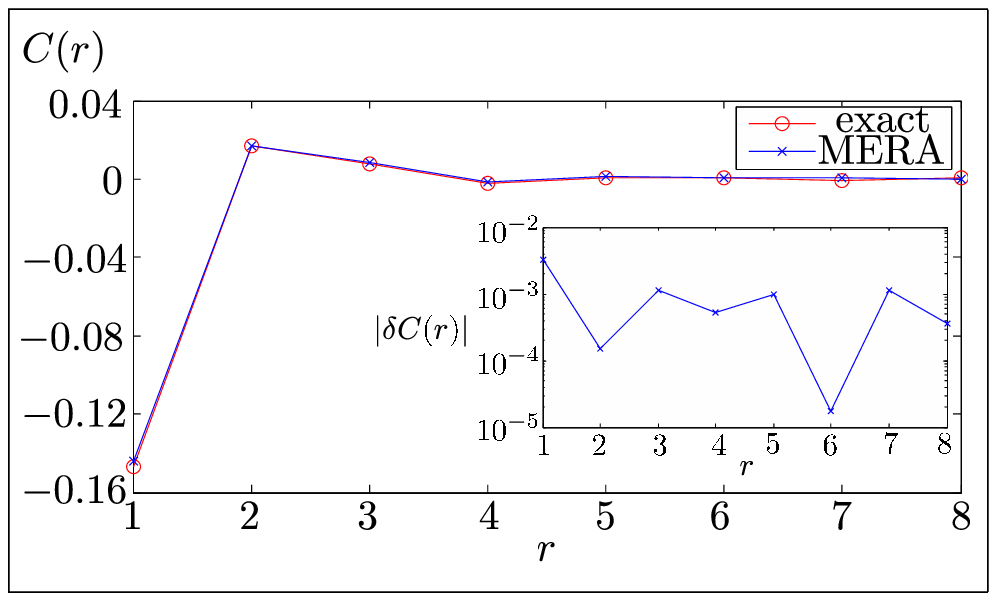,clip=}}}

\newcommand*{\allcorrfourtwo}{\raisebox{-78.543pt}{\epsfig{file=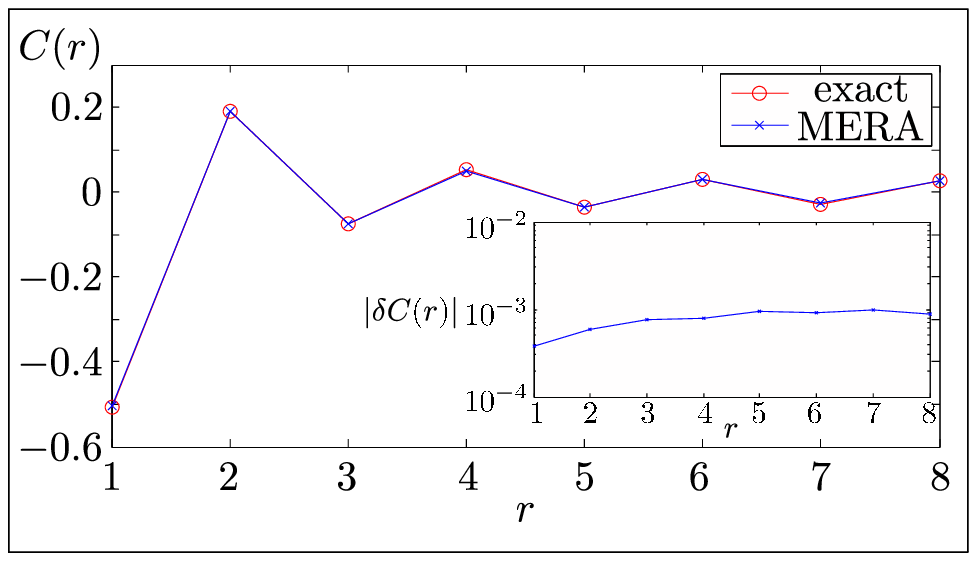,clip=}}}

\newcommand*{\epersite}{\raisebox{-89.883pt}{\epsfig{file=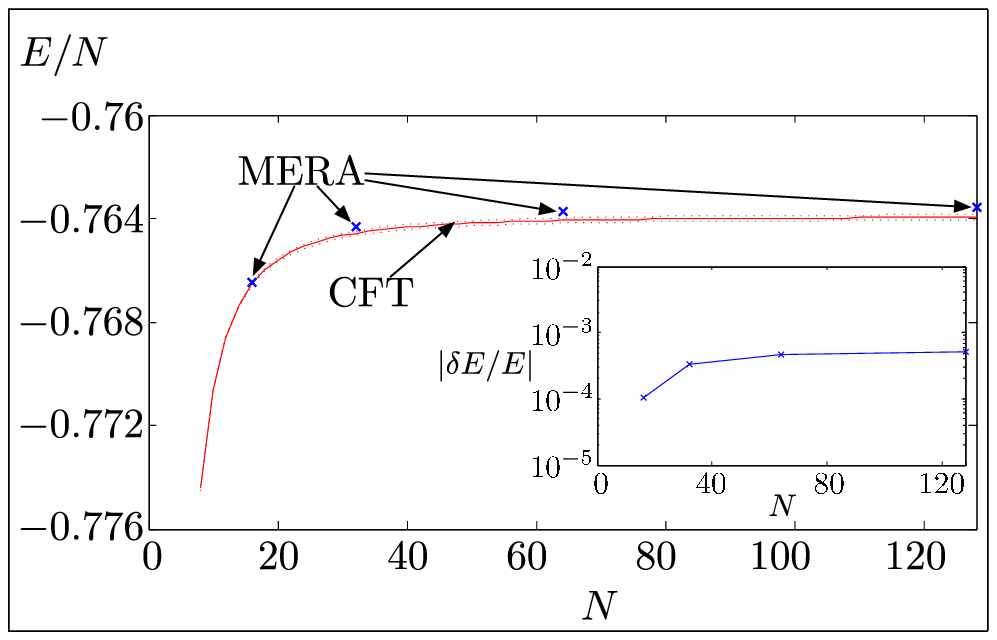,clip=}}}

\begin{document}

\title{Anyonic entanglement renormalization}
\author{Robert \surname{K{\"o}nig}\email{rkoenig@caltech.edu}}
\author{Ersen \surname{Bilgin}}
\affiliation{Institute for Quantum Information, Caltech, Pasadena CA 91125, USA}
\begin{abstract}
We introduce a family of variational ansatz states for chains of anyons which optimally exploits the structure of the anyonic Hilbert space.  This ansatz is the natural analog of  the multi-scale entanglement renormalization ansatz for spin chains. In particular, it has the same interpretation as a coarse-graining procedure and is expected to accurately describe critical systems with algebraically decaying correlations. We numerically investigate the validity of this ansatz using the anyonic golden chain and its relatives as a testbed. This demonstrates the power of entanglement renormalization in a setting with non-abelian exchange statistics, extending previous work on qudits, bosons and fermions in two dimensions. 
\end{abstract}
\maketitle

\section{Introduction} 
Strongly correlated quantum many-body systems with topological order have been proposed as a substrate for building fault-tolerant quantum computers~\cite{Kitaev03,preskill,Freedmanetal03,Nayaketal08}. Under this proposal, logical information is stored within a subspace of a fixed number of quasi-particles (anyons). Computation is performed by 
exploiting the non-abelian statistics obeyed by these anyons under braiding. Compared to more conventional implementations of quantum computers, this offers an intrinsic resilience to noise: local perturbations cannot decohere the stored information because of the non-local nature of the encoding.

At present, perhaps the most promising candidate systems exhibiting non-abelian statistics are fractional quantum Hall systems~\cite{MooreRead,readrezayia,readrezayib}. Some amount of experimental evidence is already available in this setting~\cite{dolevetal,raduetal,willettetal}. Other proposed systems include topological insulators~\cite{FuKane05,MooreBalents07,LiangKaneMele07,Roy09} and lattice spin systems~\cite{Kitaev03,KitaevAnyons,LevinWen} whose local interactions could be enginereed artificially~\cite{duanetal03,michelietal06}. Independently of the proposed physical realization, 
 the stability of the topologically ordered phase with respect to (local) perturbations is of great  interest for topological quantum computation. This presents a formidable theoretical challenge. Possible approaches range from the study of concrete physical models (see e.g.,~\cite{trebstetal07,Tupitsynetal08,JVidal09,JVidal09b}) or bounds on the gap for general families of models~\cite{klich09,bravyihastingsmichalakis10,bravyihastings10} to the investigation of effective Hamiltonians describing the inter-anyon interactions (see e.g.,~\cite{bonderson09}). In the context of the latter, it is natural to study paradigmatic models such as 1-dimensional anyonic chains. These are the natural counterpart of  spin chains. The study of such systems has led, e.g., to an exactly solved model called the golden chain~\cite{Feiguinetal07}, and novel realizations of  infinite randomness critical phases~\cite{BonesteelYang07,Fidkowskietalrandom,Fidkowskietal08b}.

A peculiar aspect of anyonic systems is the structure of the Hilbert space of $N$~anyons. In contrast to the space of $N$ qudits, this space does not decompose into an $N$-fold tensor product. Instead, its dimension scales as~$d^N$, where the quantum dimension~$d$ is generally not an integer. For analytical studies, it is sometimes convenient to embed this space into a larger tensor product space. However, this approach is inconvenient when using variational methods:  with the most straightforward encoding, it may be unclear how to vary (numerically) over the subset of physical states. Furthermore, this leads to a significant increase in computational complexity, particularly because locality is only approximately preserved by such an embedding. Even for one-dimensional anyonic systems, these issues complicate  the direct application numerical methods such as DMRG~\cite{white92}. While such methods may presently be the most powerful and successful tools available, these difficulties  motivate the search for alternative approaches.

There is  another more profound reason why the na\"ive application of numerical methods for qudits may be suboptimal: local anyonic operators preserve the total topological charge in their support.  Since a realisitic Hamiltonian consists of locally acting terms,  topological charge conservation severly constrains the action of the Hamiltonian both at the local level as well as on larger scales. Incorporating this fact into the method of study should therefore be highly beneficial. With a qudit embedding, the meaning of this constraint is obscured and may be hard to make use of.

The purpose of this paper is to introduce a variational method for anyonic systems which avoids using unphysical additional states and optimally exploits the special structure of the Hilbert space. This is motivated in part by the goal of facilitating numerical studies. Perhaps more importantly, variational families of ansatz states can provide significant insight into the nature of quantum correlations in a given system. Even in this regard, a qudit embedding  of anyons is generally undesirable, as a  physically motivated ansatz for qudit systems may lose its significance when applied directly to  the anyonic setting.

Our scheme is inspired by renormalization group studies of anyonic chains~\cite{BonesteelYang07,Fidkowskietalrandom,Fidkowskietal08b},  composite anyon distillation~\cite{Koe09} and  the  multi-scale entanglement renormalization ansatz (MERA)~\cite{Vidal07,Vidal08} for (non-anyonic) spin chains. In fact, it can be understood as the natural anyonic counterpart of the latter and shares many of its properties. In particular, it can be seen as a renormalization group scheme and is thus especially suited for describing scale-invariant systems.   The scheme allows -- in principle -- to extract, e.g., critical exponents when considering fixed points of the renormalization group flow.  More generally, it provides  procedures for computing parameters of the corresponding conformal field theory  (CFT) in the continuum limit.  While such procedures are also available for other variational methods (e.g., matrix product states~\cite{VerstraeteCirac06,Rom97,Vid04,Verstr04,VerstrMurgCirac08}, by transfer matrix methods), they are particularly natural  in the present context due to the scale-invariant form of the ansatz.

We formulate our ansatz for one-dimensional (periodic) chains of anyons. This allows us to numerically test its validity  for the golden chain~\cite{Feiguinetal07} and its relatives~\cite{Trebstetal08}. Ultimately, though, it is desirable to find methods for two-dimensional systems. Our work makes some progress towards this goal: the non-anyonic MERA extends naturally to two dimensions and this is also the case for its anyonic counterpart. In the appendix, we briefly comment on such generalizations.

Entanglement renormalization, while originally defined for qudit systems, has been extended  to free bosons~\cite{EvenblyVidal10} and interacting fermions~\cite{CorbozVidal09}. Anyonic statistics are peculiar to two dimensions, encompassing both fermions and bosons (which are abelian) in addition to interesting non-abelian generalizations. Our ansatz applies to all such models, but its use is especially suggestive in the non-abelian case. This is due to the special form of the  Hilbert space mentioned earlier. Interestingly, the exact details of the exchange statistics play essentially no role in the definition of the ansatz,  though they become important when evaluating expectation values of local operators in the case of e.g., two-dimensional arrangements of anyons. 
 
In a wider context, entanglement renormalization is a special instance of the class of tensor network states, which also includes e.g., MPS~\cite{VerstraeteCirac06,Rom97,Vid04,Verstr04,VerstrMurgCirac08}, PEPS~\cite{VerstraeteCirac04PEPS} as well as the closely related TERG-states~\cite{GuLevinWen08}. These have wide applicability beyond qudit systems. In particular, a general framework for fermions has been developed~\cite{PizornVerstraete10,Kraus10,Corbozorusetal10,Pinedaetal10,Bartheletal09}. Our focus is on entanglement renormalization because of its unique operational interpretation, as well as the possibility of efficiently computing expectation values without approximations, even in two dimensions (in contrast to e.g., PEPS~\cite{SchuchetalPEPScomp07}). Furthermore, anyonic entanglement renormalization is conceptually related to previous analytical studies for anyonic chains~\cite{BonesteelYang07,Fidkowskietalrandom,Fidkowskietal08b}.   However, our work suggests that anyonic generalizations of other tensor network states should also be possible along similar lines, though further work may be needed to evaluate their descriptive power. 
  
The structure of this paper is as follows. In Section~\ref{sec:MERA}, we review the formulation of the MERA for spin chains.  In Section~\ref{sec:anyonicformalism}, we  give some background on anyonic systems and their description in terms of fusion diagrams.  We then present the anyonic entanglement renormalization ansatz in Section~\ref{sec:anyonicMERA} and  show how to efficiently evaluate expectation values of local observables and correlation functions. We also discuss an operational interpretation in the context of composite anyon distillation.   In Section~\ref{sec:goldenchain}, we apply the ansatz to the golden chain and identify a renormalization group fixed point. We conclude in Section~\ref{sec:conclusions}.

\section{Multi-scale entanglement renormalization\label{sec:MERA}}
To motivate our ansatz, we  recapitulate the definition of  the MERA for 1D~qudit chains and that of
related tensor network states~\cite{Vidal07,Vidal08}. For more details, we refer to~\cite{evenblyvidal09}.  The  assumption underlying this ansatz is that certain local degrees of freedom are insignificant for the long-range character of the target state.  The MERA is a quantum circuit which completely decouples and discards these degrees of freedom using local unitaries. This procedure is subsequently repeated for the resulting  coarse-grained description of the state. As a variational family of states, the MERA consists of those states  which are turned into ($N$-fold) product states by such quantum circuits.

It is convenient to use the diagrammatic formalism of tensor networks to describe the  unitaries and the local isometries  constituting a MERA (the latter correspond to the elimination of local degrees of freedom, but the division into ``disentangling'' unitaries and isometries is often arbitrary). In this formalism, 
the identity on~$\mathbb{C}^d$ is represented by a single directed edge, and operators $\Xsingleoperator:(\mathbb{C}^d)^{\otimes n}\rightarrow (\mathbb{C}^d)^{\otimes m}$ are represented by shaded boxes with $n$~ingoing  and $m$~outgoing (ordered) edges. Labels on the edges from~$1,\ldots,d$ correspond to the elements of a fixed orthonormal basis of~$\mathbb{C}^d$. Products of operators are taken by connecting outgoing with ingoing edges, and summing over labels of  edges
with no free ends. The trace of an operator on~$(\mathbb{C}^d)^{\otimes n}$ is taken by connecting each outgoing strand with the corresponding ingoing strand, and then contracting the tensor network (i.e., summing over all edge labelings). Partial traces are computed analogously by connecting up subsets of edges. Tensor products are obtained by placing diagrams next to each other. 

With these conventions, the property
$\Xisometry^\dagger \Xisometry =\id_{(\mathbb{C}^d)^{\otimes n}}$ of an isometry  $\Xisometry: (\mathbb{C}^d)^{\otimes n}\rightarrow (\mathbb{C}^d)^{\otimes m}$ (for $m\geq n$) takes the simple form
\begin{align} 
\scalebox{0.9}{\isometryfirst}&=\scalebox{0.9}{\unitarysecond} \label{eq:mainMERA}
\end{align} 
Identity~\eqref{eq:mainMERA} is crucial for the definition of the MERA ansatz.

\begin{figure}
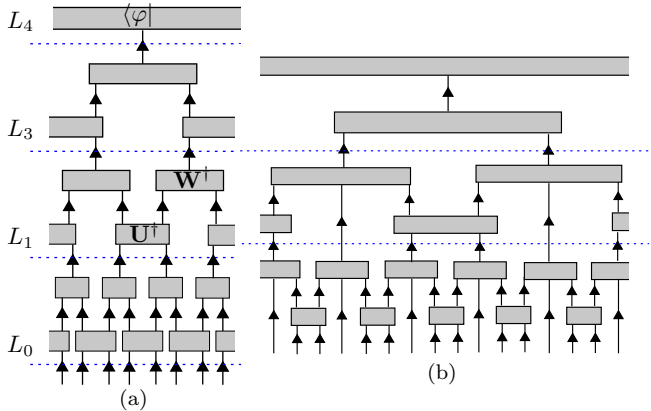

\centering
\begin{tabular}{cc }
\hspace{-0.5cm}\subfigure[]{\label{f:merafirst}\scalebox{0.9}{\merafirst}}
&
\hspace{-0.5cm}\subfigure[]{\label{f:merasecond}\scalebox{0.9}{\merasecond}}
\end{tabular}
\caption{Two examples of MERA structures: For a periodic qudit chain (at the bottom of the figure), a variational family of states is obtained by varying over the contents of the boxes. In~(a), these are (adjoints of) isometries $\Xisometry:(\mathbb{C}^{d})^{\otimes 2}\rightarrow\mathbb{C}^d$, unitaries $\Xunitary:(\mathbb{C}^d)^{\otimes 2}\rightarrow (\mathbb{C}^d)^{\otimes 2}$ and (the adjoint of) a state $\ket{\varphi}\in \mathbb{C}^d$ for the strip at the top. 
Tree-like structures may also be considered, and lead to particularly simple expressions in the homogenous case~\cite{silvietal09}.  Note that we choose to represent the top tensor by a (periodic) strip; the reason for this will become obvious once we move to the anyonic setting.\label{fig:MERA}}
\end{figure}

Consider a system of $N$~qudits arranged on  a line with periodic boundary condtions. For any tensor network as shown in Fig.~\ref{fig:MERA}, one obtains a variational family of states $\ket{\Psi}=\ket{\Psi_{\ket{\varphi},\{\Xisometry\}}}\in(\mathbb{C}^d)^{\otimes N}$ parametrized by  the isometries $\{\Xisometry\}$ and the state~$\ket{\varphi}\in\mathbb{C}^d$ at the top of the structure. This is the MERA ansatz.  Fig.~\ref{fig:MERA} represents a map $(\mathbb{C}^d)^{\otimes N}\rightarrow\mathbb{C}$ which can be understood as the bra~$\bra{\Psi}$ of the represented state. The fact that the recipe specifies~$\bra{\Psi}$ instead of $\ket{\Psi}$ is a matter of preference, but the chosen convention has a natural operational interpretation: One may think of Fig.~\ref{fig:MERA} as a renormalization prescription by decomposing the tensor network along different horizontal cuts~$L_i$. The strips $[L_{i},L_{i+1}[$ represent coarse-graining maps which reduce the number of degrees of freedom at each level.

\begin{figure} 
\centering
\begin{tabular}{ccc}
\subfigure[]{\hspace{-0.6cm}\label{f:state}\scalebox{0.9}{\merastate}}
&
\subfigure[]{\hspace{-0.6cm}\label{f:expectation}\scalebox{0.9}{\meraexpectation}}
&
\subfigure[]{\hspace{-0.6cm}\label{f:expectationsimplified}\scalebox{0.9}{\meraexpectationsimplified}}
\end{tabular}
\caption{Tensor networks associated with  the MERA of Fig.~\ref{fig:MERA}~(a). The network~(a) represents the projection~$\proj{\Psi}$ onto the state described by the MERA. The contraction of the network~(b) gives the expectation value $\bra{\Psi}\Xsingleoperator\ket{\Psi}$ of a local operator~$\Xsingleoperator$. Fig.~(c) gives the same value as~(b), and is obtained from it by using~\eqref{eq:mainMERA} repeatedly. \label{fig:MERAstateandexpectation}}
\end{figure}

Importantly, local expectation values and pair correlation functions can be computed efficiently for such a state. This is because the expectation value of an observable~$\Xsingleoperator$ is given by the contraction of  the tensor network obtained by sandwiching~$\Xsingleoperator$ between the MERA network and its adjoint as in Fig.~\ref{fig:MERAstateandexpectation}~(b). Eq.~\eqref{eq:mainMERA} then allows to  simplify the network  resulting in a significantly smaller network corresponding to the ``causal cone'' of the operator, see Fig.~\ref{fig:MERAstateandexpectation}~(c). This network can be efficiently contracted, as the number of tensors scales logarithmically with the number of sites~$N$. For example, for the case of Fig.~\ref{fig:MERAstateandexpectation}~(b), we get 
\begin{align}
\bra{\Psi}\Xsingleoperator\ket{\Psi}&=\bra{\varphi}\cE_{L_{\max}}\circ\cF_{L_{\max}}\cdots\cE_{L_0}\circ \cF_{L_0} (\Xsingleoperator)\ket{\varphi} \label{eq:vdf}
\end{align}
where $L_{\max}=\log_2 N-1$ and the superoperators~$\{\cE_L,\cF_L\}$
associated to a level~$L$ are 
\begin{align}
\cE_L\left(\ \supopzero\ \right)&=\supopsecond\ ,\  \cF_L\left(\ \supopzero\ \right)&=\supopfirst\label{eq:extraction}
\end{align}
for $L<L_{\max}$ and 
\begin{align}
\hspace{-1ex}\cE_{L_{\max}}\left(\ \supopzero\ \right)&=\supopthird,\ \cF_{L_{\max}}\left(\ \supopzero\ \right)&=\supopfirst\ .\label{eq:extractionb}
\end{align}
In particular, the reduced two-site density operator of the state~$\ket{\Psi}$ can be computed using the adjoint superoperators as
\begin{align*}
\tr_{n-2}\proj{\Psi}=\cF^\dagger_1\circ\cE^\dagger_1\cdots\cF^{\dagger}_{L_{\max}}\circ\cE^\dagger_{L_{\max}}(\proj{\varphi})\ .
\end{align*}
Similar expressions can be  obtained for translates as well as for pair correlation functions evaluated at certain specific distances~\cite{Giovannettietal08,Pfeiferetal09,silvietal09}.

Given a MERA-structure as in Fig.~\ref{fig:MERA}, we have described a recipe giving a variational family of states for a chain of~$N$ $d$-dimensional qudits parametrized by isometries of the form  $\Xisometry:(\mathbb{C}^{d})^{\otimes n}\rightarrow(\mathbb{C}^d)^{\otimes m}$ (the constants~$n\geq m$ depend on the MERA-structure). There is a natural way of enlarging this family: the edges (in higher levels of the  network) may be interpreted as corresponding to $\mathbb{C}^\chi$, where the refinement parameter $\chi>d$ is larger than the dimension~$d$ of the physical qudits. Increasing this so-called {\em bond dimension}~$\chi$ amounts  to using isometries $\Xisometry:(\mathbb{C}^{\chi})^{\otimes n}\rightarrow(\mathbb{C}^\chi)^{\otimes m}$ (and correspondingly a state $\ket{\varphi}\in\mathbb{C}^\chi$ at the top). To motivate a similar refinement in the anyonic setting, we point out that if~$\chi=d^s$ is an integer power of the physical qudit dimension~$d$, then this enlargement of the variational family is equivalent to replacing each edge in the MERA-structure by $s$~edges. Note also that, since the number of isometries is of order $O(N\log N)$ and each isometry is described by fewer than~$\chi^{n+m}$ parameters, the total number of parameters describing a MERA is of order $O(poly(\chi) N\log N)$.

We conclude this short exposition  by mentioning two subclasses of MERA-ansatz states which are of particular interest: a {\em translation-invariant} MERA has identical isometries within every level. As a consequence,  the number of parameters  of such a MERA scales as~$O(poly(\chi)\log N)$. A {\em scale-invariant} MERA is one where all isometries (with identical domain and range) are chosen to be the same: Here the number of parameters is only $O(poly(\chi))$ independent of~$N$ (see~\cite{Vidal08} for a more detailed discussion of the complexity of MERA and~\cite{evenblyvidal09} for concrete examples). For a scale-invariant MERA, it is possible to write down two-point and three-point correlators explicitly in terms of eigenvalues of a certain coarse-graining superoperator defined in terms of the isometries as in~\eqref{eq:extraction}. This allows to numerically extract critical exponents and paramters from the associated CFT, as described e.g., in~\cite{Giovannettietal08,Pfeiferetal09,silvietal09}.

\section{Anyonic states and operators\label{sec:anyonicformalism}}
In this section, we provide a short introduction to anyons, emphasizing the aspects relevant to the definition of the anyonic entanglement renormalization ansatz: the anyonic Hilbert space and the isotopy-invariant formalism for describing states and operators. We also discuss the origin of the dynamics, that is, the definition of Hamiltonians for anyons. For a thorough and accessible introduction to anyons and topological quantum computation, we recommend~\cite{preskill} (see also~\cite{Nayaketal08} for a recent review).

\subsection{A unified treatment of topological order}
Anyons arise as localized quasi-particle excitations in  what can roughly be referred to as  two-dimensional topologically ordered quantum media, e.g., qudit lattice systems with certain Hamiltonians~\cite{Kitaev03,KitaevAnyons,LevinWen}, quantum Hall systems~\cite{MooreRead,readrezayia,readrezayib} or topological insulators~\cite{FuKane05,MooreBalents07,LiangKaneMele07,Roy09}. Independently of their physical realization,  their state space and exchange statistics are described by the axioms of a topological quantum field theory (TQFT) (see e.g.,~\cite{Walker91}). This formalism is extraordinarily useful for  studying low-energy processes~\cite{bonderson09} as well as for the application to quantum computation~\cite{preskill}, as it abstracts out the relevant physics: it specifies the particle content, i.e., what particle types occur, describes what their internal degrees of freedom are, and how they are affected by braiding (exchanges) and fusion (which corresponds to bringing particles together). The relation between states of (and operations on) the physical system  and the abstract anyonic state space is discussed extensively in the literature, see e.g.,~\cite{Kitaev03,KitaevAnyons,LevinWen,KoeKupRei,preskill,readrezayia}.

The algebraic object underlying such an anyonic theory is a modular category. Roughly, this consists of (i)~a finite set of particle types~$\Omega$ equipped with an involution $^*:\Omega\rightarrow\Omega$ and containing a distinguished trivial particle~$1\in\Omega$, (ii)~fusion rules, i.e., a set of allowed triples of particles, (iii)~a quantum dimension $d_a>0$ associated to every particle~$a$, (iv)~a tensor~$F$ indexed by $6$~particles and (v)~a $3$-index tensor~$R$. These are required to satisfy a number of consistency conditions (see e.g.,~\cite{preskill}) the most important of which  express associativity of fusion and compatibility of fusion with braiding. 

In more physical terms, the involution associates an antiparticle~$a^*$ to every particle~$a$, with $1^*=1$~corresponding to the absence of a particle. The fusion rule summarizes the possible outcomes when bringing two particles together.  The quantum dimensions give a rough measure of the growth of the anyonic state space when adding particles, and the $F$-tensor relates different bases of this space (as explained below). Finally, the $R$-tensor encodes braiding of pairs of particles. This will be mostly irrelevant for our discussion, but becomes relevant for 2-dimensional arrangements of anyons as explained in Appendix~\ref{app:braidingetc}.

\subsection{The anyonic Hilbert space and anyon diagrams}
 We proceed by explaining the construction of the anyonic Hilbert space using the data~(i)-(iv) of a modular category. We will mostly follow the detailed exposition in~\cite{Bondersonetal09}, but will require slightly more general definitions when dealing with operators  (see e.g., appendix of~\cite{KitaevAnyons} for more details). 

 The state space of a set of anyons depends on their types and on the surface the quantum medium is embedded in. We will discuss two cases in detail below: anyons pinned to fixed locations on a disc and on a torus. Usually,  we assume that these are arranged on a chain, though one may also consider e.g., regular two-dimensional lattices; their particular geometric arrangement only becomes important when considering Hamiltonians, but does not affect the definition of the Hilbert space.

 Starting point  are certain trivalent graphs with directed edges. These correspond to pants decomposition of the surface with punctures inserted at the anyons' positions. Fixing such a graph, a basis of the anyonic Hilbert space is given by all labelings of the edges with  particle labels from~$\Omega$ satisfying the fusion rules at every vertex. We will give explicit examples for the punctured sphere and the torus below. 

 Labeled graphs related by reversing the direction of an edge and simultaneously replacing its particle label by the associated antiparticle represent the same vectors. This is somewhat analogous to the formalism of Feynman diagrams. Indeed, anyon diagrams may, to some extent, be interpreted as particle world-lines, though this analogy has its limitations. Note also that in many anyon theories of interest, such as the Fibonacci category considered below, every particle is its own antiparticle, $a^*=a$, and it is sufficient to work with undirected graphs.

Dividing up trivalent graphs into neighborhoods of their vertices, one arrives at the following  alternative description: the total anyonic Hilbert space is the direct sum of tensor products of  two-anyon fusion spaces~$V^{c}_{ab}$ (respectively their dual splitting space $V_c^{ab}$) corresponding to every vertex, where $a,b,c\in \Omega$, and where the sum is taken over all fusion-consistent labelings (see below).  The space~$V^{ab}_c$ can be thought of as the internal degrees of freedom of two anyons of type~$a$ and~$b$ whose combined topological charge is~$c$. Equivalently, it is the space of two anyons~$a$ and~$b$ on a disc with total topological charge~$c$ at the boundary. The latter is --  in principle -- a measurable quantity.  We assume for simplicity that fusion is multiplicity-free, i.e., $\dim V^c_{ab}\in\{0,1\}$, but our techniques directly generalize to models with fusion multiplicities (see~\cite{Bondersonetal09} for the necessary adaptations in the diagrammatic calculus).

We pick a normalized vector $\ket{ab;c}\in V^{ab}_c$ in each splitting space and represent it using the isotopy-invariant formalism~\cite{Bondersonetal09} as
\begin{align*}
\ket{ab;c}&=\left(\frac{d_c}{d_ad_b}\right)^{\textfrac{1}{4}}\scalebox{0.9}{\twoanyonsecond} \in V^{ab}_c\\
\bra{ab;c}&=\left(\frac{d_c}{d_ad_b}\right)^{\textfrac{1}{4}}\scalebox{0.9}{\twoanyonfirst}\in V^c_{ab}\ ,
\end{align*}
where a scalar~$d_a$ called quantum dimension is associated to every label~$a$.
More generally, going from a vector to its dual vector corresponds to flipping the diagram along the horizontal axis and reversing all the arrows.  Isotopy invariance means that diagrams may be continuously deformed as long as endpoints are held fixed and edges are not passed through each other or around open endpoints.

\subsubsection{Anyons on a disc}
Consider the space $V^{\vec{a}}_c$  of $n$~anyons of types $\vec{a}\in\Omega^n$
pinned to fixed locations on a disc with total charge~$c\in\Omega$ at the boundary. For this space,  we will often use the standard basis
 given by the vectors
\begin{align}
\hspace{-1ex}\ket{\vec{a},\vec{b};c}_n&=\left(\frac{d_c}{\prod_i d_{a_i}}\right)^{\textfrac{1}{4}}\scalebox{0.9}{\standardbasisvector}\  \label{eq:standardbasis}
\end{align}
where   $\vec{b}\in\Omega^{n-1}$  is such that this diagram is fusion-consistent. This corresponds to the decomposition
\begin{align*}
V^{\vec{a}}_c &\cong \bigoplus_{\vec{b}\in\Omega^{n-1}} V^{a_1a_2}_{b_1}\otimes V^{b_1a_3}_{b_2}\otimes\cdots\otimes V^{b_{n-3}a_{n-1}}_{b_{n-2}}\otimes V^{b_{n-2}a_n}_c\ .
\end{align*}
One may switch between different bases using the (unitary) $F$~move, i.e., the isomorphism between the two decompositions of 
\begin{align*}
V^{abc}_d \cong \bigoplus_{e}V^{ab}_e\otimes V^{ec}_d\cong \bigoplus_f V^{bc}_f\otimes V^{af}_d\ ,
\end{align*}
which is specified by the coefficients $[F^{abc}_d]_{e,f}$ in
\begin{align}
\scalebox{0.9}{\threeanyonsecond} &=\sum_{f} [F^{abc}_d]_{e,f}\scalebox{0.9}{\threeanyonthird} .\label{eq:Fmove}
\end{align}
The matrices $[F^{abc}_d]$ collectively constitute the $F$-tensor of the modular category mentioned earlier.
Matrix elements of related isomorphisms such as
\begin{align*}
V^{ab}_{cd}\cong\bigoplus_e V_{ce}^a\otimes V_{d}^{eb}\cong \bigoplus_{f}V^{f}_{cd}\otimes V^{ab}_f
\end{align*}
can be expressed as functions of these. Diagrammatically, all basis changes can be computed using~\eqref{eq:Fmove}, the identity
\begin{align}
\scalebox{0.9}{\bubblefirst}=\delta_{cc'}\sqrt{\frac{d_ad_b}{d_c}}\scalebox{0.9}{\bubblesecond}\label{eq:bubbleremoval}
\end{align}
and the convention that lines with the trivial label may be added and removed arbitrarily, i.e.,
\begin{align}
\scalebox{0.9}{\vacuumfirst} &=\scalebox{0.9}{\vacuumsecond}\ .\label{eq:vacuumlineaddition}
\end{align}
An operator $U_{\vec{a}}^{\vec{a'}}(c):V^{\vec{a}}_c\rightarrow  V^{\vec{a}'}_{c}$ taking the  fusion space of $m$~anyons of types  $\vec{a}=(a_1,\ldots,a_m)\in \Omega^m$ with  total charge~$c\in\Omega$ to~$n$ anyons of types~$\vec{a'}=(a_1',\ldots,a_n')\in \Omega^n$ with total charge~$c$ can be represented as a  linear combination of trivalent graphs with $m$~ingoing edges and $n$~outgoing edges attached to open endpoints and carrying the corresponding labels. For example, using the  standard basis~\eqref{eq:standardbasis},
such an operators can be written as
\begin{align}
U^{\vec{a'}}_{\vec{a}}(c)&=\sum_{\vec{b},\vec{b'}}[U^{\vec{a'}}_{\vec{a}}(c)]_{\vec{b'},\vec{b}}\ket{\vec{a'},\vec{b'};c}_{nm}\bra{\vec{a},\vec{b};c}\\
&\hspace{-1cm}=\sum_{\vec{b},\vec{b'}}[U^{\vec{a'}}_{\vec{a}}(c)]_{\vec{b'},\vec{b}}\alpha_{\vec{a},\vec{a'},c}\hspace{-1cm}\scalebox{0.9}{\totalchargeprojection}\label{eq:exampleoperator}
\end{align}
with the normalization factor 
\begin{align*}
\alpha_{\vec{a},\vec{a'},c}=\frac{1}{\sqrt{d_c}(\prod_i d_{a_i})^{1/4}(\prod_j d_{a_j'})^{1/4}}\ .
\end{align*}
Any operator $U^{\vec{a'}}_{\vec{a}}:\bigoplus V^{\vec{a}}_c\rightarrow \bigoplus_{c'} V^{\vec{a'}}_{c'}$ taking $m$~anyons of types $\vec{a}=(a_1,\ldots,a_m)\in\Omega^m$ to $n$ anyons of type~$\vec{a'}=(a_1',\ldots,a_n')\in \Omega^{n}$  can also be represented in this fashion if it conserves the total charge, i.e., if it has block-diagonal form
\begin{align}
U^{\vec{a}'}_{\vec{a}}\cong \bigoplus_c \left(U^{\vec{a'}}_{\vec{a}}(c):V^{\vec{a}}_c\rightarrow V^{\vec{a'}}_c\right)\ .\label{eq:linearmapdecomposition}
\end{align}
Such operators act locally on subsets of $n$~anyons of specified types~$\vec{a}\in\Omega^n$ and map them to a state of anyons of specified types~$\vec{a'}\in\Omega^m$.  In fact,~\eqref{eq:exampleoperator} (respectively~\eqref{eq:linearmapdecomposition}) represent the most general locally acting operator, their single most important property being charge preservation. Additional properties such as unitarity impose further conditions on the form of the map/matrix~$U^{\vec{a'}}_{\vec{a}}(c)$ for each~$c$.

More generally, we are interested in operators which act between spaces of the form $\bigoplus_c\left(\bigoplus_{\vec{a}\in\Gamma_n} V_c^{\vec{a}}\right)$, where $\Gamma_n\subset \Omega^n$ specifies a set of $n$-tuples of available (spatial) particle configurations on a chain. A total charge-preserving operator of this kind takes the form
\begin{align} 
U^{\Gamma_n}_{\Gamma_m}=\bigoplus_c U^{\Gamma_n}_{\Gamma_m}(c)\label{eq:chargepreserving}
\end{align}
where $U^{\Gamma_n}_{\Gamma_m}(c)\in\End\left(\bigoplus_{\vec{a}\in\Gamma_m} V^{\vec{a}}_c,\bigoplus_{\vec{a'}\in\Gamma_n}V^{\vec{a}'}_c\right)$.
This can be understood as the projection of a general charge-preserving operator onto inputs and outputs from~$\Gamma_m$ and~$\Gamma_n$, respectively. More precisely, let
\begin{align*}
\id_{\Gamma_n}:\bigoplus_c\left(\bigoplus_{\vec{a}\in\Omega^n} V_c^{\vec{a}}\right)\rightarrow\bigoplus_c\left(\bigoplus_{\vec{a}\in\Omega^n} V_c^{\vec{a}}\right)
\end{align*}
denote the projection onto the subset of states with anyon labels from $\Gamma_n$ defined by
\begin{align*}
\id_{\Gamma_n} &:=\sum_{\vec{a}\in\Gamma_n}\sum_{\substack{c\in\Omega\\ \vec{b}\in\Omega^{n-1}}} \ket{\vec{a},\vec{b};c}_{nn}\bra{\vec{a},\vec{b};c}\\
&=\sum_{\vec{a}\in\Gamma_n}\identityoperator
\end{align*}
Then
\begin{align}
U^{\Gamma_m}_{\Gamma_n}=\id_{\Gamma_{m}} U^{\Gamma_m}_{\Gamma_n}\id_{\Gamma_n}\ ,
\end{align}
which expresses the fact that the domain and range of the operator is restricted to states with certain particle configurations. 

Similarly as for tensor network states, we represent charge-preserving operators of the form~\eqref{eq:chargepreserving}
taking the space of $m$~anyons to the space of $n$~anyons by shaded boxes  with $m$~ingoing and $n$~outgoing (unlabeled) edges; summation over these edge labels is left implicit.  It is important to note, however, that such a box represents a different object compared to the case of tensor networks: it is defined by  a family of maps~$\{U^{\Gamma_m}_{\Gamma_n}(c)\}_{c\in \Omega}$ which specify a  weighted superposition of certain labeled trivalent graphs embedded in the box, with $m$~in- and $n$~outgoing edges as in~\eqref{eq:exampleoperator}.

The diagrammatic representation of anyonic operators and vectors satisfies
simple rules with respect to composition, tensor products and (partial) tracing. Adjoint operators are obtained by flipping the diagram, reversing the arrows and complex conjugating all coefficients. Operators are multiplied or applied to vectors by stacking their representations on top of each other and connecting up out- with ingoing edges.   Tensor products are obtained by placing diagrams next to each other. Traces and partial traces are the result of connecting up  in- and outgoing strands of an operator, with an additional minor modification~(see~\appref{app:disctrace}).

These rules are remarkably similar to the contraction of tensor networks, although their origin is distinct. In particular, the notion of evaluating of a diagram is very different: A tensor network associates a scalar quantity to every labeling of the edges. This means that contraction, i.e.,  summing over all labelings, results in a scalar. In contrast, anyonic diagrams associate a trivalent graph to every labeling, and the contraction  results in a formal linear combination of equivalence classes of labeled trivalent graphs. Equivalence is defined by isotopy and the local rules~\eqref{eq:Fmove},~\eqref{eq:bubbleremoval} and~\eqref{eq:vacuumlineaddition}.

\subsubsection{Anyons on a torus}
Here we are interested in periodic chains of anyons arranged on a line, and  a few modifications of the above formalism are necessary. 
 As in~\cite{Feiguinetal07,Trebstetal08,Gilcollective09}, we consider a chain of anyons arranged on a topologically non-trivial cycle wrapping around a torus.  We denote the space of  $n$~anyons of types~$\vec{a}=(a_1,\ldots,a_n)$  arranged on such a chain with periodic boundary conditions by~$V^{\vec{a}}_{\textrm{periodic}}$. This space does not naturally decompose into subspaces with specified total charge.  An orthonormal  basis is given by the basis vectors
\begin{align}
\ket{\vec{a},\vec{b}}_n&=\frac{1}{(\prod_i d_{a_i})^{\textfrac{1}{4}}}\scalebox{0.9}{\chainfirst}\label{eq:anyonicchainbasisvector}
\end{align}
where $\vec{b}\in\Omega^n$ is such that each vertex satisfies the fusion rules.
The basis specified by Eq.~\eqref{eq:anyonicchainbasisvector} corresponds to a decomposition of the Hilbert space as
\begin{align*}
V^{\vec{a}}_{\textrm{periodic}}\cong \bigoplus_{\vec{b}=(b_1,\ldots,b_N)}V_{b_N}^{a_1b_1}\otimes V_{b_1}^{a_1b_2}\otimes\cdots\otimes V_{b_{N-1}}^{a_Nb_N}\ .
\end{align*}
The representation of the  bra $\bra{\vec{a},\vec{b}}_n$ of   the vector~\eqref{eq:anyonicchainbasisvector} is again obtained by flipping the diagram and reversing the arrows, i.e., 
\begin{align}
\bra{\vec{a},\vec{b}}_n &=\frac{1}{(\prod_i d_{a_i})^{1/4}}\scalebox{0.9}{\chainbra}\ .\label{eq:bravectorchain}
\end{align}
Local operators acting on a subset of the $n$~anyons are represented as before by shaded boxes,  but global operators can not be represented as superpositions of graphs embedded  in such a planar surface. Instead, it is convenient to embed the chain and associated anyonic diagrams along one of the fundamental nontrivial cycles of the torus. (This is  already implicit in Eq.~\eqref{eq:anyonicchainbasisvector}.) A global operator~$\Xsingleoperator:\bigoplus_{\vec{a}\in\Omega^m}V^{\vec{a}}_{\textrm{periodic}}\rightarrow \bigoplus_{\vec{a}\in\Omega^n}V^{\vec{a}}_{\textrm{periodic}}$ mapping between periodic chains of (possibly different) lengths~$m$ and~$n$ is then represented by a shaded strip parallel to the chain on the torus, with $m$~ingoing and $n$~outgoing edges, i.e.,
\begin{align*}
\scalebox{0.9}{\chainoperatorsecond}\ .
\end{align*}
Application of such an operator is again equivalent to attaching it to a diagram. Finally, (partial) traces are computed simply by connecting up edges. For example, for an operator~$\Xsingleoperator$ acting on  $\bigoplus_{\vec{a}}V^{\vec{a}}_{\textrm{periodic}}$, the partial trace over the $n$-th anyon is
\begin{align*}
\tr_n\ \scalebox{0.9}{\chainoperator} &=\scalebox{0.9}{\chainoperatorsecondtrace}
\end{align*}
Here we used isotopy on the torus to get a convenient expression on the rhs. The (complete) trace is the result of connecting up all strands, and then computing the coefficient of the emtpy graph, i.e., 
\begin{align}
\hspace{-2ex}\tr\scalebox{0.9}{\chainoperator} &=\left[\scalebox{0.9}{\chainoperatorthirdtrace}\right]_{\textrm{vac}}.\hspace{-2ex}\label{eq:traceofopperiodic}
\end{align}
Here we denote the coefficient of the empty graph in a formal superposition~$X$ by $[X]_{\textrm{vac}}$. That is, it is obtained by writing~$X$ as a superposition of states with flux~$a$, i.e., with a line labeled $a$ going around the torus, and taking  the coefficient for $a=1$. In~\appref{app:innproduct}, we show that  the (partial) trace defined in this fashion  is equivalent to the orthonormality of the set of vectors~\eqref{eq:anyonicchainbasisvector}.

\subsection{Anyonic Hamiltonians: long-range effective theories\label{sec:longrangeffective}}
A remarkable feature of the state space of anyons is its topological degeneracy: the Hamiltonian of the quantum medium assigns equal energy to each state. Furthermore, this degeneracy is stable under local perturbations, a feature which makes anyons particularly suited for encoding and processing quantum information. These properties hold up to exponentially small corrections in the inter-anyon distances.

If the inter-anyon separation falls below a certain length scale, the microscopic details of the system become relevant and the topological degeneracy is generally lifted. Such a degeneracy lifting has been examined in various quantum media~\cite{Lahtinen08,Baraban09,cheng09}.  In the system-independent anyonic formalism, Bonderson~\cite{bonderson09} has shown that a general interaction between two anyons can be interpreted as tunneling of topological charge, and that generic tunnelling fully lifts the topological degeneracy.

Hamiltonian terms responsible for such tunneling and more generally arbitrary multi-anyon interactions take the form of Hermitian operators as in~\eqref{eq:exampleoperator} with~$m=n$. The exact form of the effective Hamiltonian 
governing the energy splitting depends on the geometric arrangement of the anyons. In a regular lattice, nearest neighbor~($m=2$) and next-to-nearest neighbor ($m=3$) interactions are most relevant physically as the interaction strength decays exponentially with distance. Lattice-like  arrangments of anyons arise when certain spatial distributions are energetically favored, e.g., by inserting defects into the quantum medium which couple to additional quantum numbers  (such as electric charge) of the anyons.

Paradigmatic models of such effective Hamiltonians have been considered extensively in the literature. They can be thought of as describing the dynamics of the  internal degrees of freedom of anyons pinned to fixed sites.  
 We refer to~\cite{Trebstetal08} for an introductory discussion of such models. We discuss explicit examples for Fibonacci anyon chains in Section~\ref{sec:goldenchain}.

\section{Anyonic entanglement renormalization\label{sec:anyonicMERA}}
\subsection{The setting\label{sec:settinganyonicMERA}}
We have described the origin of anyonic Hamiltonians as long-range effective descriptions of quantum media in Section~\ref{sec:longrangeffective}. We now turn to the problem of defining a variational ansatz for such systems. For concreteness, we restrict our attention to one-dimensional chains of anyons, arranged in a periodic fashion along a topologically non-trivial cycle on the torus. We discuss  more general~$2$-dimensional arrangements in Appendix~\ref{app:twodmerageneralization}.

One may consider different spatial distributions of anyons on the chain. For example, in a setting  with several non-trivial anyon types, one may be interested in the effective behavior of a staggered, i.e., alternating arrangement of anyons. While our ansatz could in principle be adapted to such cases, here we consider the simplest non-trivial setting. We assume that  a subset~$\Omega_{\textrm{eff}}\subset \Omega$ of anyons is allowed in each site. The Hilbert space of a (periodic) chain of length~$n$ on the torus is then given by
  \begin{align}
\cH_{\textrm{chain},n}\cong \bigoplus_{\vec{a}\in\Omega_{\textrm{eff}}^n}V_{\textrm{periodic}}^{\vec{a}}\ .\label{eq:Hilbertspacechain}
\end{align}
The most commonly considered case (e.g., the golden chain~\cite{Feiguinetal07}) is when  $\Omega_{\textrm{eff}}=\{a\}$ consists of a single anyon~$a$, that is, each site is occupied by a particle of type~$a$.  Our formulation is slightly more general, as it allows to consider Hamiltonians which can create and destroy particles on sites of the chain (resp. change particle types) when~$\Omega_{\textrm{eff}}=\Omega$. This is important when the quantum medium assigns nearly degenerate energies to different distributions of anyons. Further intermediate cases could be considered.

\subsection{The ansatz}
Consider a MERA-structure for a periodic chain with~$N$ sites as in Fig.~\ref{fig:MERA}. We associate to this structure a family of states in the anyon Hilbert space~$\cH_{\textrm{chain},N}$ (cf.~\eqref{eq:Hilbertspacechain}) as follows:
\begin{enumerate}
\item To the strip at the top, associate the bra $\bra{\varphi}$ of a normalized state~
\begin{align}\ket{\varphi}\in\cH_{\textrm{chain},n}\ ,\label{eq:chainstatetop}
\end{align} 
that is,
\begin{align} 
\topbox\mapsto\ \bra{\varphi}\ ,\label{eq:topstate}
\end{align} 
where $n$ is the number of ingoing edges. 
\item
To every intermediate box with $m$~ingoing and $n$~outgoing strands ($m\geq n$), associate the adjoint~$\Xisometry^\dagger$ of an isometric
charge-conserving map
\begin{align}
\Xisometry\in\End\left(\bigoplus_{\vec{a}\in\Omega_{\textrm{eff}}^n}V^{\vec{a}},\bigoplus_{\vec{a'}\in\Omega_{\textrm{eff}}^m} V^{\vec{a'}}\right)\ ,\label{eq:isometrictensor}
\end{align}
that is,  
\begin{align*}
\isometrybox\mapsto\Xisometry^\dagger
\end{align*}
\item Regard the state $\ket{\varphi}$ (Eq.~\eqref{eq:topstate}) and the family of maps~$\{\Xisometry\}$ (Eq.~\eqref{eq:isometrictensor}) as variational parameters specifying a state $\ket{\Psi}=\ket{\Psi_{\ket{\varphi},\{\Xisometry\}}}\in\cH_{\textrm{chain,N}}$ of the chain. The state is determined by the following recipe: in the MERA-structure of Fig.~\ref{fig:MERA}, replace every box by the superposition of trivalent labeled graphs representing the associated object (i.e., Eq.~\eqref{eq:topstate} and Eq.~\eqref{eq:isometrictensor}). The result is a superposition of labeled graphs, each with $N$~ingoing edges. This superposition represents~$\bra{\Psi}$.
\end{enumerate}

\noindent More explicitly, the maps~\eqref{eq:isometrictensor} are of the form 
\begin{align}
\Xisometry=\bigoplus_c \Xisometry(c)\ ,\label{eq:chargeconservisom}
\end{align}
with $\Xisometry(c)\in\End(\bigoplus_{\vec{a}\in\Omega_{\textrm{eff}}^n}V^{\vec{a}}_c,\bigoplus_{\vec{a'}\in\Omega_{\textrm{eff}}^m}V^{\vec{a'}}_c)$ satisfying
\begin{align}
\Xisometry(c)^\dagger\Xisometry(c)=\id_{\bigoplus_{\vec{a}\in\Omega_{\textrm{eff}}^n}V^{\vec{a}}_c}\ .\label{eq:isometrypropertyexplsec}
\end{align}
Eqs.~\eqref{eq:chargeconservisom} and~\eqref{eq:isometrypropertyexplsec} severly constrain the set of allowed maps $\Xisometry$ for certain~$(m,n)$ and~$\Omega_{\textrm{eff}}$.

This ansatz is clearly motivated by entanglement renormalization for qudits. In fact, it has the same  operational interpretation: the MERA-structure of Fig.~\ref{fig:MERA}, after replacing each box by the superposition of graphs specified by~$\{\Xisometry\}$, is a procedure for successively mapping the chain to a coarse-grained chain by local gates and isometries. Indeed, due to charge conservation, boxes with the same number of in- and outgoing edges correspond to local unitaries on the anyons, while boxes with fewer outputs than inputs correspond to local isometries reducing the number of degrees of freedom. Such  reductions  preserve the total charge in their support. Importantly, since the range of each operator~$\Xisometry^\dagger$  is contained in $\bigoplus_{\vec{a}\in\Omega^m_{\textrm{eff}}} V^{\vec{a}}$, states supported on $\cH_{\textrm{chain},N}$ are mapped to coarse-grained chains of the same type at each level in Fig~\ref{fig:MERA}, i.e., with particles from the subset~$\Omega_{\textrm{eff}}\subset\Omega$ on all sites.

\subsection{Efficient evaluation of physical quantites}
Having introduced a set of variational ansatz states parametrized by~$(\ket{\varphi},\{\Xisometry\})$, we argue that quantities of physical interest such as expectation values of local observables and correlation functions can be efficiently computed from these parameters. Here we use the formal equivalence of the manipulation rules of anyonic states and operators with usual tensor contractions. 

The anyonic analog of Eq.~\eqref{eq:mainMERA} is 
\begin{align} 
\scalebox{0.9}{\unitarythird} =\scalebox{0.9}{\isometryfirst}\ ,\label{eq:mainMERAanyonic} 
\end{align} 
for any operator $\Xisometry\in\End\left(\bigoplus_{\vec{a}\in\Omega_{\textrm{eff}}^n}V^{\vec{a}},\bigoplus_{\vec{a'}\in\Omega_{\textrm{eff}}^m} V^{\vec{a'}}\right)$ with the required isometry property (where $m\geq n$). Each of the small dark boxes represents the projection~$\id_{\Omega_{\textrm{eff}}}$ onto the subset of  allowed anyon labels. Note that, if the edges on the lhs.~of Eq.~\eqref{eq:mainMERAanyonic} are connected to an operator~$\Xisometry'$ as specified in the ansatz, these  may be omitted. This is because both the support and range of $\Xisometry'$ are already restricted to  the set of allowed anyons~$\Omega_{\textrm{eff}}$ (cf.~\eqref{eq:isometrictensor}). In particular, we formally recover the rule Eq.~\eqref{eq:mainMERA} in this case.

The second important ingredient is formula~\eqref{eq:traceofopperiodic} for the trace of an operator on the chain. Eq.~\eqref{eq:mainMERAanyonic} and~\eqref{eq:traceofopperiodic} immediately imply that  expectation values of local operators can be efficiently evaluated for a state~$\ket{\Psi}=\ket{\Psi_{\ket{\varphi},\{\Xisometry\}}}$ in essentially the same manner as for MERA-states of qudits. The same is true for two-point correlation functions for certain distances of the points related to the MERA structure.

Consider for example a local observable~$\Xsingleoperator$ acting on two sites of the chain.
The expectation value of this operator, given a  density matrix~$\rho$, is equal to the diagrammatic expression
\begin{align}
\tr(\Xsingleoperator \rho)=\left[\scalebox{0.9}{\traceexpectation}\right]_{\textrm{vac}}\ .\label{eq:localexpectationvalue}
\end{align}
If $\rho=\proj{\Psi}$ is an anyonic MERA-state corresponding e.g., to the structure of Fig.~\ref{fig:MERA}~(a), we conclude from~\eqref{eq:localexpectationvalue} that
\begin{align*}
\bra{\Psi}\Xsingleoperator\ket{\Psi}=\left[X\right]_{\textrm{vac}}\ ,
\end{align*}
where~$X$ is the superposition of trivalent labeled graphs specified by the diagram in Fig.~\ref{fig:MERAstateandexpectation}~(b). Using~\eqref{eq:mainMERAanyonic}, this immediately reduces to $\left[X'\right]_{\textrm{vac}}$, where~$X'$ is the superposition in Fig.~\ref{fig:MERAstateandexpectation}~(c). This can be efficiently evaluated using the superoperators defined in~\eqref{eq:extraction} and~\eqref{eq:extractionb} (again interpreted as anyonic expressions). Two-point correlations functions can be computed analogously.

\subsection{Computational cost and refinements of the ansatz}
 To count the number of parameters needed to describe the anyonic ansatz states, let $\cD=\max_{a\in\Omega_{\textrm{eff}}} d_a$ be the maximal quantum dimension of the particles used. Since the number of states of the form~\eqref{eq:standardbasis} can be upper bounded by~$O(\cD^n)$, a charge-conserving map as in~\eqref{eq:isometrictensor} is described by fewer than $O(\cD^{n+m})$ parameters. Similarly, a state~$\ket{\varphi}$ as in~\eqref{eq:chainstatetop} is described by $O(\cD^n)$~parameters.
 
As with the MERA for spin chains, the family of ansatz states may be enlarged by replacing  an edge by $s>1$~edges; this is analogous to increasing the bond dimension. In this case, isometries are described by $O(\cD^{s(n+m)})$ parameters as opposed to $O(d^{s(n+m)})$ in the qudit case.  (Note that, by definition, $\cD<d$~for  any embedding of anyonic states into qudits: for example, for the Fibonacci anyons considered below, $\cD\approx 1.618$.) In summary, a general anyonic MERA is described by $O(poly(\cD^s)N\log N)$~parameters, and translation-invariant and scale-invariant MERAs by $O(poly(\cD^s)\log N)$ and $O(poly(\cD^s))$~parameters, respectively.

The remainder of this  story is the same as that of the MERA for spin chains, and we refer to the extensive literature (e.g.,~\cite{evenblyvidal09}) on this subject. For example, the computational cost of computing local expectation values is roughly the same as for qudit chains (with~$\chi$ replaced by~$\cD^s$), and methods used, e.g., for varying over the isometries in Fig.~\ref{fig:MERA} can directly be adapted to the anyonic framework. Compared to the qudit chain setting, an additional advantage stems from the fact that the isometries are charge-preserving and thus take a block-diagonal form. Therefore,  matrix multiplication and singular value decompositions can be performed on matrices whose dimensions are a constant factor smaller than with a na\"ive qudit ansatz. Similarly, methods for extracting critical exponents from scale-invariant MERA states~\cite{Montangeroetal09,Pfeiferetal09} may be applied in the anyonic setting.

\subsection{Example:  Fibonacci anyons} 
The Fibonacci theory has one nontrivial particle~$\tau$ with quantum dimension $d_\tau=\phi=\frac{\sqrt{5}+1}{2}$ equal to the golden ratio and fusion rule $\tau\times\tau=1+\tau$. The $F$-matrix is given by
\begin{align}
\begin{matrix}
\fibn&=\frac{1}{\phi}\fibq+\frac{1}{\sqrt{\phi}}\fibp\\
\fibo&=\frac{1}{\sqrt{\phi}}\fibq-\frac{1}{\phi}\fibp\ ,
\end{matrix}\label{eq:fmatrixfib}
\end{align}
where we use the convention that a solid line represents the~$\tau$-label, while a dotted line represents the trivial label~$1$. 
Consider a periodic chain of $\tau$-anyons. To get a corresponding family of ansatz states, we set $\Omega_{\textrm{eff}}=\{\tau\}$. Using the convention that a solid line represents an edge with label~$\tau$ and a dotted line represents an edge with label~$1$, the Hilbert spaces $\cH_{\textrm{chain},n}$ of a periodic chains with $n\in\{1,2\}$~particles are spanned by (cf.~\eqref{eq:anyonicchainbasisvector})
\begin{align}
\cH_{\textrm{chain},1}&=\mathbb{C}\statea=:\mathbb{C}\ket{\tau}\label{eq:chainonefib}\\
\cH_{\textrm{chain},2}&=\textsf{span}\left\{\stateb,\statec,\stated\right\}\nonumber\\
&=:\textsf{span}\left\{\ket{\tau\tau},\ket{1\tau},\ket{\tau 1}\right\} .\nonumber\end{align} 
In general, the space $\cH_{\textrm{chain},n}$ is spanned by vectors corresponding to (periodic) sequences  of $1$s and $\tau$s, with the fusion constraint forbidding neighboring~$1$s (this defines an embedding into a subspace of~$(\mathbb{C}^2)^{\otimes n}$). This determines the form of the state $\ket{\varphi_{n}}\in\cH_{\textrm{chain},n}$ corresponding to the top box in structures as in Fig.~\ref{fig:MERA}~(a) and~(b), respectively. Next, we consider the constraints on the maps $\Xisometry=\Xisometry_{(n,m)}$ in~\eqref{eq:chargeconservisom} with $n$~input- and $m$~output anyons, for small~$(n,m)$.  The standard form of  charge preserving isometries/unitaries  is
\begin{align}
\hspace{-3ex}\Xisometry_{(1,2)} &=\frac{e^{i\theta}}{\phi^{1/4}}\fiba\label{eq:wonetwo}\\
\nonumber\\
\Xisometry_{(2,2)} &=\frac{e^{i\theta_1}}{\phi^{\textfrac{1}{2}}}\fibb+\frac{e^{i\theta_2}}{\phi}\fibc\label{eq:wtwotwo}\\
\nonumber\\
\Xisometry_{(1,3)} &=\frac{\alpha}{\phi^{\textfrac{1}{2}}}\fibd+\frac{\beta}{\phi^{\textfrac{1}{2}}}\fibe\label{eq:wonethree}\\
\nonumber\\
\Xisometry_{(2,3)} &=\frac{e^{i\theta}}{\phi^{\textfrac{5}{4}}}\fibf+\frac{1}{\phi^{3/4}}\left(\alpha\fibg+\beta\fibh\ \right)\label{eq:wtwothree}
\end{align} 
where $\theta\in [0,2\pi[$ and $|\alpha|^2+|\beta|^2=1$. In all these cases, the map is completely specified by a phase and/or a qubit state. A less trivial case is~$\Xisometry_{(3,3)}$, which is equal to a phase times the projection onto the span of~$\fibm$ plus a two-by-two unitary~$\Xisometry(\tau)$ on the span of $\{\fibn,\fibo\}$.  
For later reference, we also state the most general form of an isometry with two input and four output strands:
\begin{align}
\begin{matrix}
\Xisometry_{(2,4)}&=\frac{1}{\phi^{\textfrac{3}{2}}}\left(\alpha\ffiba+\beta\ffibb\ \right)\\
&\ +\frac{1}{\phi}\left(\gamma \ffibc+\delta\ffibd+\varepsilon\ffibe\ \right)
\end{matrix}\label{eq:generaltwofour}
\end{align}
where $|\alpha|^2+|\beta|^2+|\gamma|^2+|\delta|^2+|\varepsilon|^2=1$.

From~\eqref{eq:chainonefib},~\eqref{eq:wonetwo} and~\eqref{eq:wtwotwo}, we conclude that the family of states associated to Fig.~\ref{fig:MERA}~(a) is rather uninteresting as the variational parameters $(\ket{\varphi},\{\Xisometry\})$ are merely a set of phases. In contrast, the structure in Fig.~\ref{fig:MERA}~(b) gives rise to a less trivial family of states due to~\eqref{eq:wonethree}. 

We will give additional  non-trivial explicit examples in Section~\ref{sec:goldenchain}.

\subsection{Distillable states for composite anyon coding}
The MERA ansatz for qudits is motivated by quantum circuits. Indeed, a MERA-description of a state provides a circuit preparing the state starting from  the top-level state~$\ket{\varphi}$. This is achieved by realizing isometries using ancillas prepared in pure states; it corresponds to running the coarse-graining procedure in reverse. This use of the MERA has been proposed for example as a way of efficiently preparing topologically ordered states~\cite{AguadoVidal08}. 

The one-to-one correspondence between preparation circuits of a certain form and states described by the entanglement renormalization ansatz clearly extends to anyons. However, there is an additional relation in the anyonic setting which corresponds to running the coarse-graining forwards (instead of backwards as in the qudit case): a subclass of the anyonic ansatz is in one-to-one correspondence with  distillation procedures preparing a logical state~$\ket{\varphi}$ of composite anyons.  The anyonic renormalization ansatz therefore provides an alternative characterization of ``distillable'' states in the framework of composite anyon coding~\cite{Koe09}.

 The goal of composite anyon coding is to prepare a suitable  state for computation starting from some unknown initial state, but without using measurements. In the terminology of anyonic entanglement renormalization, composite anyons are anyons at higher levels in the coarse-graining scheme.  Initial states which allow to prepare a given target state~$\ket{\varphi}$ of composite anyons can be characterized as follows. They can be represented by an anyonic renormalization ansatz (corresponding to the preparation scheme) with the following properties:  the state at the top is fixed to~$\ket{\varphi}$, and all adjoints of isometries~$\Xisometry^\dagger$ are implementable by braiding and fusion. The latter condition
 means that the renormalization scheme only consists of unitaries effected by braiding, and coarse-graining, that is, any reduction in the number of anyons, is achieved by bunching together some particles. Such fusion-based coarse-graining is given by a product of the isometry 
\begin{align*}
\sum_{c}\ket{a,b;c}\bra{c}&=\sum_{c}\left(\frac{d_c}{d_ad_b}\right)^{\textfrac{1}{4}}\scalebox{0.9}{\twoanyonsecond}\ a=a(c), b=b(c)
\end{align*}
whose adjoint describes the fusion of a pair of particles. An anyonic renormalization ansatz with these properties can  directly be implemented using the operations commonly envisioned to be  available for manipulating anyons. Thus we can regard such schemes as  state preparation circuits for  topological quantum computers.

\section{Application to the golden chain and the Majumdar-Ghosh chain\label{sec:goldenchain}}
\subsection{The model}
In this section, we consider the use of the anyonic MERA ansatz in the context of the golden chain~\cite{Feiguinetal07} and its relatives. An introduction to these models can be found in~\cite{Trebstetal08}. Specifically, we  consider the Fibonacci-anyonic analog of the Heisenberg and the Majumdar-Gosh (MG) spin chains.  The former was introduced in~\cite{Feiguinetal07} and consists of a uniform chain of~$N$ Fibonacci anyons with Hamiltonian terms favoring one of either possible total charge of two neighboring~$\tau$-particles. Concretely, the Hamiltonian is given by
\begin{align}
J_2\cdot H^{\textrm{golden}} &= -J_2 \sum_{i=1}^N H^i_2\ ,\label{eq:hgolden}
\end{align}
where each term~$H^i_2$ is a projection onto trivial charge of the anyons~$i$ and~$i+1$, i.e., it has the diagrammatic representation~$\frac{1}{\phi}\fibc$. In analogy with the familiar $SU(2)$~spin chains, the case $J_2>0$ energetically favoring trivial total charge is referred to as `antiferromagnetic' (AFM) coupling, whereas~$J_2<0$ is called `ferromagnetic' (FM) coupling. 

The Majumdar-Gosh (MG) chain~\cite{MajumdarGosh69} is a model of~$SU(2)$ spin-$1/2$~particles arranged on a chain, with three-particle interactions favoring either total spin~$3/2$ (called ferromagnetic/FM) or~$1/2$ (called antiferromagnetic/AFM). Its anyonic analog~\cite{trebstetalcollective08} takes the form
\begin{align}
J_3\cdot H^{\textrm{MG}} &=-J_3\sum_{i=1}^N H^i_3\ ,\label{eq:hmajumdargosh}
\end{align}
where~$H^i_3$ is the projection onto trivial charge of three neighboring anyons.  Using the F-matrix~\eqref{eq:fmatrixfib}, it is straightforward to rewrite the  terms~$H^i_2$ and~$H^i_3$
in the standard basis~\eqref{eq:anyonicchainbasisvector}. Corresponding expressions can be found in~\cite{trebstetalcollective08}. Expressed in the standard embedding into~$(\mathbb{C}^2)^{\otimes n}$, this leads to $3$- and $4$-qubit terms, respectively.

\begin{figure}
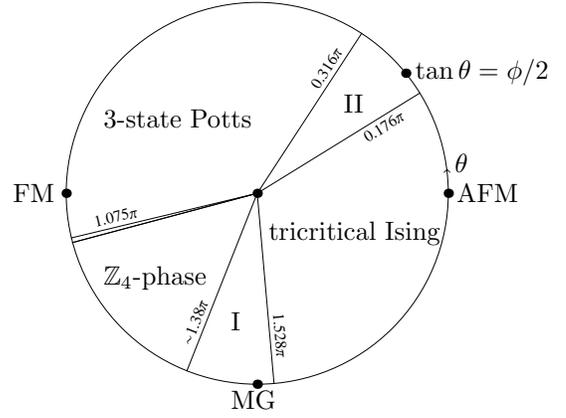

\phasediagram
\caption{Phase diagram of the model~\eqref{eq:oneparam} as obtained in~\cite{trebstetalcollective08}. Phases~I and~II are gapped, with the exact ground states known at the Majumdar-Gosh (MG) point $\theta=3\pi/2$ (see Section~\ref{sec:exactfixedpoint}) and at $\tan\theta=\phi/2$. There are two extended criticial phases for which an exact mapping was established to the $3$-state Potts and the tricritical Ising model at the FM- and AFM-golden chain points, respectively~\cite{Feiguinetal07}. A small sliver of an incommensurate phase is found near~$\theta=1.075\pi$ next to a phase with $\mathbb{Z}_4$-symmetry; both these phases are believed to be critical. See~\cite{trebstetalcollective08} for a detailed discussion.\label{fig:phasediagram}}
\end{figure}

The Hamiltonian~\eqref{eq:hgolden} was studied in detail in~\cite{Feiguinetal07}. Criticality and a two-dimensional CFT description were established numerically. Furthermore, an exact mapping to a standard integrable 2D classical lattice model~\cite{andrewsbaxterforrester83} known as the RSOS model was given. These studies were extended to the one-parameter family of Hamiltonians
\begin{align}
H_\theta &:=\cos\theta \cdot H^{\textrm{golden}}+\sin\theta\cdot H^{\textrm{MG}}\ ,\label{eq:oneparam}
\end{align}
which exhibits a rich phase diagram as discussed extensively in~\cite{trebstetalcollective08}, see Fig.~\ref{fig:phasediagram}. Subsequent work~\cite{Gilcollective09,LudwPoil10} considered generalizations to $su(2)_k$~anyons and established a connection between the  gapless modes of these anyon chains to edge states of topological liquids~\cite{Gilcollective09}, which then provides some insight into the collective behavior of anyons in a two-dimensional setting~\cite{Gilcollective09,LudwPoil10}. We refer to these references for further details, and do not attempt to give a complete account here.  Instead, we restrict  ourselves to a few example computations illustrating the power of the anyonic MERA ansatz. Specifically, we consider the model~\eqref{eq:oneparam}.

\subsection{Exact RG fixed point at the Majumdar-Gosh point\label{sec:exactfixedpoint}}
We first consider the FM (i.e., $J_3<0$) case of the MG chain, the anyonic analog of the Majumdar-Gosh point of the spin-1/2 Heisenberg chain for the point~$\theta=3\pi/2$ in~\eqref{eq:oneparam}. This point lies in a gapped phase extending from $\theta\approx 1.38\pi$ to $\theta\approx 1.528\pi$, with
four-fold degeneracy throughout the phase (for chains with even length)~\cite{trebstetalcollective08}. The ground space at this point is  spanned by the states
\begin{align}
\begin{matrix}
\ket{1\tau 1\tau 1\tau\cdots }&\propto \goshbbasis\\
\ket{\tau\tau_x\tau\tau_x\tau\tau_x\cdots}&\propto\goshdbasis
\end{matrix}\label{eq:MGgroundspace}
\end{align}
and their translates by one site. Here,  a wiggly line denotes a superposition
\begin{align*}
\tau_x&=\goshwiggle=\frac{1}{\phi}\goshdotted+\frac{1}{\sqrt{\phi}}\goshstraight\ .
\end{align*}
We will now argue that the two-dimensional subspace spanned by~\eqref{eq:MGgroundspace} is {\em exactly} described by a scale-invariant anyonic MERA of a simple form. In other words, these states are fixed points under the corresponding renormalization group procedure. This provides an encoding of a subspace of the ground space, with the property that the identity of the encoded state is  revealed only at the top level in the MERA structure.  This is analogous to MERA-descriptions of topologically ordered systems which are exact fixed points of a renormalization group flow: information encoded in a topologically degenerate ground space can be recovered at the top level of the MERA-hierarchy~\cite{AguadoVidal08,KoeReiVid09}. However, in the  case considered here, there is a local order parameter given by the density of $\tau$-labels on the chain.

To specify the scale-invariant MERA-ansatz describing the ground space~\eqref{eq:MGgroundspace}, consider the MERA-structure of  Figure~\ref{f:merafirst} with refinement parameter $s=2$, i.e., with every strand replaced by two. Due to this doubling and scale-invariance, the corresponding ansatz then is described by  an isometry~$\Xisometry=\Xisometry_{(2,4)}$ of the form~\eqref{eq:generaltwofour} and a unitary~$\Xunitary$ acting on four $\tau$-anyons. We set the unitary equal to the identity, and  the isometry equal to
\begin{align}
\Xisometry:=\frac{1}{\sqrt{\phi}}\goshisom\ \label{eq:MGisometry}
\end{align}
which corresponds to  the parameters
$(\alpha,\beta,\gamma,\delta,\epsilon) =(\frac{1}{\phi},\frac{1}{\sqrt{\phi}},0,\frac{1}{\phi},\frac{1}{\sqrt{\phi}})$
in~\eqref{eq:generaltwofour}. This completes the specification of the MERA up to translation (which we fix later), as the top-level state $\ket{\varphi}$ depends on the actual state considered. Note that the MERA-structure of  Figure~\ref{f:merafirst} with unitaries equal to the identity is known as a tree-tensor network~\cite{ShiDuanVidal06,TagliazzoEvenblyVidal09,Murgetaltree10} in the non-anyonic setting.

Let us argue that the renormalization group scheme  has the ground states~\eqref{eq:MGgroundspace} as fixed points.  This is most easily seen  for the state~$\ket{1\tau 1\tau 1\tau\cdots}$ using the diagrammatic calculus: applying a layer of (adjoints of) isometries corresponds to stacking~$N/4$ parallel copies of~\eqref{eq:MGisometry} on top of the state. We assume that the isometries are aligned in such as way that this takes the form
\begin{align*}
\goshb& \propto \goshbY\
\end{align*}
for the state~$\ket{1\tau 1\tau 1\tau\cdots}$, where we suppressed scalar factors. Here we used isotopy invariance and~\eqref{eq:bubbleremoval}.
 With~\eqref{eq:anyonicchainbasisvector}, it is easy to verify that the operation~$(\Xisometry^\dagger)^{\otimes N/4}$ also preserves the norm of the state. This establishes the claimed fixed-point property for $\ket{1\tau 1\tau 1\tau\cdots}$. To verify the claim for the second state $\ket{\tau\tau_x\tau\tau_x\tau\tau_x\cdots}$, it suffices to observe that this state is essentially equivalent to the former but with $\tau$-flux,
i.e.,
\begin{align*}
\goshdbasis\ \  &\propto &\goshdbasismodified
\end{align*}
because of the identity 
\begin{align*}\goshbmove= \goshamove\ .
\end{align*}
This immediately implies that this state is fixed by the same renormalization group scheme.

\subsection{Numerical variation over ansatz states}
To assess the suitability of the anyonic entanglement renormalization ansatz as a numerical method, we have implemented an algorithm for numerically miniziming the energy by varying over the parameters of a translation-invariant ansatz.  The algorithm is based on iterative optimization of the (identical) isometries at each level and the top-level state. It is described in detail in~\cite{evenblyvidal09} for non-anyonic spin chains.  For a fixed isometry, it proceeds by computing its {\em environment}, that is, the contraction of the MERA-network with the isometry omitted. The resulting tensor can be interpreted as a linear map whose singular value decomposition dictates how the isometry is updated. Adapting this to the anyonic setting is straightforward: here the environment always has a block-diagonal form with respect to total charge. Compared to the algorithm of~\cite{evenblyvidal09}, the only significant difference lies in the implementation of the ascending and descending superoperators (see e.g.,~\eqref{eq:extraction} and~\eqref{eq:extractionb}) used to compute the environments.  As with all anyonic operations, they require  applying basis changes into compatible tree-like bases (cf.~\eqref{eq:Fmove}). These basis changes can be precomputed. 

This randomized optimization algorithm is susceptible to local minima, and its convergence depends on the choice of initial points. In practice, these issues appear to be minor and  can be addressed by starting with a large number of initial points and postselecting after a few iterations.

\subsubsection*{Ground state energy and correlation functions}
We have applied the variational algorithm to periodic chains of 12~and 16~Fibonacci anyons governed by the Hamiltonian~\eqref{eq:oneparam}. These system sizes were chosen to allow for comparison with exact diagonalization data and to test the suitability of different MERA-structures. 

\begin{figure}
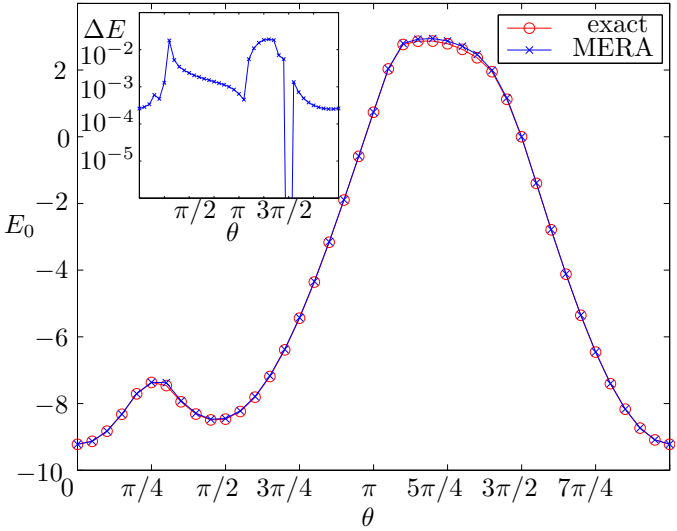
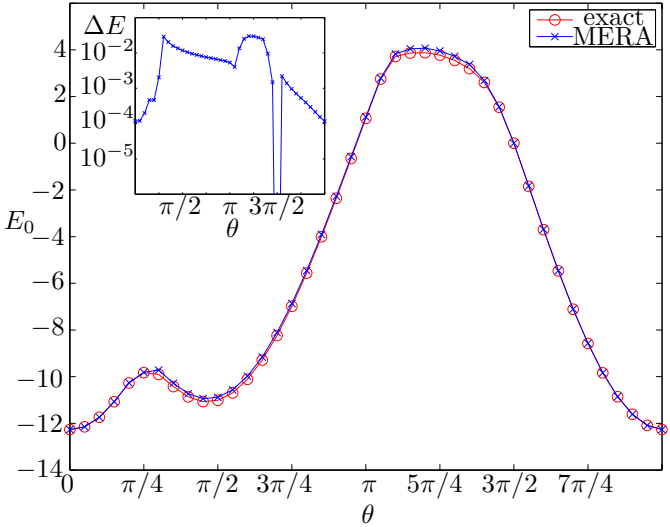

\subfigure[~Ternary MERA-structure applied to $12$~anyons]{\hspace{-0.6cm}\label{f:sixtwo}\scalebox{1.0}{\sixtotwomera}}
\subfigure[~Binary MERA-structure applied to $16$~anyons]{\hspace{-0.6cm}\label{f:fourtwo}\scalebox{1.0}{\fourtotwomera}}
\caption{Ground state energy  approximated variationally by an anyonic entanglement renormalization ansatz. The inset in each figure shows the relative error $\Delta E=(E_{\textrm{MERA}}-E_0)/(E_{\max}-E_0)$. Fig.~(a) is based on a ternary anyonic MERA for a system of 12~anyons, while Fig.~(b) shows the result of a binary anyonic MERA for a chain of 16~anyons. At the Majumdar-Gosh point ($\theta=3\pi/2$), the variational procedure recovers the exact fixed-point  discussed in Section~\ref{sec:exactfixedpoint} for the binary MERA. In~(b), the approximation is best around around the golden chain-point with AFM-couplings~$\theta=0$. This is consistent with the fact that the ground state has a $\mathbb{Z}_2$-sublattice ordering~\cite{Feiguinetal07} which is compatible with the coarse-graining structure of a binary MERA.  For $\theta=\pi$, the ground state has a $\mathbb{Z}_3$-sublattice ordering for which the binary MERA structure  is less suited. However, the approximation is still better than in intermediate regions where the Hamiltonian has both $2$-local and $3$-local terms. Similarly, Fig.~(a) shows that  a ternary structure appears to be suitable for capturing the $\mathbb{Z}_3$-sublattice ordering at the golden chain-FM-point $\theta=\pi$. \label{fig:gsenergy}}
\end{figure} 

For the 12- and 16-anyon chains, we use the ``ternary'' and ``binary'' MERA-structures
\begin{align}
\scalebox{0.9}{\merafourth}\qquad \scalebox{0.9}{\merathird}\ ,\label{eq:merastructurenumerical}
\end{align}
with $s=2$ (i.e., every strand is doubled). The former consists of a single level of coarse-graining isometries, while the latter has two such levels. The variationally obtained ground state energy is compared to its exact value in Fig.~\ref{fig:gsenergy}. As shown, we find good agreement between the variationally estimated ground state energy and its exact value, over a wide range of values of the parameter~$\theta$.

\begin{figure}
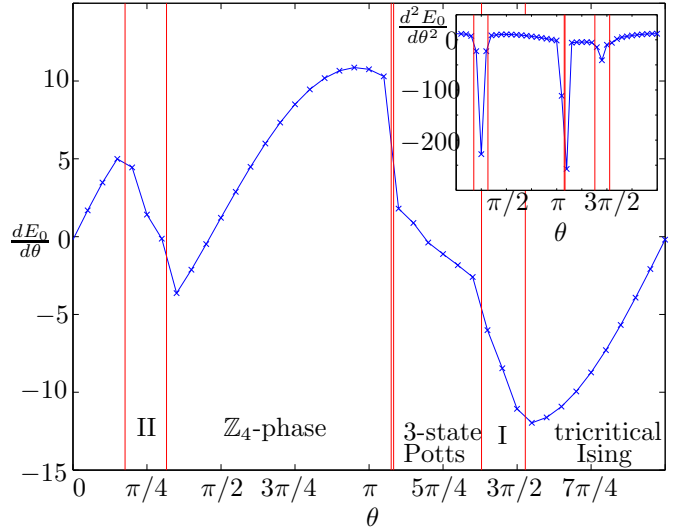

\fourtwomerapt
\caption{The first and second (inset) derivative of the ground state energy with respect to the parameter~$\theta$ approximately reveals  the location of the phase boundaries (red vertical lines). This data was obtained from the binary MERA for $16$~anyons used to produce Fig.~\ref{fig:gsenergy}~(b). For each of the $40$ points $\theta_0\in [0,2\pi]$, the ground state energies $E_0(\theta_0\pm \Delta \theta)$ at two neighboring points at distance $\Delta\theta=10^{-3}$ were approximated using the MERA. The plots show the corresponding discrete approximation to the first and second derivative at each point. We stress that only limited information can be gained from this plot. In particular, it does not reveal the nature of the phase transition. We refer to the detailed discussion in~\cite{Trebstetal08}, where e.g., the CFT descriptions of the transitions out of the tricritial Ising phase have been identified.\label{fig:phaseboundaries}}
\end{figure}

Fig.~\ref{fig:phaseboundaries} shows that ground state energies computed using the binary MERA are sufficient to obtain a rough estimate for the location of the phase boundaries.

\begin{figure}
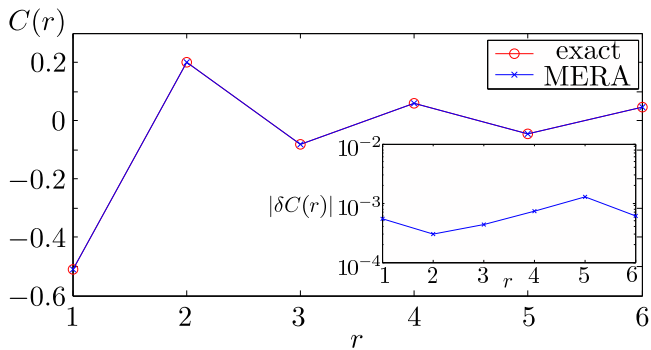
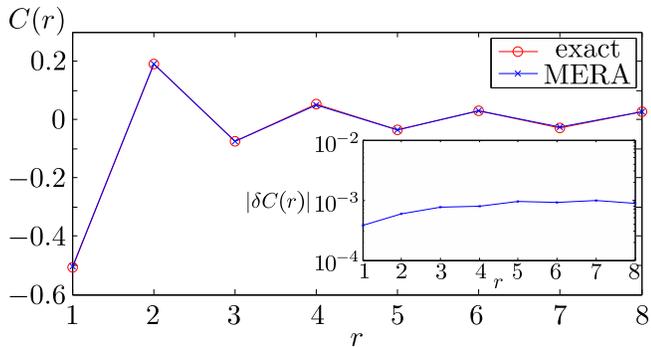

\centering 
\begin{tabular}{cc }
\hspace{-0.5cm}\subfigure[ $C(r)$ with the ternary MERA at the AFM-point]{\label{f:AFMpointcorr}\scalebox{0.9}{\allcorrsixtwo}}\\
\hspace{-0.5cm}\subfigure[ $C(r)$ with the binary MERA at the AFM-point]{\label{f:AFMpointcorrb}\scalebox{0.9}{\allcorrfourtwo}}
\end{tabular}
\caption{Two-point correlation function~$C(r)$ (cf.~\eqref{eq:chargecorrelationfct}) of the local topological charge density at the AFM  point. They reveal a~$\mathbb{Z}_2$-sublattice ordering of the ground state wave function. The inset shows the absolute error  $|\delta C(r)|=|C_{\textrm{MERA}}(r)-C_{\textrm{exact}}(r)|$. \label{fig:afmcorrelations}}
\end{figure}

\begin{figure}
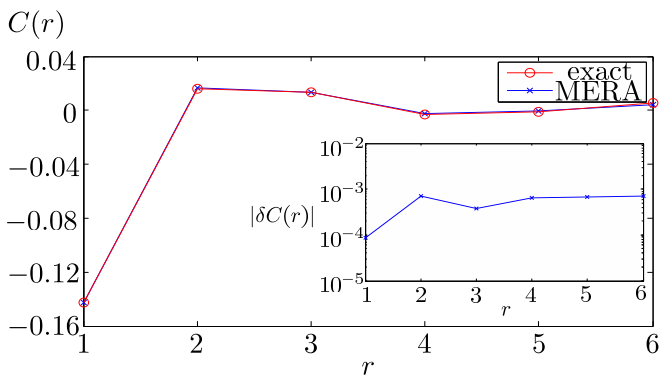
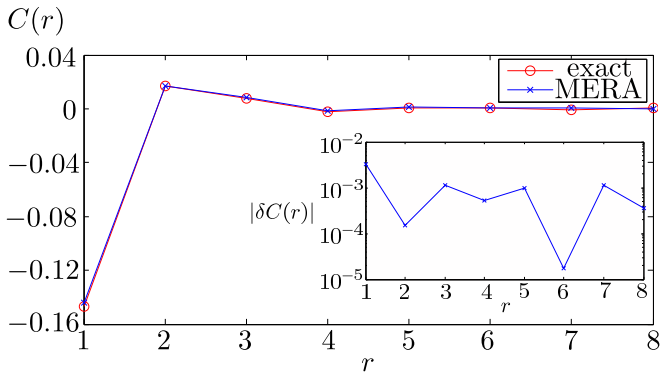

\centering 
\begin{tabular}{cc }
\hspace{-0.5cm}\subfigure[ $C(r)$ with the ternary MERA at the FM-point]{\label{f:FMpointcorr}\scalebox{0.9}{\allcorrsixtwoFM}}\\
\hspace{-0.5cm}\subfigure[ $C(r)$ with the binary MERA at the FM-point]{\label{f:FMpointcorrb}\scalebox{0.9}{\allcorrfourtwoFM}}
\end{tabular}
\caption{Two-point correlation function~$C(r)$ (cf.~\eqref{eq:chargecorrelationfct}) of the local topological charge density at the FM  point.  The inset shows the absolute error  $|\delta C(r)|=|C_{\textrm{MERA}}(r)-C_{\textrm{exact}}(r)|$. \label{fig:fmcorrelations}}
\end{figure}

To study whether the anyonic MERA correctly reproduces correlations in the ground state, we have computed the (translation-averaged) 
two-point correlation functions
\begin{align}
C(r)&= \frac{1}{N}\sum_{i=1}^N \left(
\langle H^i_2H^{i+r}_2\rangle-\langle H^i_2\rangle\cdot\langle H^{i+r}_2\rangle\right)\label{eq:chargecorrelationfct}
\end{align} 
of the local topological charge density (as measured in terms of the local projection~$H^i_2$ onto trivial charge for a pair).  
The result of this computation for the AFM-point $\theta=0$ and the FM-point $\theta=\pi$ are shown in Fig.~\ref{fig:afmcorrelations} and Fig.~\ref{fig:fmcorrelations}, respectively.  They exhibit a remarkably good agreement with the exact values.

\begin{figure}
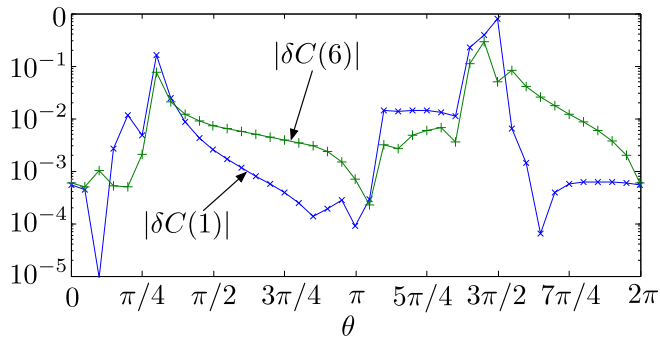
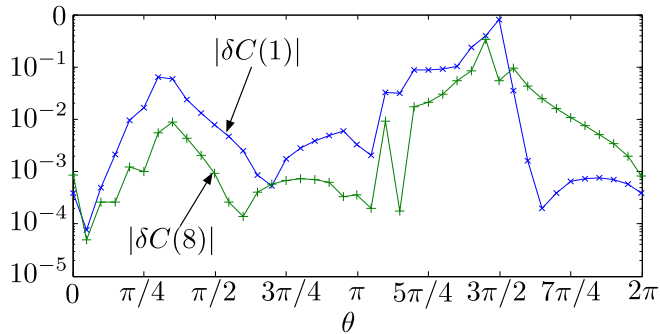

\begin{tabular}{cc } 
\hspace{-0.5cm}\subfigure[$|\delta C(1)|$ and $|\delta C(6)|$ for the ternary MERA]{\label{f:correlationfirst}\scalebox{0.9}{\onesixcorrsixtwo}}\\
\hspace{-0.5cm}\subfigure[$|\delta C(1)|$ and $|\delta C(8)|$ for the binary MERA]{\label{f:correlationsecond}\scalebox{0.9}{\oneeightcorrfourtwo}}
\end{tabular}
\caption{The absolute error $|\delta C(r)|=|C_{\textrm{MERA}}(r)-C_{\textrm{exact}}(r)|$ of the nearest-neighbor~($r=1$) and long-range ($r=6$ respectively $r=8$) charge-charge correlation functions. Qualitatively, these deviations agree with the errors in the ground state energy (see Fig.~\ref{fig:gsenergy}). Note that at the Majumdar-Gosh point $(\theta=3\pi/2)$, the plotted~$|\delta C|$ is large even though the MERA accurately represents one of the ground states. This is because of the degeneracy and the fact that $C_{\textrm{exact}}$ is computed from the completely mixed state on the ground space.\label{fig:twopoint}}
\end{figure}

In Fig.~\ref{fig:twopoint}, we show the  error when computing nearest-neighbor and long-range correlations as a function of the Hamiltonian parameter~$\theta$. We observe good accuracy in regions where the ground state energy is well approximated (compare Fig.~\ref{fig:gsenergy}). As expected, the considered MERA-structures are less suited for describing ground state correlations at intermediate values of~$\theta$, where the Hamiltonian has both $2$-anyon and $3$-anyon interactions.

We emphasize that the structures~\eqref{eq:merastructurenumerical} with~$s=2$ are two of the simplest possible leading to  non-trivial families of ansatz states for Fibonacci anyons.  The support of the unitaries/isometries ($4$~anyons) is only marginally larger than that of the terms in the Hamiltonian~\eqref{eq:oneparam}. This suggests that  the approximation by such states may be rather coarse. Nevertheless, the ansatz  provides reasonable approximations to the ground state energies (for all~$\theta$), and correlation functions at the AFM- and FM-points where the Hamiltonian consists of nearest-neighbor-terms. This illustrates that anyonic entanglement renormalization successfully exploits the constraints imposed by conservation of topological charge. Future work may go beyond this proof of principle by considering refined families with parameters~$s>2$. This should lead to significant improvements as in the non-anyonic setting: Here accurate results for correlation functions are usually obtained only for high bond dimension (e.g., $\chi=22$ for Ising and Potts chains at criticality~\cite{evenblyvidal09}).

\subsubsection*{Larger systems}
To test the scalability of the method, we have additionally computed the ground state energy of chains of length~$N\in\{32,64,128\}$ using the binary MERA-structure (with $s=2$) obtained by adding levels to~\eqref{eq:merastructurenumerical}. Since this is beyond the reach of exact diagonalization, we consider the AFM-point, which allows us to compare our results to the CFT-predictions of~\cite{Feiguinetal07}.

Explicitly, we use the fact that the low-lying spectrum of a periodic $1D$~critical quantum systems of length~$N$ takes the form
\begin{align}
E=\varepsilon N+\frac{2\pi v}{N}\left(h_L+h_R-\frac{c}{12}\right)\ .\label{eq:CFTspectrum}
\end{align}
Here $\varepsilon$ and $v$ are non-universal constants, $c$~is the central charge of the CFT, and~$h_L$ and~$h_R$ are the conformal weights of the holomorphic and antiholomorphic part of the local  field associated with the energy level.  The latter parameters  are defined in terms of a representation of the  Virasoro algebra and are tabulated for unitary minimal CFTs.  In~\cite{Feiguinetal07}, the CFT corresponding to the AFM-point was unambiguously identified as that describing the tricritical Ising model at its critical point, with central charge~$c=7/10$. The ground state energy~$E_0$ corresponds to $h_L=h_R=0$, whereas the first excited energy~$E_1$ is determined by $h_L=h_R=3/80$.

Using  the exact values of $E_0$ and $E_1$ for $N=16$, we determine the non-universal constants in~\eqref{eq:CFTspectrum} (approximately). The resulting prediction for the ground state energy density~$E_0/N$ for system sizes $N\leq 16$ differs by only~$10^{-2}\%$ from that obtained in the same way using the exact spectrum at $N=8$. This suggest that finite-size effects are negligible, and we use the prediction~\eqref{eq:CFTspectrum} obtained in this fashion to study the anyonic MERA for larger systems.  

\begin{figure}
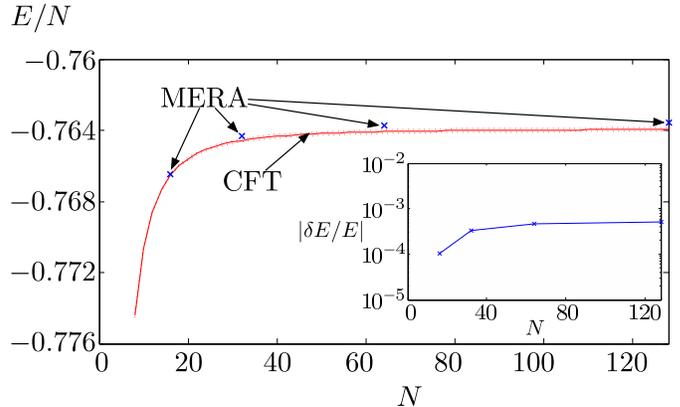

\scalebox{0.9}{\epersite}
\caption{Ground state energy per site obtained for  $N=16,32,64$ and $128$ anyons at the AFM-point using the anyonic MERA (blue crosses). The red line shows the CFT prediction for the ground state  energy  obtained by extrapolating from $N=16$ (as discussed in the text). The dotted curves indicate the (modulus of the) deviation from this prediction when using~$N=8$ instead. The inset illustrates the relative error of the MERA with respect to the CFT prediction.
\label{fig:scalinglarge}}
\end{figure}

Fig.~\ref{fig:scalinglarge} shows the result of this computation. We find that the ground state energy density is well approximated by the anyonic MERA ansatz even for larger systems. 

Throughout this paper, we have considered finite-size systems. A modification of the optimization algorithm for scale-invariant qudit MERA allows to numerically estimate data of the associated CFT in the continuum limit~\cite{Giovannettietal08,Pfeiferetal09,silvietal09}. Adapting this to the anyonic setting,
this can provide an additional benchmark when comparing to the predictions of~\cite{Feiguinetal07,trebstetalcollective08}. Indeed, results of this kind were recently presented in~\cite{Pfeiferetal10} (see note below).

\section{Conclusions\label{sec:conclusions}}
We have proposed a variational ansatz for periodic chains of interacting anyons. It is the natural counterpart of entanglement renormalization for spin chains. Based on the empirical evidence for the numerical accuracy of latter, it is reasonable to expect that the ansatz is a powerful tool for describing critical anyonic systems. Indeed, we have obtained numerical evidence for its suitability   by comparing with exact diagonalization data in the case of (a variation of) the golden chain.

Our ansatz  makes optimal use of the anyonic structure of the Hilbert space by incorporating conservation of topological charge at different scales. We expect this to lead to  significant computational savings and improved accuracy compared to more conventional methods based on embedding the anyons into qubits.  It may be more pronounced for more general anyon models than the Fibonacci anyon chains numerically studied here.

Beyond providing an  efficient numerical tool, the proposed ansatz is  a starting point for interesting generalizations. For example, along the lines of~\cite{evenblyvidal09}, one may define a Hamiltonian renormalization group flow based on the ansatz. This  flow generalizes the perturbative renormalization prescription analytically considered in~\cite{BonesteelYang07,Fidkowskietalrandom,Fidkowskietal08b} in the context of random couplings.  More importantly, the current work may serve as a stepping stone for the development of variational methods for 2D~systems of interacting anyons. We give a rough sketch of corresponding adaptations in Appendix~\ref{app:twodmerageneralization}.

The transition from entanglement renormalization for qudits  to anyons is remarkably simple on a conceptual level: it boils down to the replacement of isometries by topological charge-preserving isometries, and a reinterpretation of networks in terms of the isotopy-invariant calculus. This suggest that more generally, tensor network states such as MPS or PEPS may also be adapted to the anyonic setting using similar substitutions. This should add significantly to the repertoire of variational methods for topologically ordered systems.

\vspace{2ex}
\begin{acknowledgments}
The authors thank Simon Trebst and the anonymous referees for numerous useful suggestions. They also thank Robert Pfeifer, Guifr\'e Vidal and Julien Vidal for comments, and Jens Eisert for bringing~\cite{Bartheletal09,Pinedaetal10} to their attention. RK acknowledges support by the Swiss National Science Foundation (SNF) under grant PA00P2-126220. He also thanks Station Q for their hospitality and Lukas Fidkowski, John Preskill and Ben Reichardt for discussions.  EB is supported by DoE under grant no.~DE-FG03-92-E40701 and NSF under grant no.~PHY-0803371.
\end{acknowledgments} 

\vspace{2ex}
\noindent
{\em Note added:} Shortly after posting this work to the arXiv, Pfeifer et al. have also presented work on the anyonic MERA~\cite{Pfeiferetal10}. They use the scale-invariant MERA to  numerically extract CFT data
in the thermodynamic limit, studying the Fibonacci-golden chain  at the AFM point. 
\appendix   
\section{The inner product\label{app:innproduct}}
\subsection{Anyons on a torus\label{app:periodicchain}}
In this appendix, we show that the 
basis states $\{\ket{\vec{a},\vec{b}}\}$ of $\bigoplus_{\vec{a}}V^{\vec{a}}_{\textrm{periodic}}$ (cf.~\eqref{eq:anyonicchainbasisvector}) are indeed orthonormal with respect to the Hilbert-Schmidt inner product defined by the trace~\eqref{eq:traceofopperiodic}. By definition, we have 
\begin{align*}
|\langle\vec{a'},\vec{b'}|\vec{a},\vec{b}\rangle|^2=\prod_i \frac{\delta_{a_i,a_i'}}{d_{a_i}^{1/2}}\left[\chainsecond\right]_{\textrm{vac}}
\end{align*}
Inserting the projection  $\id_{\{(e_1,e_2)\}}$  onto a pair of anyons~$(e_1,e_2)$ decomposed into total charge~$c$ as
\begin{align*}
\id_{\{(e_1,e_2)\}}&=\parallelfirst=\sum_c \sqrt{\frac{d_c}{d_{e_1}d_{e_2}}}\parallelsecond
\end{align*}
into the horizontal lines leads to 
\begin{align}
|\langle\vec{a'},\vec{b'}|\vec{a},\vec{b}\rangle|^2=\delta_{\vec{a},\vec{a'}}\sum_{\vec{c}}\left(\prod_i \frac{1}{\sqrt{d_{a_i}}}\right)\left[X\right]_{\textrm{vac}}\label{eq:innerproducttoproveeq}
\end{align}
where 
\begin{align*}
X=\sum_{\vec{c}}\sqrt{\prod_j \frac{d_{c_j}}{d_{b_j}d_{b_j'}}}\chainthird
\end{align*}
Since we are interested in the vacuum coefficient of $X$, we can set all $c_i=1$,  getting
\begin{align}
\hspace{-1ex}[X]_{\textrm{vac}}&=\prod_j\frac{\delta_{b_j,b_j'}}{d_{b_j}}\left[\chainfourth\right]_{\textrm{vac}}\hspace{-2ex}.\label{eq:Xvacuumdef}
\end{align}
But each $\theta$-like graph on the rhs.~is proportional to the empty graph, with scalar
\begin{align}
\left[\chainfifth\right]_{\textrm{vac}} &=\sqrt{d_{a}d_bd_c}\ .\label{eq:thetagraph}
\end{align}
This can be verified by applying~\eqref{eq:bubbleremoval} twice.
Inserting~\eqref{eq:thetagraph} into~\eqref{eq:Xvacuumdef} gives
\begin{align*}
[X]_{\textrm{vac}} &=\prod_{j} \delta_{b_j,b_j'} \sqrt{d_{a_j}}\ .
\end{align*}
When combined with~\eqref{eq:innerproducttoproveeq}, this implies the claim. 

\vspace{4ex}

\subsection{Anyons on a disc\label{app:disctrace}}
For the spaces $\bigoplus_{\vec{a}\in\Omega^n, c\in\Omega}V^{\vec{a}}_c$ of $n$~anyons on a disc, the (partial) trace  can be defined in a similar diagrammatic manner as for the space~$\bigoplus_{\vec{a}} V^{\vec{a}}_{\textrm{periodic}}$ of a periodic chain. This leads to a slight modification when defining the anyonic MERA for anyons on a disc.  For completeness, we include a short description here. 

A partial trace corresponds to connecting up in- and outgoing strands of an operator, while  simultaneously inserting the operator 
\begin{align}
\scalebox{0.9}{\factoropfirst}=\scalebox{0.9}{\factoropsecond}=\scalebox{0.9}{\factoropthird}=\sum_c d^{-1}_c \proj{c}\ \label{eq:aperiodiccorrection}
\end{align}
into each line.  (Here we use the convention that 
either input- or output  of an operator which acts diagonally on anyon labels  may be represented by an undirected edge.) For example, the partial  trace over  the $n$-th anyon for an operator~$U$ acting on $n$~anyons takes the form
\begin{align*}
\tr_{n}\scalebox{0.9}{\operatorfirst} &= \scalebox{0.9}{\operatorthird}\ .
\end{align*} 
The (complete) trace is obtained by connecting up all the strands in this way and computing the vacuum coefficient, i.e., 
\begin{align}
\tr\scalebox{0.8}{\operatorfirst} &= \left[\scalebox{0.8}{\operatorsecond}\right]_{\textrm{vac}}\hspace{-2ex}. \label{eq:completetrace} 
\end{align}
The diagram on the rhs.~in~\eqref{eq:completetrace} is in fact proportional to the vacuum graph. Using~\eqref{eq:bubbleremoval} repeatedly, it is straightforward to check that the states~\eqref{eq:standardbasis} are orthonormal.
 
\section{Braiding and more general  arrangements of anyons\label{app:braidingetc}\label{app:twodmerageneralization}}
In the main text, our focus is on one-dimensional chains of anyons. Here we briefly comment on the generalization of the anyonic entanglement renormalization ansatz to systems of anyons arranged in a more general way (e.g., a regular lattice). Such systems have been studied before (see~\cite{LudwPoil10}), and the  modifications necessary to define corresponding anyonic Hamiltonians are nicely explained in~\cite{Trebstetal08}. In fact, these modifications directly carry over to anyonic entanglement renormalization when a linear ordering of the anyons  at every level is chosen. An analogous situation arises when considering fermionic tensor networks, and corresponding techniques~\cite{Bartheletal09,Pinedaetal10} can thus be extended to anyons.

\begin{figure}
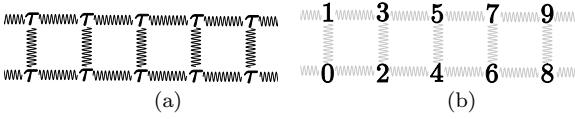
  
\centering
\begin{tabular}{ccc}
\subfigure[]{\hspace{-0.6cm}{\label{f:twochainladder}\scalebox{0.8}{\twochain}}}
&
\subfigure[]{\hspace{-0.6cm}\label{f:ladderlabeling}\scalebox{0.8}{\twochainnumbered}}
\end{tabular}
\caption{Two-chain ladder arrangement of Fibonacci anyons as discussed in~\cite{Trebstetal08}, with nearest-neighbor interactions. We consider the ordering indicated in~(b). \label{fig:twochainladder}}
\end{figure}

For concreteness, we first consider the example of  the two-leg ladder model discussed in~\cite{Trebstetal08}, closely following that presentation. This is a system of two chains of $m$~anyons each  placed next to each other as shown in Fig.~\ref{f:twochainladder} along a non-trivial cycle on the torus. The figure also indicates two-anyon nearest neighbor interactions, which are e.g., projections onto trivial charge as in the golden chain. 

The Hilbert space of this system is $V^{\vec{a}}_{\textrm{periodic}}$ with $\vec{a}=\underbrace{(\tau,\ldots,\tau)}_{2m}$, as this space depends only on the topology of the surface (an $n=2m$-punctured torus) and the boundary labels (all equal to~$\tau$).  However, the basis~\eqref{eq:anyonicchainbasisvector} is now ambiguous and we must choose a linear ordering of the anyons. A convenient choice of an ordering is shown in Fig.~\ref{f:ladderlabeling}. 

Given this ordering, some of the nearest-neighbor terms in the Hamiltonian
now act on  non-neighboring pairs $(i,i+2)$~of strands  in the diagrammatic representation. To make sense of such terms, it is necessary to introduce an additional basis change corresponding to a transposition $(i,i+1)\mapsto (i+1,i)$. Physically, such basis changes correspond to movements of the anyons, and this is where the (non-abelian) braid group statistics appears: pairs of neighboring anyons may be exchanged in either clockwise or anticlockwise fashion. Thus there are two inequivalent ways of transposing neighboring pairs.   It is natural to define the Hamiltonian  as the result of transposing, applying the charge projection, and undoing the transposition, averaged over either version of exchanging.

 In the diagrammatic formalism, clockwise- and anticlockwise exchanges of neighboring anyons are represented by over- and undercrossings, respectively. The following computational rules involving the $R$-matrix of the modular category are then added to the isotopy-invariant calculus:
\begin{align*}
\scalebox{0.9}{\braida}=R^{ab}_c\scalebox{0.9}{\trivalent}\qquad\qquad\scalebox{0.9}{\braidb}=\overline{R^{a^*b^*}_{c^*}}\scalebox{0.9}{\trivalent}
\end{align*}
With~\eqref{eq:Fmove}-\eqref{eq:vacuumlineaddition}, these rules imply that over- and undercrossings may be resolved according to
\begin{align}
\begin{matrix}
\scalebox{0.9}{\crossinga} &=\sum_{c}\sqrt{\frac{d_c}{d_ad_b}}R^{ab}_c\scalebox{0.9}{\resolvedcrossing}\\
\vspace{0.25ex}\\
\scalebox{0.9}{\crossingb}&=\sum_{c}\sqrt{\frac{d_c}{d_ad_b}}\overline{R^{a^*b^*}_{c^*}}\scalebox{0.9}{\resolvedcrossing}
\end{matrix} \label{eq:undercrossresolved}
\end{align}
where the sums are restricted to fusion-compatible labels~$c$.  Eq.~\eqref{eq:undercrossresolved} gives the matrix elements of the linear operators $B_1,B_2:\bigoplus_c V^{ab}_c\rightarrow\bigoplus V^{ba}_c$ corresponding to the two types of exchanges. They are unitary if $a=b$.

Given this definition, we can write down the interaction corresponding to preferred trivial charge of a `horizontal' pair $(2j,2j+2)$ of neighboring anyons on the bottom chain in Fig.~\ref{f:twochainladder} as follows:
\begin{align*}
\sum_{a,b}\braidproja+\sum_{a,b}\braidprojb
\end{align*}
These are $3$-anyon operators acting on the anyons indexed~$(2j,2j+1,2j+2)$. Analogous expressions apply to pairs on the upper chain. 

The two-chain ladder is one of the simplest examples where transpositions need to be used to define physically interesting Hamiltonians. In a more generally arrangment of anyons, such as a regular $2D$-lattice, the same procedure applies. Depending the chosen standard ordering, however, more transpositions may be required to apply a local operator to a subset of anyons. For example, a projection onto trivial charge of two anyons may take the form
\begin{align}
\sum_a\braidprojc\label{eq:multianyonbraid}
\end{align}
where we sum over particles for undirected edges.

 The fact that~\eqref{eq:multianyonbraid} is a multi-anyon non-local operator now appears to be an obstacle to the use of the anyonic entanglement renormalization ansatz.  But the special structure of~\eqref{eq:multianyonbraid} allows to efficiently evaluate its expectation value for suitable MERA structures in spite of this. This is because crossings can be pushed past isometries to higher levels using the fact that
\begin{align*}
\pusha=\pushab
\end{align*}
(and similarly for undercrossings), and be absorbed into the isometries/unitaries by applying the linear maps~\eqref{eq:undercrossresolved} to their inputs and outputs, respectively. This eventually reduces the evaluation to the expectation value of a local operator for a related anyonic MERA-state.   The efficiency of this procedure depends on the number of crossings that need to be resolved, and therefore on a judicious choice of orderings. This feature is identical to fermionic tensor networks/MERAs, where it is necessary to keep track of the number of swaps.

\input{main.bbl}

\begin{thebibliography}{10}%
\makeatletter
\providecommand \@ifxundefined [1]{%
 \ifx #1\undefined \expandafter \@firstoftwo
 \else \expandafter \@secondoftwo
\fi
}%
\providecommand \@ifnum [1]{%
 \ifnum #1\expandafter \@firstoftwo
 \else \expandafter \@secondoftwo
\fi
}%
\providecommand \enquote [1]{``#1''}%
\providecommand \bibnamefont  [1]{#1}%
\providecommand \bibfnamefont [1]{#1}%
\providecommand \citenamefont [1]{#1}%
\providecommand\href[0]{\@sanitize\@href}%
\providecommand\@href[1]{\endgroup\@@startlink{#1}\endgroup\@@href}%
\providecommand\@@href[1]{#1\@@endlink}%
\providecommand \@sanitize [0]{\begingroup\catcode`\&12\catcode`\#12\relax}%
\@ifxundefined \pdfoutput {\@firstoftwo}{%
 \@ifnum{\z@=\pdfoutput}{\@firstoftwo}{\@secondoftwo}%
}{%
 \providecommand\@@startlink[1]{\leavevmode\special{html:<a href="#1">}}%
 \providecommand\@@endlink[0]{\special{html:</a>}}%
}{%
 \providecommand\@@startlink[1]{%
  \leavevmode
  \pdfstartlink
   attr{/Border[0 0 1 ]/H/I/C[0 1 1]}%
   user{/Subtype/Link/A<</Type/Action/S/URI/URI(#1)>>}%
  \relax
 }%
 \providecommand\@@endlink[0]{\pdfendlink}%
}%
\providecommand \url  [0]{\begingroup\@sanitize \@url }%
\providecommand \@url [1]{\endgroup\@href {#1}{\urlprefix}}%
\providecommand \urlprefix [0]{URL }%
\providecommand \Eprint[0]{\href }%
\@ifxundefined \urlstyle {%
  \providecommand \doi [1]{doi:\discretionary{}{}{}#1}%
}{%
  \providecommand \doi [0]{doi:\discretionary{}{}{}\begingroup
  \urlstyle{rm}\Url }%
}%
\providecommand \doibase [0]{http://dx.doi.org/}%
\providecommand \Doi[1]{\href{\doibase#1}}%
\providecommand \bibAnnote [3]{%
  \BibitemShut{#1}%
  \begin{quotation}\noindent
    \textsc{Key:}\ #2\\\textsc{Annotation:}\ #3%
  \end{quotation}%
}%
\providecommand \bibAnnoteFile [2]{%
  \IfFileExists{#2}{\bibAnnote {#1} {#2} {\input{#2}}}{}%
}%
\providecommand \typeout [0]{\immediate \write \m@ne }%
\providecommand \selectlanguage [0]{\@gobble}%
\providecommand \bibinfo [0]{\@secondoftwo}%
\providecommand \bibfield [0]{\@secondoftwo}%
\providecommand \translation [1]{[#1]}%
\providecommand \BibitemOpen[0]{}%
\providecommand \bibitemStop [0]{}%
\providecommand \bibitemNoStop [0]{.\EOS\space}%
\providecommand \EOS [0]{\spacefactor3000\relax}%
\providecommand \BibitemShut [1]{\csname bibitem#1\endcsname}%
\bibitem{Kitaev03}%
  \BibitemOpen
  \bibfield{author}{%
  \bibinfo {author} {\bibfnamefont{A.~Y.}\ \bibnamefont{Kitaev}},\ }%
  \bibfield{journal}{%
  \bibinfo {journal} {Annals of Physics}\ }%
  \textbf{\bibinfo {volume} {303}},\ \bibinfo {pages} {2 } (\bibinfo {year}
  {2003})%
  \bibAnnoteFile{NoStop}{Kitaev03}%
\bibitem{preskill}%
  \BibitemOpen
  \bibfield{author}{%
  \bibinfo {author} {\bibfnamefont{J.}~\bibnamefont{Preskill}},\ }%
  \enquote{\bibinfo {title} {Topological quantum computation},}\ \bibinfo
  {howpublished} {Lecture Notes (Chapter 9)} (\bibinfo {year} {2004}),\
  \url{www.theory.caltech.edu/people/preskill/ph229/}%
  \bibAnnoteFile{NoStop}{preskill}%
\bibitem{Freedmanetal03}%
  \BibitemOpen
  \bibfield{author}{%
  \bibinfo {author} {\bibfnamefont{M.~H.}\ \bibnamefont{Freedman}}, \bibinfo
  {author} {\bibfnamefont{A.}~\bibnamefont{Kitaev}}, \bibinfo {author}
  {\bibfnamefont{M.~J.}\ \bibnamefont{Larsen}},\ and\ \bibinfo {author}
  {\bibfnamefont{Z.}~\bibnamefont{Wang}},\ }%
  \bibfield{journal}{%
  \bibinfo {journal} {Bull. Amer. Math. Soc.}\ }%
  \textbf{\bibinfo {volume} {40}},\ \bibinfo {pages} {31} (\bibinfo {year}
  {2003}),\
  \Eprint{http://arxiv.org/abs/arXiv:quant-ph/0101025}{arXiv:quant-ph/0101025}%
  \bibAnnoteFile{NoStop}{Freedmanetal03}%
\bibitem{Nayaketal08}%
  \BibitemOpen
  \bibfield{author}{%
  \bibinfo {author} {\bibfnamefont{C.}~\bibnamefont{Nayak}}, \bibinfo {author}
  {\bibfnamefont{S.~H.}\ \bibnamefont{Simon}}, \bibinfo {author}
  {\bibfnamefont{A.}~\bibnamefont{Stern}}, \bibinfo {author}
  {\bibfnamefont{M.}~\bibnamefont{Freedman}},\ and\ \bibinfo {author}
  {\bibfnamefont{S.}~\bibnamefont{Das~Sarma}},\ }%
  \bibfield{journal}{%
  \Doi{10.1103/RevModPhys.80.1083}{\bibinfo {journal} {Rev. Mod. Phys.}}\ }%
  \textbf{\bibinfo {volume} {80}},\ \bibinfo {pages} {1083} (\bibinfo {month}
  {Sep}\ \bibinfo {year} {2008})%
  \bibAnnoteFile{NoStop}{Nayaketal08}%
\bibitem{MooreRead}%
  \BibitemOpen
  \bibfield{author}{%
  \bibinfo {author} {\bibfnamefont{G.}~\bibnamefont{Moore}}\ and\ \bibinfo
  {author} {\bibfnamefont{N.}~\bibnamefont{Read}},\ }%
  \bibfield{journal}{%
  \bibinfo {journal} {Nuclear Physics B}\ }%
  \textbf{\bibinfo {volume} {360}},\ \bibinfo {pages} {362 } (\bibinfo {year}
  {1991})%
  \bibAnnoteFile{NoStop}{MooreRead}%
\bibitem{readrezayia}%
  \BibitemOpen
  \bibfield{author}{%
  \bibinfo {author} {\bibfnamefont{N.}~\bibnamefont{Read}}\ and\ \bibinfo
  {author} {\bibfnamefont{E.}~\bibnamefont{Rezayi}},\ }%
  \bibfield{journal}{%
  \Doi{10.1103/PhysRevB.59.8084}{\bibinfo {journal} {Phys. Rev. B}}\ }%
  \textbf{\bibinfo {volume} {59}},\ \bibinfo {pages} {8084} (\bibinfo {month}
  {Mar}\ \bibinfo {year} {1999})%
  \bibAnnoteFile{NoStop}{readrezayia}%
\bibitem{readrezayib}%
  \BibitemOpen
  \bibfield{author}{%
  \bibinfo {author} {\bibfnamefont{E.~H.}\ \bibnamefont{Rezayi}}\ and\ \bibinfo
  {author} {\bibfnamefont{N.}~\bibnamefont{Read}},\ }%
  \bibfield{journal}{%
  \Doi{10.1103/PhysRevB.79.075306}{\bibinfo {journal} {Phys. Rev. B}}\ }%
  \textbf{\bibinfo {volume} {79}},\ \bibinfo {pages} {075306} (\bibinfo {month}
  {Feb}\ \bibinfo {year} {2009})%
  \bibAnnoteFile{NoStop}{readrezayib}%
\bibitem{dolevetal}%
  \BibitemOpen
  \bibfield{author}{%
  \bibinfo {author} {\bibfnamefont{M.}~\bibnamefont{Dolev}}, \bibinfo {author}
  {\bibfnamefont{M.}~\bibnamefont{Heiblum}}, \bibinfo {author}
  {\bibfnamefont{V.}~\bibnamefont{Umansky}}, \bibinfo {author}
  {\bibfnamefont{A.}~\bibnamefont{Stern}},\ and\ \bibinfo {author}
  {\bibfnamefont{D.}~\bibnamefont{Mahalu}},\ }%
  \bibfield{journal}{%
  \Doi{doi:10.1038/nature06855}{\bibinfo {journal} {Nature}}\ }%
  \textbf{\bibinfo {volume} {452}},\ \bibinfo {pages} {829} (\bibinfo {year}
  {2008}),\ \Eprint{http://arxiv.org/abs/arXiv:0802.0930}{arXiv:0802.0930}%
  \bibAnnoteFile{NoStop}{dolevetal}%
\bibitem{raduetal}%
  \BibitemOpen
  \bibfield{author}{%
  \bibinfo {author} {\bibfnamefont{I.~P.}\ \bibnamefont{Radu}}, \bibinfo
  {author} {\bibfnamefont{J.~B.}\ \bibnamefont{Miller}}, \bibinfo {author}
  {\bibfnamefont{C.~M.}\ \bibnamefont{Marcus}}, \bibinfo {author}
  {\bibfnamefont{M.~A.}\ \bibnamefont{Kastner}}, \bibinfo {author}
  {\bibfnamefont{L.~N.}\ \bibnamefont{Pfeiffer}},\ and\ \bibinfo {author}
  {\bibfnamefont{K.~W.}\ \bibnamefont{West}},\ }%
  \bibfield{journal}{%
  \Doi{10.1126/science.1157560}{\bibinfo {journal} {Science}}\ }%
  \textbf{\bibinfo {volume} {320}},\ \bibinfo {pages} {899} (\bibinfo {year}
  {2008}),\ \Eprint{http://arxiv.org/abs/arXiv:0803.3530}{arXiv:0803.3530}%
  \bibAnnoteFile{NoStop}{raduetal}%
\bibitem{willettetal}%
  \BibitemOpen
  \bibfield{author}{%
  \bibinfo {author} {\bibfnamefont{R.~L.}\ \bibnamefont{Willett}}, \bibinfo
  {author} {\bibfnamefont{L.~N.}\ \bibnamefont{Pfeiffer}},\ and\ \bibinfo
  {author} {\bibfnamefont{K.~W.}\ \bibnamefont{West}},\ }%
  \bibfield{journal}{%
  \bibinfo {journal} {Proceedings of the National Academy of Sciences}\ }%
  \textbf{\bibinfo {volume} {106}} (\bibinfo {year} {2009}),\
  \Eprint{http://arxiv.org/abs/arXiv:0807.0221}{arXiv:0807.0221}%
  \bibAnnoteFile{NoStop}{willettetal}%
\bibitem{FuKane05}%
  \BibitemOpen
  \bibfield{author}{%
  \bibinfo {author} {\bibfnamefont{C.~L.}\ \bibnamefont{Kane}}\ and\ \bibinfo
  {author} {\bibfnamefont{E.~J.}\ \bibnamefont{Mele}},\ }%
  \bibfield{journal}{%
  \Doi{10.1103/PhysRevLett.95.226801}{\bibinfo {journal} {Phys. Rev. Lett.}}\
  }%
  \textbf{\bibinfo {volume} {95}},\ \bibinfo {pages} {226801} (\bibinfo {month}
  {Nov}\ \bibinfo {year} {2005})%
  \bibAnnoteFile{NoStop}{FuKane05}%
\bibitem{MooreBalents07}%
  \BibitemOpen
  \bibfield{author}{%
  \bibinfo {author} {\bibfnamefont{J.~E.}\ \bibnamefont{Moore}}\ and\ \bibinfo
  {author} {\bibfnamefont{L.}~\bibnamefont{Balents}},\ }%
  \bibfield{journal}{%
  \Doi{10.1103/PhysRevB.75.121306}{\bibinfo {journal} {Phys. Rev. B}}\ }%
  \textbf{\bibinfo {volume} {75}},\ \bibinfo {pages} {121306} (\bibinfo {month}
  {Mar}\ \bibinfo {year} {2007})%
  \bibAnnoteFile{NoStop}{MooreBalents07}%
\bibitem{LiangKaneMele07}%
  \BibitemOpen
  \bibfield{author}{%
  \bibinfo {author} {\bibfnamefont{L.}~\bibnamefont{Fu}}, \bibinfo {author}
  {\bibfnamefont{C.~L.}\ \bibnamefont{Kane}},\ and\ \bibinfo {author}
  {\bibfnamefont{E.~J.}\ \bibnamefont{Mele}},\ }%
  \bibfield{journal}{%
  \Doi{10.1103/PhysRevLett.98.106803}{\bibinfo {journal} {Phys. Rev. Lett.}}\
  }%
  \textbf{\bibinfo {volume} {98}},\ \bibinfo {pages} {106803} (\bibinfo {month}
  {Mar}\ \bibinfo {year} {2007})%
  \bibAnnoteFile{NoStop}{LiangKaneMele07}%
\bibitem{Roy09}%
  \BibitemOpen
  \bibfield{author}{%
  \bibinfo {author} {\bibfnamefont{R.}~\bibnamefont{Roy}},\ }%
  \bibfield{journal}{%
  \Doi{10.1103/PhysRevB.79.195322}{\bibinfo {journal} {Phys. Rev. B}}\ }%
  \textbf{\bibinfo {volume} {79}},\ \bibinfo {pages} {195322} (\bibinfo {month}
  {May}\ \bibinfo {year} {2009})%
  \bibAnnoteFile{NoStop}{Roy09}%
\bibitem{KitaevAnyons}%
  \BibitemOpen
  \bibfield{author}{%
  \bibinfo {author} {\bibfnamefont{A.}~\bibnamefont{Kitaev}},\ }%
  \bibfield{journal}{%
  \bibinfo {journal} {Ann. Phys.}\ }%
  \textbf{\bibinfo {volume} {321}},\ \bibinfo {pages} {2} (\bibinfo {year}
  {2006}),\
  \Eprint{http://arxiv.org/abs/arXiv:cond-mat/0506438v3}{arXiv:cond-mat/050643%
8v3}%
  \bibAnnoteFile{NoStop}{KitaevAnyons}%
\bibitem{LevinWen}%
  \BibitemOpen
  \bibfield{author}{%
  \bibinfo {author} {\bibfnamefont{M.~A.}\ \bibnamefont{Levin}}\ and\ \bibinfo
  {author} {\bibfnamefont{X.-G.}\ \bibnamefont{Wen}},\ }%
  \bibfield{journal}{%
  \bibinfo {journal} {Phys. Rev. B}\ }%
  \textbf{\bibinfo {volume} {71}},\ \bibinfo {pages} {045110} (\bibinfo {year}
  {2005}),\
  \Eprint{http://arxiv.org/abs/arXiv:cond-mat/0404617}{arXiv:cond-mat/0404617}%
  \bibAnnoteFile{NoStop}{LevinWen}%
\bibitem{duanetal03}%
  \BibitemOpen
  \bibfield{author}{%
  \bibinfo {author} {\bibfnamefont{L.-M.}\ \bibnamefont{Duan}}, \bibinfo
  {author} {\bibfnamefont{E.}~\bibnamefont{Demler}},\ and\ \bibinfo {author}
  {\bibfnamefont{M.~D.}\ \bibnamefont{Lukin}},\ }%
  \bibfield{journal}{%
  \Doi{10.1103/PhysRevLett.91.090402}{\bibinfo {journal} {Phys. Rev. Lett.}}\
  }%
  \textbf{\bibinfo {volume} {91}},\ \bibinfo {pages} {090402} (\bibinfo {month}
  {Aug}\ \bibinfo {year} {2003}),\
  \Eprint{http://arxiv.org/abs/arXiv:cond-mat/0210564}{arXiv:cond-mat/0210564}%
  \bibAnnoteFile{NoStop}{duanetal03}%
\bibitem{michelietal06}%
  \BibitemOpen
  \bibfield{author}{%
  \bibinfo {author} {\bibfnamefont{A.}~\bibnamefont{Micheli}}, \bibinfo
  {author} {\bibfnamefont{G.~K.}\ \bibnamefont{Brennen}},\ and\ \bibinfo
  {author} {\bibfnamefont{P.}~\bibnamefont{Zoller}},\ }%
  \bibfield{journal}{%
  \Doi{10.1038/nphys287}{\bibinfo {journal} {Nature Physics}}\ }%
  \textbf{\bibinfo {volume} {2}},\ \bibinfo {pages} {341} (\bibinfo {year}
  {2006}),\
  \Eprint{http://arxiv.org/abs/arXiv:quant-ph/0512222}{arXiv:quant-ph/0512222}%
  \bibAnnoteFile{NoStop}{michelietal06}%
\bibitem{trebstetal07}%
  \BibitemOpen
  \bibfield{author}{%
  \bibinfo {author} {\bibfnamefont{S.}~\bibnamefont{Trebst}}, \bibinfo {author}
  {\bibfnamefont{P.}~\bibnamefont{Werner}}, \bibinfo {author}
  {\bibfnamefont{M.}~\bibnamefont{Troyer}}, \bibinfo {author}
  {\bibfnamefont{K.}~\bibnamefont{Shtengel}},\ and\ \bibinfo {author}
  {\bibfnamefont{C.}~\bibnamefont{Nayak}},\ }%
  \bibfield{journal}{%
  \Doi{10.1103/PhysRevLett.98.070602}{\bibinfo {journal} {Phys. Rev. Lett.}}\
  }%
  \textbf{\bibinfo {volume} {98}},\ \bibinfo {pages} {070602} (\bibinfo {month}
  {Feb}\ \bibinfo {year} {2007}),\
  \Eprint{http://arxiv.org/abs/cond-mat/0609048v1}{cond-mat/0609048v1}%
  \bibAnnoteFile{NoStop}{trebstetal07}%
\bibitem{JVidal09}%
  \BibitemOpen
  \bibfield{author}{%
  \bibinfo {author} {\bibfnamefont{J.}~\bibnamefont{Vidal}}, \bibinfo {author}
  {\bibfnamefont{S.}~\bibnamefont{Dusuel}},\ and\ \bibinfo {author}
  {\bibfnamefont{K.~P.}\ \bibnamefont{Schmidt}},\ }%
  \bibfield{journal}{%
  \Doi{10.1103/PhysRevB.79.033109}{\bibinfo {journal} {Phys. Rev. B}}\ }%
  \textbf{\bibinfo {volume} {79}},\ \bibinfo {pages} {033109} (\bibinfo {month}
  {Jan}\ \bibinfo {year} {2009})%
  \bibAnnoteFile{NoStop}{JVidal09}%
\bibitem{JVidal09b}%
  \BibitemOpen
  \bibfield{author}{%
  \bibinfo {author} {\bibfnamefont{J.}~\bibnamefont{Vidal}}, \bibinfo {author}
  {\bibfnamefont{R.}~\bibnamefont{Thomale}}, \bibinfo {author}
  {\bibfnamefont{K.~P.}\ \bibnamefont{Schmidt}},\ and\ \bibinfo {author}
  {\bibfnamefont{S.}~\bibnamefont{Dusuel}},\ }%
  \bibfield{journal}{%
  \Doi{10.1103/PhysRevB.80.081104}{\bibinfo {journal} {Phys. Rev. B}}\ }%
  \textbf{\bibinfo {volume} {80}},\ \bibinfo {pages} {081104} (\bibinfo {month}
  {Aug}\ \bibinfo {year} {2009})%
  \bibAnnoteFile{NoStop}{JVidal09b}%
\bibitem{Tupitsynetal08}%
  \BibitemOpen
  \bibfield{author}{%
  \bibinfo {author} {\bibfnamefont{I.}~\bibnamefont{Tupitsyn}}, \bibinfo
  {author} {\bibfnamefont{A.}~\bibnamefont{Kitaev}}, \bibinfo {author}
  {\bibfnamefont{N.}~\bibnamefont{Prokof{'}ev}},\ and\ \bibinfo {author}
  {\bibfnamefont{P.}~\bibnamefont{Stamp}},\ }%
  \enquote{\bibinfo {title} {Topological multicritical point in the toric code
  and 3d gauge {H}iggs models},}\
  \Eprint{http://arxiv.org/abs/arXiv:0804.3175}{arXiv:0804.3175}%
  \bibAnnoteFile{NoStop}{Tupitsynetal08}%
\bibitem{bravyihastings10}%
  \BibitemOpen
  \bibfield{author}{%
  \bibinfo {author} {\bibfnamefont{S.}~\bibnamefont{Bravyi}}\ and\ \bibinfo
  {author} {\bibfnamefont{M.}~\bibnamefont{Hastings}}}%
   (\bibinfo {year} {2010}),\
  \Eprint{http://arxiv.org/abs/arXiv:1001.4363}{arXiv:1001.4363}%
  \bibAnnoteFile{NoStop}{bravyihastings10}%
\bibitem{bravyihastingsmichalakis10}%
  \BibitemOpen
  \bibfield{author}{%
  \bibinfo {author} {\bibfnamefont{S.}~\bibnamefont{Bravyi}}, \bibinfo {author}
  {\bibfnamefont{M.~B.}\ \bibnamefont{Hastings}},\ and\ \bibinfo {author}
  {\bibfnamefont{S.}~\bibnamefont{Michalakis}}}%
   (\bibinfo {year} {2010}),\
  \Eprint{http://arxiv.org/abs/arXiv:1001.0344}{arXiv:1001.0344}%
  \bibAnnoteFile{NoStop}{bravyihastingsmichalakis10}%
\bibitem{klich09}%
  \BibitemOpen
  \bibfield{author}{%
  \bibinfo {author} {\bibfnamefont{I.}~\bibnamefont{Klich}}}%
   (\bibinfo {year} {2009}),\
  \Eprint{http://arxiv.org/abs/arXiv:0912.0945}{arXiv:0912.0945}%
  \bibAnnoteFile{NoStop}{klich09}%
\bibitem{bonderson09}%
  \BibitemOpen
  \bibfield{author}{%
  \bibinfo {author} {\bibfnamefont{P.}~\bibnamefont{Bonderson}},\ }%
  \bibfield{journal}{%
  \Doi{10.1103/PhysRevLett.103.110403}{\bibinfo {journal} {Phys. Rev. Lett.}}\
  }%
  \textbf{\bibinfo {volume} {103}},\ \bibinfo {pages} {110403} (\bibinfo
  {month} {Sep}\ \bibinfo {year} {2009}),\
  \Eprint{http://arxiv.org/abs/arXiv:0905.2726}{arXiv:0905.2726}%
  \bibAnnoteFile{NoStop}{bonderson09}%
\bibitem{Feiguinetal07}%
  \BibitemOpen
  \bibfield{author}{%
  \bibinfo {author} {\bibfnamefont{A.}~\bibnamefont{Feiguin}}, \bibinfo
  {author} {\bibfnamefont{S.}~\bibnamefont{Trebst}}, \bibinfo {author}
  {\bibfnamefont{A.~W.~W.}\ \bibnamefont{Ludwig}}, \bibinfo {author}
  {\bibfnamefont{M.}~\bibnamefont{Troyer}}, \bibinfo {author}
  {\bibfnamefont{A.}~\bibnamefont{Kitaev}}, \bibinfo {author}
  {\bibfnamefont{Z.}~\bibnamefont{Wang}},\ and\ \bibinfo {author}
  {\bibfnamefont{M.~H.}\ \bibnamefont{Freedman}},\ }%
  \bibfield{journal}{%
  \Doi{10.1103/PhysRevLett.98.160409}{\bibinfo {journal} {Phys. Rev. Lett.}}\
  }%
  \textbf{\bibinfo {volume} {98}},\ \bibinfo {pages} {160409} (\bibinfo {month}
  {Apr}\ \bibinfo {year} {2007}),\
  \Eprint{http://arxiv.org/abs/arXiv:cond-mat/0612341}{arXiv:cond-mat/0612341}%
  \bibAnnoteFile{NoStop}{Feiguinetal07}%
\bibitem{Fidkowskietalrandom}%
  \BibitemOpen
  \bibfield{author}{%
  \bibinfo {author} {\bibfnamefont{L.}~\bibnamefont{Fidkowski}}, \bibinfo
  {author} {\bibfnamefont{G.}~\bibnamefont{Refael}}, \bibinfo {author}
  {\bibfnamefont{N.~E.}\ \bibnamefont{Bonesteel}},\ and\ \bibinfo {author}
  {\bibfnamefont{J.~E.}\ \bibnamefont{Moore}},\ }%
  \bibfield{journal}{%
  \Doi{10.1103/PhysRevB.78.224204}{\bibinfo {journal} {Phys. Rev. B}}\ }%
  \textbf{\bibinfo {volume} {78}},\ \bibinfo {pages} {224204} (\bibinfo {month}
  {Dec}\ \bibinfo {year} {2008})%
  \bibAnnoteFile{NoStop}{Fidkowskietalrandom}%
\bibitem{Fidkowskietal08b}%
  \BibitemOpen
  \bibfield{author}{%
  \bibinfo {author} {\bibfnamefont{L.}~\bibnamefont{Fidkowski}}, \bibinfo
  {author} {\bibfnamefont{H.-H.}\ \bibnamefont{Lin}}, \bibinfo {author}
  {\bibfnamefont{P.}~\bibnamefont{Titum}},\ and\ \bibinfo {author}
  {\bibfnamefont{G.}~\bibnamefont{Refael}},\ }%
  \bibfield{journal}{%
  \Doi{10.1103/PhysRevB.79.155120}{\bibinfo {journal} {Phys. Rev. B}}\ }%
  \textbf{\bibinfo {volume} {79}},\ \bibinfo {pages} {155120} (\bibinfo {month}
  {Apr}\ \bibinfo {year} {2009})%
  \bibAnnoteFile{NoStop}{Fidkowskietal08b}%
\bibitem{BonesteelYang07}%
  \BibitemOpen
  \bibfield{author}{%
  \bibinfo {author} {\bibfnamefont{N.~E.}\ \bibnamefont{Bonesteel}}\ and\
  \bibinfo {author} {\bibfnamefont{K.}~\bibnamefont{Yang}},\ }%
  \bibfield{journal}{%
  \Doi{10.1103/PhysRevLett.99.140405}{\bibinfo {journal} {Phys. Rev. Lett.}}\
  }%
  \textbf{\bibinfo {volume} {99}},\ \bibinfo {pages} {140405} (\bibinfo {month}
  {Oct}\ \bibinfo {year} {2007})%
  \bibAnnoteFile{NoStop}{BonesteelYang07}%
\bibitem{white92}%
  \BibitemOpen
  \bibfield{author}{%
  \bibinfo {author} {\bibfnamefont{S.~R.}\ \bibnamefont{White}},\ }%
  \bibfield{journal}{%
  \Doi{10.1103/PhysRevLett.69.2863}{\bibinfo {journal} {Phys. Rev. Lett.}}\ }%
  \textbf{\bibinfo {volume} {69}},\ \bibinfo {pages} {2863} (\bibinfo {month}
  {Nov}\ \bibinfo {year} {1992})%
  \bibAnnoteFile{NoStop}{white92}%
\bibitem{Koe09}%
  \BibitemOpen
  \bibfield{author}{%
  \bibinfo {author} {\bibfnamefont{R.}~\bibnamefont{K\"onig}},\ }%
  \bibfield{journal}{%
  \Doi{10.1103/PhysRevA.81.052309}{\bibinfo {journal} {Phys. Rev. A}}\ }%
  \textbf{\bibinfo {volume} {81}},\ \bibinfo {pages} {052309} (\bibinfo {month}
  {May}\ \bibinfo {year} {2010})%
  \bibAnnoteFile{NoStop}{Koe09}%
\bibitem{Vidal07}%
  \BibitemOpen
  \bibfield{author}{%
  \bibinfo {author} {\bibfnamefont{G.}~\bibnamefont{Vidal}},\ }%
  \bibfield{journal}{%
  \Doi{10.1103/PhysRevLett.99.220405}{\bibinfo {journal} {Phys. Rev. Lett.}}\
  }%
  \textbf{\bibinfo {volume} {99}},\ \bibinfo {pages} {220405} (\bibinfo {month}
  {Nov}\ \bibinfo {year} {2007})%
  \bibAnnoteFile{NoStop}{Vidal07}%
\bibitem{Vidal08}%
  \BibitemOpen
  \bibfield{author}{%
  \bibinfo {author} {\bibfnamefont{G.}~\bibnamefont{Vidal}},\ }%
  \bibfield{journal}{%
  \Doi{10.1103/PhysRevLett.101.110501}{\bibinfo {journal} {Phys. Rev. Lett.}}\
  }%
  \textbf{\bibinfo {volume} {101}},\ \bibinfo {pages} {110501} (\bibinfo
  {month} {Sep}\ \bibinfo {year} {2008})%
  \bibAnnoteFile{NoStop}{Vidal08}%
\bibitem{Rom97}%
  \BibitemOpen
  \bibfield{author}{%
  \bibinfo {author} {\bibfnamefont{S.}~\bibnamefont{Rommer}}\ and\ \bibinfo
  {author} {\bibfnamefont{S.}~\bibnamefont{\"Ostlund}},\ }%
  \bibfield{journal}{%
  \Doi{10.1103/PhysRevB.55.2164}{\bibinfo {journal} {Phys. Rev. B}}\ }%
  \textbf{\bibinfo {volume} {55}},\ \bibinfo {pages} {2164} (\bibinfo {month}
  {Jan}\ \bibinfo {year} {1997})%
  \bibAnnoteFile{NoStop}{Rom97}%
\bibitem{Vid04}%
  \BibitemOpen
  \bibfield{author}{%
  \bibinfo {author} {\bibfnamefont{G.}~\bibnamefont{Vidal}},\ }%
  \bibfield{journal}{%
  \Doi{10.1103/PhysRevLett.93.040502}{\bibinfo {journal} {Phys. Rev. Lett.}}\
  }%
  \textbf{\bibinfo {volume} {93}},\ \bibinfo {pages} {040502} (\bibinfo {month}
  {Jul}\ \bibinfo {year} {2004})%
  \bibAnnoteFile{NoStop}{Vid04}%
\bibitem{Verstr04}%
  \BibitemOpen
  \bibfield{author}{%
  \bibinfo {author} {\bibfnamefont{F.}~\bibnamefont{Verstraete}}, \bibinfo
  {author} {\bibfnamefont{D.}~\bibnamefont{Porras}},\ and\ \bibinfo {author}
  {\bibfnamefont{J.~I.}\ \bibnamefont{Cirac}},\ }%
  \bibfield{journal}{%
  \Doi{10.1103/PhysRevLett.93.227205}{\bibinfo {journal} {Phys. Rev. Lett.}}\
  }%
  \textbf{\bibinfo {volume} {93}},\ \bibinfo {pages} {227205} (\bibinfo {month}
  {Nov}\ \bibinfo {year} {2004})%
  \bibAnnoteFile{NoStop}{Verstr04}%
\bibitem{VerstrMurgCirac08}%
  \BibitemOpen
  \bibfield{author}{%
  \bibinfo {author} {\bibfnamefont{F.}~\bibnamefont{Verstraete}}, \bibinfo
  {author} {\bibfnamefont{V.}~\bibnamefont{Murg}},\ and\ \bibinfo {author}
  {\bibfnamefont{J.~I.}\ \bibnamefont{Cirac}},\ }%
  \bibfield{journal}{%
  \Doi{10.1080/14789940801912366}{\bibinfo {journal} {Advances in Physics}}\ }%
  \textbf{\bibinfo {volume} {57}},\ \bibinfo {pages} {143} (\bibinfo {month}
  {Mar}\ \bibinfo {year} {2008})%
  \bibAnnoteFile{NoStop}{VerstrMurgCirac08}%
\bibitem{VerstraeteCirac06}%
  \BibitemOpen
  \bibfield{author}{%
  \bibinfo {author} {\bibfnamefont{F.}~\bibnamefont{Verstraete}}\ and\ \bibinfo
  {author} {\bibfnamefont{J.~I.}\ \bibnamefont{Cirac}},\ }%
  \bibfield{journal}{%
  \Doi{10.1103/PhysRevB.73.094423}{\bibinfo {journal} {Phys. Rev. B}}\ }%
  \textbf{\bibinfo {volume} {73}},\ \bibinfo {pages} {094423} (\bibinfo {month}
  {Mar}\ \bibinfo {year} {2006})%
  \bibAnnoteFile{NoStop}{VerstraeteCirac06}%
\bibitem{Trebstetal08}%
  \BibitemOpen
  \bibfield{author}{%
  \bibinfo {author} {\bibfnamefont{S.}~\bibnamefont{Trebst}}, \bibinfo {author}
  {\bibfnamefont{M.}~\bibnamefont{Troyer}}, \bibinfo {author}
  {\bibfnamefont{Z.}~\bibnamefont{Wang}},\ and\ \bibinfo {author}
  {\bibfnamefont{A.~W.~W.}\ \bibnamefont{Ludwig}},\ }%
  \bibfield{journal}{%
  \Doi{10.1143/PTPS.176.384}{\bibinfo {journal} {Progr. of Theor. Phys.
  Suppl.}}\ }%
  \textbf{\bibinfo {volume} {176}},\ \bibinfo {pages} {384} (\bibinfo {year}
  {2008})%
  \bibAnnoteFile{NoStop}{Trebstetal08}%
\bibitem{EvenblyVidal10}%
  \BibitemOpen
  \bibfield{author}{%
  \bibinfo {author} {\bibfnamefont{G.}~\bibnamefont{Evenbly}}\ and\ \bibinfo
  {author} {\bibfnamefont{G.}~\bibnamefont{Vidal}},\ }%
  \bibfield{journal}{%
  \bibinfo {journal} {New Journal of Physics}\ }%
  \textbf{\bibinfo {volume} {12}},\ \bibinfo {pages} {025007} (\bibinfo {year}
  {2010})%
  \bibAnnoteFile{NoStop}{EvenblyVidal10}%
\bibitem{CorbozVidal09}%
  \BibitemOpen
  \bibfield{author}{%
  \bibinfo {author} {\bibfnamefont{P.}~\bibnamefont{Corboz}}\ and\ \bibinfo
  {author} {\bibfnamefont{G.}~\bibnamefont{Vidal}},\ }%
  \bibfield{journal}{%
  \Doi{10.1103/PhysRevB.80.165129}{\bibinfo {journal} {Phys. Rev. B}}\ }%
  \textbf{\bibinfo {volume} {80}},\ \bibinfo {pages} {165129} (\bibinfo {month}
  {Oct}\ \bibinfo {year} {2009})%
  \bibAnnoteFile{NoStop}{CorbozVidal09}%
\bibitem{VerstraeteCirac04PEPS}%
  \BibitemOpen
  \bibfield{author}{%
  \bibinfo {author} {\bibfnamefont{F.}~\bibnamefont{Verstraete}}\ and\ \bibinfo
  {author} {\bibfnamefont{J.~I.}\ \bibnamefont{Cirac}},\ }%
  \bibfield{journal}{%
  \Doi{10.1103/PhysRevA.70.060302}{\bibinfo {journal} {Phys. Rev. A}}\ }%
  \textbf{\bibinfo {volume} {70}},\ \bibinfo {pages} {060302} (\bibinfo {month}
  {Dec}\ \bibinfo {year} {2004})%
  \bibAnnoteFile{NoStop}{VerstraeteCirac04PEPS}%
\bibitem{GuLevinWen08}%
  \BibitemOpen
  \bibfield{author}{%
  \bibinfo {author} {\bibfnamefont{Z.-C.}\ \bibnamefont{Gu}}, \bibinfo {author}
  {\bibfnamefont{M.}~\bibnamefont{Levin}},\ and\ \bibinfo {author}
  {\bibfnamefont{X.-G.}\ \bibnamefont{Wen}},\ }%
  \bibfield{journal}{%
  \Doi{10.1103/PhysRevB.78.205116}{\bibinfo {journal} {Phys. Rev. B}}\ }%
  \textbf{\bibinfo {volume} {78}},\ \bibinfo {pages} {205116} (\bibinfo {month}
  {Nov}\ \bibinfo {year} {2008})%
  \bibAnnoteFile{NoStop}{GuLevinWen08}%
\bibitem{PizornVerstraete10}%
  \BibitemOpen
  \bibfield{author}{%
  \bibinfo {author} {\bibfnamefont{I.}~\bibnamefont{Pi\ifmmode~\check{z}\else
  \v{z}\fi{}orn}}\ and\ \bibinfo {author}
  {\bibfnamefont{F.}~\bibnamefont{Verstraete}},\ }%
  \bibfield{journal}{%
  \Doi{10.1103/PhysRevB.81.245110}{\bibinfo {journal} {Phys. Rev. B}}\ }%
  \textbf{\bibinfo {volume} {81}},\ \bibinfo {pages} {245110} (\bibinfo {month}
  {Jun}\ \bibinfo {year} {2010})%
  \bibAnnoteFile{NoStop}{PizornVerstraete10}%
\bibitem{Kraus10}%
  \BibitemOpen
  \bibfield{author}{%
  \bibinfo {author} {\bibfnamefont{C.~V.}\ \bibnamefont{Kraus}}, \bibinfo
  {author} {\bibfnamefont{N.}~\bibnamefont{Schuch}}, \bibinfo {author}
  {\bibfnamefont{F.}~\bibnamefont{Verstraete}},\ and\ \bibinfo {author}
  {\bibfnamefont{J.~I.}\ \bibnamefont{Cirac}},\ }%
  \bibfield{journal}{%
  \Doi{10.1103/PhysRevA.81.052338}{\bibinfo {journal} {Phys. Rev. A}}\ }%
  \textbf{\bibinfo {volume} {81}},\ \bibinfo {pages} {052338} (\bibinfo {month}
  {May}\ \bibinfo {year} {2010})%
  \bibAnnoteFile{NoStop}{Kraus10}%
\bibitem{Pinedaetal10}%
  \BibitemOpen
  \bibfield{author}{%
  \bibinfo {author} {\bibfnamefont{C.}~\bibnamefont{Pineda}}, \bibinfo {author}
  {\bibfnamefont{T.}~\bibnamefont{Barthel}},\ and\ \bibinfo {author}
  {\bibfnamefont{J.}~\bibnamefont{Eisert}},\ }%
  \bibfield{journal}{%
  \Doi{10.1103/PhysRevA.81.050303}{\bibinfo {journal} {Phys. Rev. A}}\ }%
  \textbf{\bibinfo {volume} {81}},\ \bibinfo {pages} {050303} (\bibinfo {month}
  {May}\ \bibinfo {year} {2010})%
  \bibAnnoteFile{NoStop}{Pinedaetal10}%
\bibitem{Bartheletal09}%
  \BibitemOpen
  \bibfield{author}{%
  \bibinfo {author} {\bibfnamefont{T.}~\bibnamefont{Barthel}}, \bibinfo
  {author} {\bibfnamefont{C.}~\bibnamefont{Pineda}},\ and\ \bibinfo {author}
  {\bibfnamefont{J.}~\bibnamefont{Eisert}},\ }%
  \bibfield{journal}{%
  \Doi{10.1103/PhysRevA.80.042333}{\bibinfo {journal} {Phys. Rev. A}}\ }%
  \textbf{\bibinfo {volume} {80}},\ \bibinfo {pages} {042333} (\bibinfo {month}
  {Oct}\ \bibinfo {year} {2009})%
  \bibAnnoteFile{NoStop}{Bartheletal09}%
\bibitem{Corbozorusetal10}%
  \BibitemOpen
  \bibfield{author}{%
  \bibinfo {author} {\bibfnamefont{P.}~\bibnamefont{Corboz}}, \bibinfo {author}
  {\bibfnamefont{R.}~\bibnamefont{Or\'us}}, \bibinfo {author}
  {\bibfnamefont{B.}~\bibnamefont{Bauer}},\ and\ \bibinfo {author}
  {\bibfnamefont{G.}~\bibnamefont{Vidal}},\ }%
  \bibfield{journal}{%
  \Doi{10.1103/PhysRevB.81.165104}{\bibinfo {journal} {Phys. Rev. B}}\ }%
  \textbf{\bibinfo {volume} {81}},\ \bibinfo {pages} {165104} (\bibinfo {month}
  {Apr}\ \bibinfo {year} {2010})%
  \bibAnnoteFile{NoStop}{Corbozorusetal10}%
\bibitem{SchuchetalPEPScomp07}%
  \BibitemOpen
  \bibfield{author}{%
  \bibinfo {author} {\bibfnamefont{N.}~\bibnamefont{Schuch}}, \bibinfo {author}
  {\bibfnamefont{M.~M.}\ \bibnamefont{Wolf}}, \bibinfo {author}
  {\bibfnamefont{F.}~\bibnamefont{Verstraete}},\ and\ \bibinfo {author}
  {\bibfnamefont{J.~I.}\ \bibnamefont{Cirac}},\ }%
  \bibfield{journal}{%
  \Doi{10.1103/PhysRevLett.98.140506}{\bibinfo {journal} {Phys. Rev. Lett.}}\
  }%
  \textbf{\bibinfo {volume} {98}},\ \bibinfo {pages} {140506} (\bibinfo {month}
  {Apr}\ \bibinfo {year} {2007})%
  \bibAnnoteFile{NoStop}{SchuchetalPEPScomp07}%
\bibitem{evenblyvidal09}%
  \BibitemOpen
  \bibfield{author}{%
  \bibinfo {author} {\bibfnamefont{G.}~\bibnamefont{Evenbly}}\ and\ \bibinfo
  {author} {\bibfnamefont{G.}~\bibnamefont{Vidal}},\ }%
  \bibfield{journal}{%
  \Doi{10.1103/PhysRevB.79.144108}{\bibinfo {journal} {Phys. Rev. B}}\ }%
  \textbf{\bibinfo {volume} {79}},\ \bibinfo {pages} {144108} (\bibinfo {month}
  {Apr}\ \bibinfo {year} {2009}),\
  \Eprint{http://arxiv.org/abs/arXiv:0707.1454v4}{arXiv:0707.1454v4}%
  \bibAnnoteFile{NoStop}{evenblyvidal09}%
\bibitem{silvietal09}%
  \BibitemOpen
  \bibfield{author}{%
  \bibinfo {author} {\bibfnamefont{P.}~\bibnamefont{Silvi}}, \bibinfo {author}
  {\bibfnamefont{V.}~\bibnamefont{Giovannetti}}, \bibinfo {author}
  {\bibfnamefont{S.}~\bibnamefont{Montangero}}, \bibinfo {author}
  {\bibfnamefont{M.}~\bibnamefont{Rizzi}}, \bibinfo {author}
  {\bibfnamefont{J.~I.}\ \bibnamefont{Cirac}},\ and\ \bibinfo {author}
  {\bibfnamefont{R.}~\bibnamefont{Fazio}},\ }%
  \enquote{\bibinfo {title} {Critical properties of homogeneous binary
  trees},}\  (\bibinfo {year} {2009}),\
  \Eprint{http://arxiv.org/abs/arXiv:0912.0466}{arXiv:0912.0466}%
  \bibAnnoteFile{NoStop}{silvietal09}%
\bibitem{Giovannettietal08}%
  \BibitemOpen
  \bibfield{author}{%
  \bibinfo {author} {\bibfnamefont{V.}~\bibnamefont{Giovannetti}}, \bibinfo
  {author} {\bibfnamefont{S.}~\bibnamefont{Montangero}},\ and\ \bibinfo
  {author} {\bibfnamefont{R.}~\bibnamefont{Fazio}},\ }%
  \bibfield{journal}{%
  \Doi{10.1103/PhysRevLett.101.180503}{\bibinfo {journal} {Phys. Rev. Lett.}}\
  }%
  \textbf{\bibinfo {volume} {101}},\ \bibinfo {pages} {180503} (\bibinfo
  {month} {Oct}\ \bibinfo {year} {2008})%
  \bibAnnoteFile{NoStop}{Giovannettietal08}%
\bibitem{Pfeiferetal09}%
  \BibitemOpen
  \bibfield{author}{%
  \bibinfo {author} {\bibfnamefont{R.~N.~C.}\ \bibnamefont{Pfeifer}}, \bibinfo
  {author} {\bibfnamefont{G.}~\bibnamefont{Evenbly}},\ and\ \bibinfo {author}
  {\bibfnamefont{G.}~\bibnamefont{Vidal}},\ }%
  \bibfield{journal}{%
  \Doi{10.1103/PhysRevA.79.040301}{\bibinfo {journal} {Phys. Rev. A}}\ }%
  \textbf{\bibinfo {volume} {79}},\ \bibinfo {pages} {040301} (\bibinfo {month}
  {Apr}\ \bibinfo {year} {2009}),\
  \Eprint{http://arxiv.org/abs/arXiv:0810.0580}{arXiv:0810.0580}%
  \bibAnnoteFile{NoStop}{Pfeiferetal09}%
\bibitem{Walker91}%
  \BibitemOpen
  \bibfield{author}{%
  \bibinfo {author} {\bibfnamefont{K.}~\bibnamefont{Walker}},\ }%
  \enquote{\bibinfo {title} {On {W}itten's 3-manifold invariants},}\  (\bibinfo
  {year} {1991}),\ \bibinfo {note} {available at http://canyon23.net/math/}%
  \bibAnnoteFile{NoStop}{Walker91}%
\bibitem{KoeKupRei}%
  \BibitemOpen
  \bibfield{author}{%
  \bibinfo {author} {\bibfnamefont{R.}~\bibnamefont{K\"onig}}, \bibinfo
  {author} {\bibfnamefont{G.}~\bibnamefont{Kuperberg}},\ and\ \bibinfo {author}
  {\bibfnamefont{B.~W.}\ \bibnamefont{Reichardt}},\ }%
  \enquote{\bibinfo {title} {Quantum computation with {T}uraev-{V}iro codes},}\
  \Eprint{http://arxiv.org/abs/arXiv:1002.2816}{arXiv:1002.2816}%
  \bibAnnoteFile{NoStop}{KoeKupRei}%
\bibitem{Bondersonetal09}%
  \BibitemOpen
  \bibfield{author}{%
  \bibinfo {author} {\bibfnamefont{P.}~\bibnamefont{Bonderson}}, \bibinfo
  {author} {\bibfnamefont{M.}~\bibnamefont{Freedman}},\ and\ \bibinfo {author}
  {\bibfnamefont{C.}~\bibnamefont{Nayak}},\ }%
  \bibfield{journal}{%
  \Doi{DOI: 10.1016/j.aop.2008.09.009}{\bibinfo {journal} {Annals of Physics}}\
  }%
  \textbf{\bibinfo {volume} {324}},\ \bibinfo {pages} {787 } (\bibinfo {year}
  {2009}),\ \Eprint{http://arxiv.org/abs/0808.1933}{0808.1933}%
  \bibAnnoteFile{NoStop}{Bondersonetal09}%
\bibitem{Gilcollective09}%
  \BibitemOpen
  \bibfield{author}{%
  \bibinfo {author} {\bibfnamefont{C.}~\bibnamefont{Gils}}, \bibinfo {author}
  {\bibfnamefont{E.}~\bibnamefont{Ardonne}}, \bibinfo {author}
  {\bibfnamefont{S.}~\bibnamefont{Trebst}}, \bibinfo {author}
  {\bibfnamefont{A.~W.~W.}\ \bibnamefont{Ludwig}}, \bibinfo {author}
  {\bibfnamefont{M.}~\bibnamefont{Troyer}},\ and\ \bibinfo {author}
  {\bibfnamefont{Z.}~\bibnamefont{Wang}},\ }%
  \bibfield{journal}{%
  \Doi{10.1103/PhysRevLett.103.070401}{\bibinfo {journal} {Phys. Rev. Lett.}}\
  }%
  \textbf{\bibinfo {volume} {103}},\ \bibinfo {pages} {070401} (\bibinfo
  {month} {Aug}\ \bibinfo {year} {2009})%
  \bibAnnoteFile{NoStop}{Gilcollective09}%
\bibitem{Lahtinen08}%
  \BibitemOpen
  \bibfield{author}{%
  \bibinfo {author} {\bibfnamefont{V.}~\bibnamefont{Lahtinen}}, \bibinfo
  {author} {\bibfnamefont{G.}~\bibnamefont{Kells}}, \bibinfo {author}
  {\bibfnamefont{A.}~\bibnamefont{Carollo}}, \bibinfo {author}
  {\bibfnamefont{T.}~\bibnamefont{Stitt}}, \bibinfo {author}
  {\bibfnamefont{J.}~\bibnamefont{Vala}},\ and\ \bibinfo {author}
  {\bibfnamefont{J.~K.}\ \bibnamefont{Pachos}},\ }%
  \bibfield{journal}{%
  \Doi{DOI: 10.1016/j.aop.2007.12.009}{\bibinfo {journal} {Annals of Physics}}\
  }%
  \textbf{\bibinfo {volume} {323}},\ \bibinfo {pages} {2286 } (\bibinfo {year}
  {2008}),\ ISSN \bibinfo {issn} {0003-4916}%
  \bibAnnoteFile{NoStop}{Lahtinen08}%
\bibitem{Baraban09}%
  \BibitemOpen
  \bibfield{author}{%
  \bibinfo {author} {\bibfnamefont{M.}~\bibnamefont{Baraban}}, \bibinfo
  {author} {\bibfnamefont{G.}~\bibnamefont{Zikos}}, \bibinfo {author}
  {\bibfnamefont{N.}~\bibnamefont{Bonesteel}},\ and\ \bibinfo {author}
  {\bibfnamefont{S.~H.}\ \bibnamefont{Simon}},\ }%
  \bibfield{journal}{%
  \Doi{10.1103/PhysRevLett.103.076801}{\bibinfo {journal} {Phys. Rev. Lett.}}\
  }%
  \textbf{\bibinfo {volume} {103}},\ \bibinfo {pages} {076801} (\bibinfo
  {month} {Aug}\ \bibinfo {year} {2009})%
  \bibAnnoteFile{NoStop}{Baraban09}%
\bibitem{cheng09}%
  \BibitemOpen
  \bibfield{author}{%
  \bibinfo {author} {\bibfnamefont{M.}~\bibnamefont{Cheng}}, \bibinfo {author}
  {\bibfnamefont{R.~M.}\ \bibnamefont{Lutchyn}}, \bibinfo {author}
  {\bibfnamefont{V.}~\bibnamefont{Galitski}},\ and\ \bibinfo {author}
  {\bibfnamefont{S.}~\bibnamefont{Das~Sarma}},\ }%
  \bibfield{journal}{%
  \Doi{10.1103/PhysRevLett.103.107001}{\bibinfo {journal} {Phys. Rev. Lett.}}\
  }%
  \textbf{\bibinfo {volume} {103}},\ \bibinfo {pages} {107001} (\bibinfo
  {month} {Aug}\ \bibinfo {year} {2009})%
  \bibAnnoteFile{NoStop}{cheng09}%
\bibitem{Montangeroetal09}%
  \BibitemOpen
  \bibfield{author}{%
  \bibinfo {author} {\bibfnamefont{S.}~\bibnamefont{Montangero}}, \bibinfo
  {author} {\bibfnamefont{M.}~\bibnamefont{Rizzi}}, \bibinfo {author}
  {\bibfnamefont{V.}~\bibnamefont{Giovannetti}},\ and\ \bibinfo {author}
  {\bibfnamefont{R.}~\bibnamefont{Fazio}},\ }%
  \bibfield{journal}{%
  \Doi{10.1103/PhysRevB.80.113103}{\bibinfo {journal} {Phys. Rev. B}}\ }%
  \textbf{\bibinfo {volume} {80}},\ \bibinfo {pages} {113103} (\bibinfo {month}
  {Sep}\ \bibinfo {year} {2009}),\
  \Eprint{http://arxiv.org/abs/arXiv:0810.1414}{arXiv:0810.1414}%
  \bibAnnoteFile{NoStop}{Montangeroetal09}%
\bibitem{AguadoVidal08}%
  \BibitemOpen
  \bibfield{author}{%
  \bibinfo {author} {\bibfnamefont{M.}~\bibnamefont{Aguado}}\ and\ \bibinfo
  {author} {\bibfnamefont{G.}~\bibnamefont{Vidal}},\ }%
  \bibfield{journal}{%
  \Doi{10.1103/PhysRevLett.100.070404}{\bibinfo {journal} {Phys. Rev. Lett.}}\
  }%
  \textbf{\bibinfo {volume} {100}},\ \bibinfo {pages} {070404} (\bibinfo
  {month} {Feb}\ \bibinfo {year} {2008})%
  \bibAnnoteFile{NoStop}{AguadoVidal08}%
\bibitem{MajumdarGosh69}%
  \BibitemOpen
  \bibfield{author}{%
  \bibinfo {author} {\bibfnamefont{C.~K.}\ \bibnamefont{Majumdar}}\ and\
  \bibinfo {author} {\bibfnamefont{D.~K.}\ \bibnamefont{Gosh}},\ }%
  \bibfield{journal}{%
  \bibinfo {journal} {J. Math. Phys.}\ }%
  \textbf{\bibinfo {volume} {10}},\ \bibinfo {pages} {1388} (\bibinfo {year}
  {1969})%
  \bibAnnoteFile{NoStop}{MajumdarGosh69}%
\bibitem{trebstetalcollective08}%
  \BibitemOpen
  \bibfield{author}{%
  \bibinfo {author} {\bibfnamefont{S.}~\bibnamefont{Trebst}}, \bibinfo {author}
  {\bibfnamefont{E.}~\bibnamefont{Ardonne}}, \bibinfo {author}
  {\bibfnamefont{A.}~\bibnamefont{Feiguin}}, \bibinfo {author}
  {\bibfnamefont{D.~A.}\ \bibnamefont{Huse}}, \bibinfo {author}
  {\bibfnamefont{A.~W.~W.}\ \bibnamefont{Ludwig}},\ and\ \bibinfo {author}
  {\bibfnamefont{M.}~\bibnamefont{Troyer}},\ }%
  \bibfield{journal}{%
  \Doi{10.1103/PhysRevLett.101.050401}{\bibinfo {journal} {Phys. Rev. Lett.}}\
  }%
  \textbf{\bibinfo {volume} {101}},\ \bibinfo {pages} {050401} (\bibinfo
  {month} {Jul}\ \bibinfo {year} {2008}),\
  \Eprint{http://arxiv.org/abs/arXiv:0801.4602}{arXiv:0801.4602}%
  \bibAnnoteFile{NoStop}{trebstetalcollective08}%
\bibitem{andrewsbaxterforrester83}%
  \BibitemOpen
  \bibfield{author}{%
  \bibinfo {author} {\bibfnamefont{G.~E.}\ \bibnamefont{Andrews}}, \bibinfo
  {author} {\bibfnamefont{R.~J.}\ \bibnamefont{Baxter}},\ and\ \bibinfo
  {author} {\bibfnamefont{P.~J.}\ \bibnamefont{Forrester}},\ }%
  \bibfield{journal}{%
  \bibinfo {journal} {Journal of Statistical Physics}\ }%
  \textbf{\bibinfo {volume} {35}},\ \bibinfo {pages} {193} (\bibinfo {month}
  {May}\ \bibinfo {year} {1984})%
  \bibAnnoteFile{NoStop}{andrewsbaxterforrester83}%
\bibitem{LudwPoil10}%
  \BibitemOpen
  \bibfield{author}{%
  \bibinfo {author} {\bibfnamefont{A.~W.~W.}\ \bibnamefont{Ludwig}}, \bibinfo
  {author} {\bibfnamefont{D.}~\bibnamefont{Poilblanc}}, \bibinfo {author}
  {\bibfnamefont{S.}~\bibnamefont{Trebst}},\ and\ \bibinfo {author}
  {\bibfnamefont{M.}~\bibnamefont{Troyer}}\ }%
  \Eprint{http://arxiv.org/abs/arXiv:1003.3453}{arXiv:1003.3453}%
  \bibAnnoteFile{NoStop}{LudwPoil10}%
\bibitem{KoeReiVid09}%
  \BibitemOpen
  \bibfield{author}{%
  \bibinfo {author} {\bibfnamefont{R.}~\bibnamefont{K\"onig}}, \bibinfo
  {author} {\bibfnamefont{B.~W.}\ \bibnamefont{Reichardt}},\ and\ \bibinfo
  {author} {\bibfnamefont{G.}~\bibnamefont{Vidal}},\ }%
  \bibfield{journal}{%
  \Doi{10.1103/PhysRevB.79.195123}{\bibinfo {journal} {Phys. Rev. B}}\ }%
  \textbf{\bibinfo {volume} {79}},\ \bibinfo {pages} {195123} (\bibinfo {month}
  {May}\ \bibinfo {year} {2009})%
  \bibAnnoteFile{NoStop}{KoeReiVid09}%
\bibitem{ShiDuanVidal06}%
  \BibitemOpen
  \bibfield{author}{%
  \bibinfo {author} {\bibfnamefont{Y.-Y.}\ \bibnamefont{Shi}}, \bibinfo
  {author} {\bibfnamefont{L.-M.}\ \bibnamefont{Duan}},\ and\ \bibinfo {author}
  {\bibfnamefont{G.}~\bibnamefont{Vidal}},\ }%
  \bibfield{journal}{%
  \Doi{10.1103/PhysRevA.74.022320}{\bibinfo {journal} {Phys. Rev. A}}\ }%
  \textbf{\bibinfo {volume} {74}},\ \bibinfo {pages} {022320} (\bibinfo {month}
  {Aug}\ \bibinfo {year} {2006})%
  \bibAnnoteFile{NoStop}{ShiDuanVidal06}%
\bibitem{TagliazzoEvenblyVidal09}%
  \BibitemOpen
  \bibfield{author}{%
  \bibinfo {author} {\bibfnamefont{L.}~\bibnamefont{Tagliacozzo}}, \bibinfo
  {author} {\bibfnamefont{G.}~\bibnamefont{Evenbly}},\ and\ \bibinfo {author}
  {\bibfnamefont{G.}~\bibnamefont{Vidal}},\ }%
  \bibfield{journal}{%
  \Doi{10.1103/PhysRevB.80.235127}{\bibinfo {journal} {Phys. Rev. B}}\ }%
  \textbf{\bibinfo {volume} {80}},\ \bibinfo {pages} {235127} (\bibinfo {month}
  {Dec}\ \bibinfo {year} {2009})%
  \bibAnnoteFile{NoStop}{TagliazzoEvenblyVidal09}%
\bibitem{Murgetaltree10}%
  \BibitemOpen
  \bibfield{author}{%
  \bibinfo {author} {\bibfnamefont{V.}~\bibnamefont{{Murg}}}, \bibinfo {author}
  {\bibfnamefont{{\"O}.}~\bibnamefont{{Legeza}}}, \bibinfo {author}
  {\bibfnamefont{R.~M.}\ \bibnamefont{{Noack}}},\ and\ \bibinfo {author}
  {\bibfnamefont{F.}~\bibnamefont{{Verstraete}}}\ }%
  \Eprint{http://arxiv.org/abs/arXiv:1006.3095}{arXiv:1006.3095}%
  \bibAnnoteFile{NoStop}{Murgetaltree10}%
\bibitem{Pfeiferetal10}%
  \BibitemOpen
  \bibfield{author}{%
  \bibinfo {author} {\bibfnamefont{R.}~\bibnamefont{Pfeifer}}, \bibinfo
  {author} {\bibfnamefont{P.}~\bibnamefont{Corboz}}, \bibinfo {author}
  {\bibfnamefont{O.}~\bibnamefont{Buerschaper}}, \bibinfo {author}
  {\bibfnamefont{M.}~\bibnamefont{Aguado}}, \bibinfo {author}
  {\bibfnamefont{M.}~\bibnamefont{Troyer}},\ and\ \bibinfo {author}
  {\bibfnamefont{G.}~\bibnamefont{Vidal}},\ }%
  \enquote{\bibinfo {title} {Simulation of anyons using entanglement
  renormalisation},}\
  \Eprint{http://arxiv.org/abs/arXiv:1006.3532v1}{arXiv:1006.3532v1}%
  \bibAnnoteFile{NoStop}{Pfeiferetal10}%
\end{thebibliography}%
\end{document}